\documentclass[pre,aps,twocolumn,nopacs,floatfix,superscriptaddress]{revtex4-1}

\usepackage{amsfonts,amssymb,amsmath,array}
\usepackage{floatrow}
\usepackage{graphicx}
\usepackage[labelfont=bf,labelformat=simple,subrefformat=simple]{subfig}
\usepackage[font=small]{caption}
\usepackage{amsbsy}
\usepackage{color}
\usepackage{hyperref} 
\usepackage{enumerate}
\usepackage{mathtools}
\usepackage{lipsum} 
\usepackage{cleveref} 
\usepackage{nicematrix}
\usepackage{bbold}

\begin{document}

\title{Thermal Fluctuations For a Three-Beads Swimmer}

\author{R. Ferretta}
\affiliation{Department of Physics, University of Rome Sapienza, P.le Aldo Moro 2, 00185, Rome, Italy}
\affiliation{Institute for Complex Systems - CNR, P.le Aldo Moro 2, 00185, Rome, Italy}
\author{R. Di Leonardo}
\affiliation{Department of Physics, University of Rome Sapienza, P.le Aldo Moro 2, 00185, Rome, Italy}
\affiliation{Nanotec - CNR, P.le Aldo Moro 2, 00185, Rome, Italy}
\author{A. Puglisi}
\affiliation{Department of Physics, University of Rome Sapienza, P.le Aldo Moro 2, 00185, Rome, Italy}\affiliation{Institute for Complex Systems - CNR, P.le Aldo Moro 2, 00185, Rome, Italy}
\affiliation{INFN, University of Rome Tor Vergata, Via della Ricerca Scientifica 1, 00133, Rome, Italy}

\begin{abstract}
We discuss a micro-swimmer model made of three spheres actuated by an internal active time-periodic force, tied by an elastic potential and submitted to hydrodynamic interactions with thermal noise. The dynamical approach we use, replacing the more common kinetic one, unveils the instability of the original model and the need of a confining potential to prevent the evaporation of the swimmer. We investigate the effect of the main parameters of the model, such as the frequency and phase difference of the periodic active force, the stiffness of the confining potential, the length of the swimmer and the temperature and viscosity of the fluid. Our observables of interest are the averages of the swim velocity, of the energy consumption rate, the diffusion coefficient and the swimming precision, which is limited by the energy consumption through the celebrated Thermodynamic Uncertainty Relations. An optimum for velocity and precision is found for an intermediate frequency. Reducing the potential stiffness, the viscosity or the length, is also beneficial for the swimming performance, but these parameters are limited by the consistency of the model. Analytical approximation for many of the interesting observables is obtained for small deformations of the swimmer. We also discuss the efficiency of the swimmer in terms of its maximum precision and of the hydrodynamic, or Lighthill, criterion, and how they are connected.
\end{abstract}

\maketitle

\section{Introduction}

The physics of active particles is a prominent application of non-equilibrium statistical mechanics~\cite{marchetti2013hydrodynamics,bechinger2016active}. Several interesting challenges in this field belong to the category of collective phenomena and emergent order, such as motility induced phase separation and flocking, which are usually investigated theoretically through simplified models where each particle has no internal structure but only a "self-propulsion" force with simple properties~\cite{fily2012athermal,cates2015motility,cavagna2018physics}. However a fascinating aspect of the physics of active systems is the mechanism of self-propulsion itself~\cite{elgeti2015physics}. Self-propulsion  typically originates from complex mechanisms and - even when considering microscopic systems - may be based upon the conspiracy of a  number of sub-units, such as molecular motors~\cite{riedel2007molecular}. For this reason, in some cases, statistical physics becomes useful also for understanding what happens in a single self-propelling unit~\cite{maggi2022thermodynamic,yasuda23}.

Swimming at the micro-scale requires strategies to break time-reversal invariance in the absence of relevant inertia, i.e. circumventing the "scallop" theorem~\cite{purcell1977life,lauga2011life}. A simple mechanism is to perform a periodic non-reciprocal change of shape: the swimmer exploits some source of energy in order to make internal changes and explore a sequence of configurations which is periodic without being time-symmetric, the minimal example being $A \to B \to C \to A$. This sequence becomes a limit cycle in the case of a system whose shape is characterised by continuous parameters. Biological systems offer a plethora of examples for such a behavior, the most common instances being self-propelling cells equipped with pushing or pulling flagella or cilia, e.g. motile cells, sperms, bacteria and {\em C. reinhardtii} algae~\cite{klindt2015flagellar}. The sperm tail, for instance, excited by a myriads of molecular motors innervating the axoneme, and under the constraints of elasticity and hydrodynamics, displays a wave deformation continuously travelling from the head to the end tip, that pushes forward the sperm~\cite{maggi2022thermodynamic}. More simple models have been introduced in the literature, in order to pinpoint the essential mechanism behind the strategies invented by nature and potentially useful for the design of artificial swimming micro-machines. Historically the ancestor of simplified, analytically treatable, models, is the Lighthill squirmer~\cite{lighthill1952squirming}. A more recent model is the so-called Three-Beads swimmer, whose analytical treatment is even more manageable, at least in some limits~\cite{najafi2004simple,golestanian2008analytic}. This model, which has also been realised experimentally, has been rarely studied in the presence of noise. Noise is not only a realistic and interesting perturbation to be added to the limit cycle of a micro-swimmer, but may offer a new test ground for the so-called Thermodynamic Uncertainty Relations (TUR) which establish a bound to the signal-to-noise ratio based upon the energy consumption of the system~\cite{Barato2015,Gingrich2016}. 

Here we analyse the behavior of the three-beads swimmer when it feels also the effect of thermal noise. For the purpose of this study we modified the original Three-Beads swimmer model in order to be described dynamically instead of kinetically. The result of this change of perspective unveils a crucial problem of the original system, i.e. the lack of stability of its limit cycle. To solve this problem we have introduced an additional elastic force keeping together the three beads, and avoiding the observed instability. We underline that noise for a real or model micro-swimmer may have two distinct origins: thermal noise originated in the molecular agitation of the surrounding fluid or of the swimmer's body, and active noise originated in the randomness of the dynamics of the sub-units, such as molecular motors, actuating the swimmer. It is clear that, ultimately, active noise can be traced back to thermal noise, but in a mesoscopic model it would be distinct from it and, for the purpose of the present study, it is ignored. A discussion of the effect of stochastic active forces on the three-beads swimmer model can be found in~\cite{golestanian2009stochastic}. The effect of noise induced by fluctuations in the dynamic of molecular motors actuating the swimmer (such as in a biological microscopic flagellum) is the subject of a future publication~\cite{maggi2022thermodynamic}. After introducing the model and showing the results of its numerical solutions, we also pose the question of the efficiency in terms of different definitions, one being the distance from the bound dictated by the TUR. We show that spatial correlations in the noise due to hydrodynamic interactions (originated in the fluid momentum conservation) modify such an efficiency, a fact which is usually neglected in models of engines or swimmer constituted by several coupled motors~\cite{izumida2016energetics,lee2018thermodynamic,hasegawa2018thermodynamics,hong2020thermodynamic,zhang2020energy,leighton2022performance,maggi2022thermodynamic}. The consequence also for other figure of merits, such as the Lighthill efficiency, are discussed.

The structure of the paper is the following. In Section~\ref{sec:model} we describe the model, briefly review previous studies concerning it, explain the differences between a kinematic and dynamic approach, introduce the properties of hydrodynamic thermal noise and finally describe in details the observables of interest. In Section~\ref{sec:numerical} we report the results of the numerical simulations. In Section~\ref{sec:tur} we discuss the Thermodynamic Uncertainty Relation and the Lighthill efficiency. In Section~\ref{sec:analyt} we discuss analytical results in the linearised limit for the velocity of the swimmer and sketch some approximate formula for its diffusivity and noise-to-signal ratio. Finally in Section~\ref{sec:concl} we draw conclusions and discuss perspective.

\section{The model}
\label{sec:model}

The Three-Beads model we consider here is represented by the equations of motion for three spherical particles (index $i=1,2,3$) of radius $a$ immersed in a fluid of viscosity $\eta$ - with no rotation or internal degrees of freedom - each particle $i$ described by velocity vector ${\bf v}_i$ and submitted to internal force vector ${\bf f}_i$ and external noise vector ${\bf f}_i^R$. The condition that the force vector is internal reads $\sum_i {\bf f}_i=0$. The system, in the Stokesian regime (high viscosity, negligible inertia), is described by the instantaneous balance between viscous drags and applied  forces, which can be written in the usual form:
\begin{equation}
\label{eq:langintegr}
    {\bf v}_i = \sum_{j}H_{ij}({\bf r}_i-{\bf r}_j)({\bf f}_j + {\bf f}_j^R)\,.
\end{equation}
Here $H_{ij}({\bf r})=\frac{1}{6\pi \eta a}\left[\delta_{ij}+(1-\delta_{ij})\frac{3a}{4 r}\left(\mathbb{1}+{\bf \hat r}{\bf \hat r}\right)\right]$ is the Oseen mobility tensor which comes from the solution of the Stokes equation. We recall that the Oseen tensor describes the part of the sphere-sphere hydrodynamic interaction that decreases inversely as the first power of the distance between the spheres and becomes inaccurate at short ranges~\cite{kynch1959slow}.

The three beads are allowed to move only along one direction, say $\hat{x}$, and all the forces act only along that direction~\footnote{This is of course not realistic in the case of thermal noise, but we adopted such an assumption in order too keep the problem simple.}. 
Assuming that this constraint does not affect the validity of Eqs.~\eqref{eq:langintegr}, the $x_i$ coordinates obey the following relation between velocities and forces:
\begin{equation}
\label{eq:genericmodel}
    \frac{dx_i(t)}{dt} = \sum_j T_{ij}({\underline x})[F_j({\underline x},t)+F_j^R({\underline x},t)],
\end{equation}
where $T_{ij}$ represents the mobility coefficient coupling the $x$-component of the force acting on particle $j$ and the $x$-component of the velocity of the particle $i$, while the $F_j$ represent the $x$-components of the forces ${\bf f}_j$. We use $\underline x$ and $\underline F$ to represent the 3-ple $\{x_1,x_2,x_3\}$ and $\{F_1,F_2,F_3\}$. From the Oseen tensor, the matrix ${\bf T}$ is obtained by the relation $T_{ij}(\underline x) = H_{ij}({\bf r}_i-{\bf r}_j)_{xx}$. Therefore it takes the form
\begin{equation}
\label{eq:oseentensor1}{\bf T}=
    \begin{pNiceMatrix}
        \frac{1}{6\pi\eta a} & \frac{1}{4\pi\eta L_1} & \frac{1}{4\pi\eta (L_1+L_2)} \\
        \frac{1}{4\pi\eta L_1} & \frac{1}{6\pi\eta a} & \frac{1}{4\pi\eta L_2} \\
        \frac{1}{4\pi\eta (L_1+L_2)} & \frac{1}{4\pi\eta L_2} & \frac{1}{6\pi\eta a}
    \end{pNiceMatrix}\,.
\end{equation}
where $L_1=x_2-x_1$ and $L_2=x_3-x_2$.

The deterministic forces $\underline F$ are the sum of internal conservative (a confining potential to avoid instabilities) and internal non-conservative  forces (a time-dependent periodic perturbation that conserves total momentum, also denoted as "active force"), i.e.  $\underline{F}=\underline{F}^{act}+\underline{F}^{pot}$, all detailed below.

\subsection{Original setup}
\begin{figure}
    \centering
    \includegraphics[width=\textwidth]{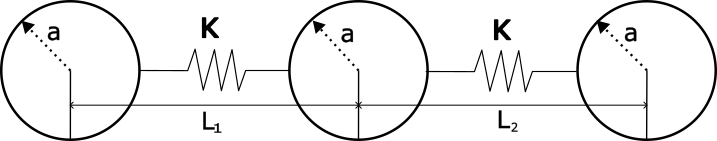} 
    \caption{Sketch of the swimmer model made of three spheres of radius $a$, distances $L_1$ and $L_2$ and springs with stiffness $K$.}
    \label{NG_model_sketch}
\end{figure}
The original three-beads model was described by the same equations in~\eqref{eq:genericmodel}, without noise i.e. $F_j^R \equiv 0$, and with $F_j({\underline x},t)$ reduced to only internal non-conservative forces: $F_j({\underline x},t)=F^{act}_j(t)$ with
\begin{equation}
\label{forcefree}
    F^{act}_1(t)\,+\,F^{act}_2(t)\,+\,F^{act}_3(t)\, = \,0 .
\end{equation}


The instantaneous velocity of the center of mass $V = \frac{1}{3}\left(v_1\,+\,v_2\,+\,v_3 \right)$ was computed in the limit of small deformation of the swimmer, i.e. by assuming
\begin{equation}
    L_1=l_1\,+\,u_1\,, \qquad L_2=l_2\,+\,u_2\,,
\end{equation}
and, for small $u_1,u_2$ the following formula was obtained
\begin{equation}
\label{V0expr}
    v=\overline{V} = \frac{\alpha}{2}\overline{\left(u_1\dot{u}_2\,-\,u_2\dot{u}_1 \right)}\,, 
\end{equation}
with
\begin{equation} \label{alphadef}
    \alpha = \frac{a}{3}\Big[\frac{1}{l_1^2}\,+\,\frac{1}{l_2^2}\,-\,\frac{1}{(l_1+l_2)^2} \Big]\,.
\end{equation}
For instance, when $u_{1,2}$ obey the following oscillation laws
\begin{equation}
\label{harmdefu}
    \begin{split}
        u_1(t) = d_1\,\cos{(\omega t\,+\,\phi_1)}\,,\\
        u_2(t) = d_2\,\cos{(\omega t\,+\,\phi_2)}\,.
    \end{split}
\end{equation}
one has that the average swimming velocity reads
\begin{equation}
\label{eq:avg_vel_gol_2008}
    v=\overline{V} = \frac{\alpha}{2}\,d_1\,d_2\,\omega \sin{(\phi_1-\phi_2)}.
\end{equation}
It is important to realise that the above formula are based upon the knowledge of $u_i(t)$ and not from the knowledge of the forces $F_i(t)$. This is what we call a kinematic approach and implies that the distances among the particles are prescribed by definition. In the rest of the paper we change the point of view and start from the knowledge of the forces.

\subsection{Difference between kinematic and dynamical approach: instability of the relative distances}

In the rest of the paper we adopt a dynamic point of view, i.e. we set the forces acting on the first and third particles:
\begin{equation}
\label{eq:forcesform}
    F^{act}_i(t) = F_0\,\cos{(\omega t+\phi_i)}\,, \qquad i=1,3\,,
\end{equation}
with the constraint in Eq.~\eqref{forcefree}. The reason for this different, in principle more complicate, approach is that when out-of-equilibrium, i.e. under the presence of non-conservative forces, the properties of the  fluctuations of the relative positions of the particles - due to interaction with the molecules of the fluid - are not known in general. On the contrary, one may reasonably describe the effects of these interactions in terms of known fluctuating forces, that is exactly what we call $F_i^R$, see later for explicit formula.

Interestingly, when changing the point of view from kinematic to dynamic, it is possible to
unveil a weakness of the original model, i.e. an intrinsic instability toward the evaporation of the swimmer: the particles relative distances do not remain limited and the velocity of the swimmer - for this reason - tends to vanish.

\begin{figure}[htb]
    \centering
    \sidesubfloat[]
        {\includegraphics[width=0.95\textwidth]{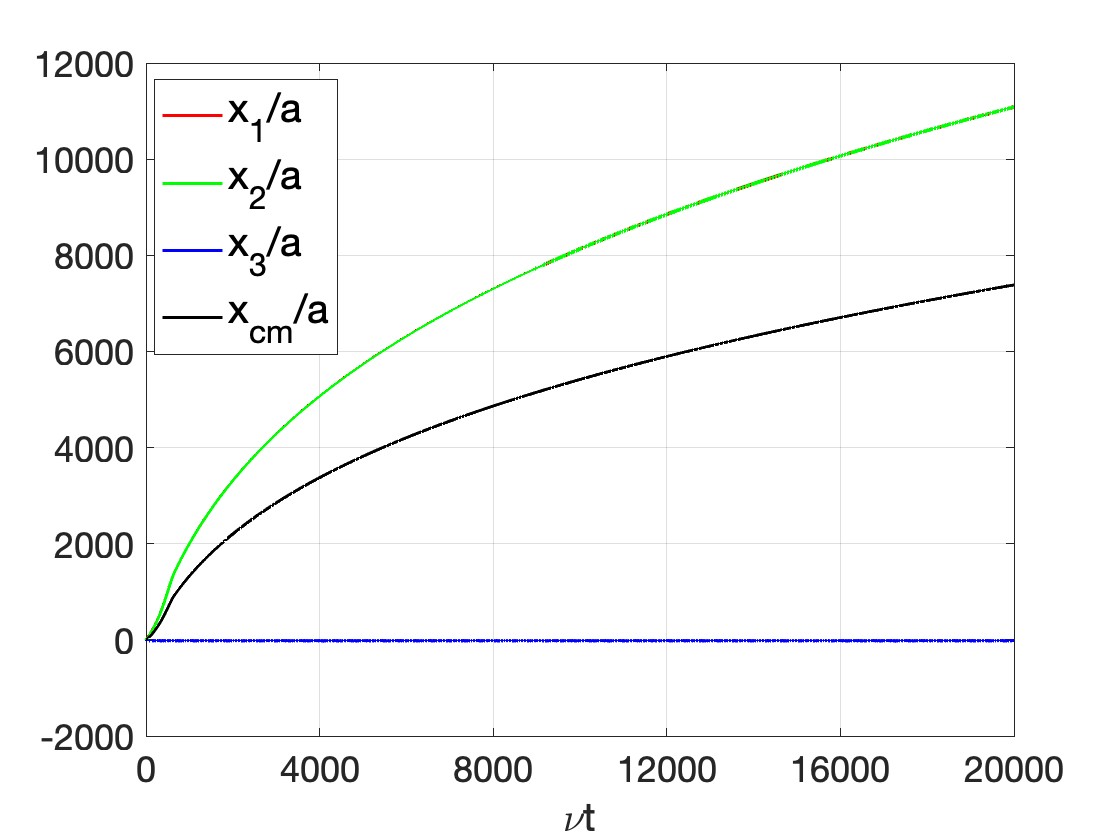}
        \label{}}
    \\
    \sidesubfloat[]
        {\includegraphics[width=0.95\textwidth]{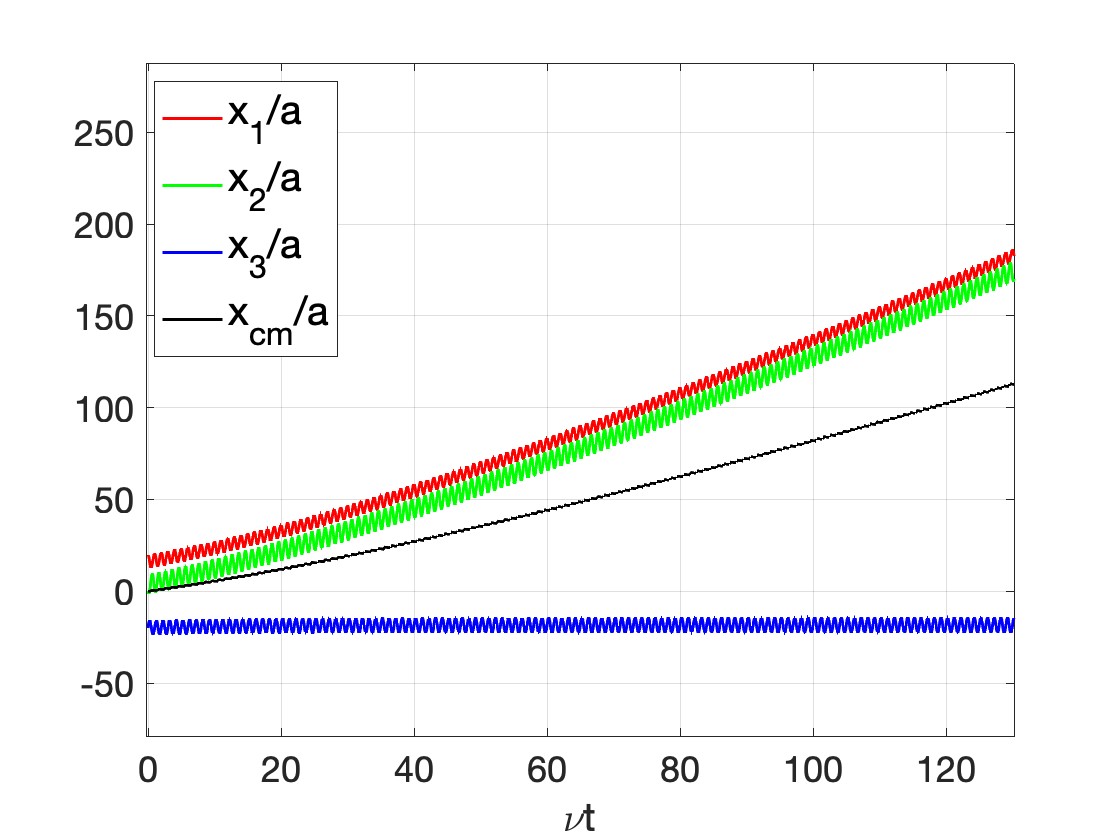}
        \label{}}
    \caption{Positions of the three spheres when there is no noise and no elastic potential. Initial arms length $x_2(0)-x_1(0)=x_3(0)-x_2(0)=20$, spheres' radius $a=1$, $\eta=1$,  $\Delta\phi=\pi/2$, $\omega=2\pi/50$, $F_0=10$.}
    \label{fig:instability}
\end{figure}
An example of the numerical solution of the model in Eq.~\eqref{eq:genericmodel} with zero temperature ($F^R=0$) and $F_i=F^{act}_i(t)$ expressed by Eqs.~\eqref{eq:forcesform} is shown in Fig.~\ref{fig:instability}. In Section~\ref{sec:analyt} we discuss the linear stability of the limit cycle (in the plane $L_1,L_2$), showing the presence of a positive eigenvalue for all the values of the oscillating forces. In the whole paper we set $\phi_1=\Delta \phi$ and $\phi_3=0$ (the force $F_2(t)$ is entirely deduced by the knowledge of $F_1$ and $F_3$ through the force constraint of internal forces, Eq.~\eqref{forcefree}).

For this reason we introduce a harmonic attractive potential among each couple of adjacent spheres, i.e.
\begin{equation}
    {\underline F}(\underline x,t)={\underline F}^{act}(t)+{\underline F}^{pot}(\underline x)
\end{equation}
with
\begin{align}
F^{pot}_1&=K(x_2-x_1-l_1)\\
F^{pot}_2&=K(x_3-x_2-l_2)-K(x_2-x_1-l_1)\\
F^{pot}_3&=-K(x_3-x_2-l_2),
\end{align}
with $K$ the elastic constant and $l_1,l_2$ the lengths at rest of the harmonic springs.
\begin{figure}[H]
    \centering
    \includegraphics[width=\textwidth]{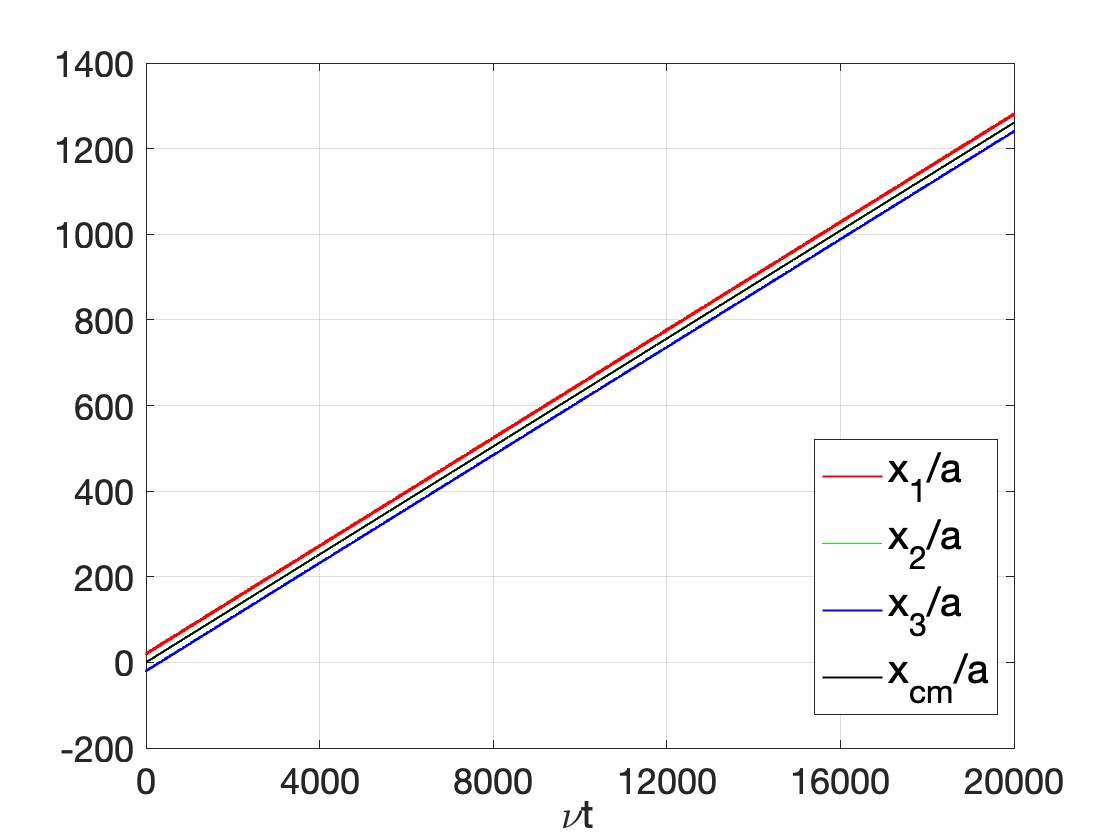}
    \caption{Positions of the three spheres with the elastic potential and without noise. Parameters: $l_1=l_2=L=20$, $a=1$, $\eta=1$,  $\Delta\phi=\pi/2$, $\omega=2\pi/50$, $K=2$, $F_0=10$.}
    \label{fig:stable}
\end{figure}
When $K>0$ we observe that the instability disappear and the relative positions of the $3$ spheres remain limited, see Fig.~\ref{fig:stable}.

It is interesting to note that an experimental realisation of the three-beads swimmer has been obtained by applying a time-modulated magnetic field on permanently magnetised spheres {\em linked} by elastic rods~\cite{box2017motion}. A theoretical study of a swimming system of beads with the presence of elastic confining potential can be found in~\cite{yasuda2017elastic}, by prescribing a cyclical law for the length at rest of the potentials, $l_1,l_2$ and ignoring the effect of thermal noise as well as  energetic considerations. In the rest of the paper we assume that the lengths at rest are equal ("symmetric swimmer") $l_1=l_2=L$, but in the Appendix we consider the most general case $l_1 \neq l_2$.

\subsection{Details of the noise}

A passive colloidal particle moves under the stochastic effect of the molecules of the surrounding fluid, i.e. what is called thermal noise and is described by the physics of Brownian motion~\cite{gardiner1985handbook}. When more colloidal particles are present, spatial correlations appear in the thermal forces acting on each particle~\cite{doi1988theory}. The simplest explanation for such correlations is the consistency between noise and dissipation dictated by the Fluctuation-Dissipation relation of the second kind, that generalises the Einstein relation between mobility and diffusivity: since dissipation appears as a matrix of correlated mobilities (the Oseen tensor), also noise must be described by a matrix of correlated diffusivities~\cite{kubo1966fluctuation,marconi2008fluctuation}. The physical counterpart of this consistency argument is the fact that an incompressible viscous fluid transmits forces acting in a point over a long distance , and this principle holds for both dissipative and fluctuating forces~\cite{mazur1982motion}. 

The Fluctuation Dissipation relation of the second kind which relates the diffusivity matrix ${\bf D}$ and the mobility matrix ${\bf T}$ in our terms reads ${\bf D}={\bf T}/\beta$ where $\beta=1/(k_B T)$ is the inverse temperature (in the simulations we assume unitary Boltzmann constant $k_B=1$)~\cite{kubo1966fluctuation}. Since ${\bf T}$ depends upon the coordinates $\underline x$, the Fokker-Planck equation for the process is not uniquely determined by $D$. The ambiguity is removed asking for detailed balance (i.e. absence of physical currents) when the  non-conservative forces are absent. Then, following for instance~\cite{lau2007state}, this condition implies the following form for the probability current 
\begin{equation}
    J_i(\underline x,t)= \sum_j D_{ij}\left[\beta F_j(\underline x,t) P(\underline x,t)-\partial_{x_j}P(\underline x,t)\right],
\end{equation}
appearing in the Fokker-Planck equation
\begin{equation}
    \partial_t P(\underline x,t)=-\partial_{x_i} J_i(\underline x,t).
\end{equation}
Such Fokker-Planck equation implies the following anti-Ito stochastic differential equation to hold
\begin{equation}
\label{eq:langtotalintegrationAI}
    dx_i = \sum_j \left[T_{ij}F_j\right]\, dt + \sqrt{\frac{2}{\beta}}\,\sum_j T_{ij} \Big(\sqrt{\zeta}\cdot d\mathbf{W}(t) \Big)_j\,,
\end{equation}
or conversely the following Ito stochastic differential equation:
\begin{equation}
\label{eq:langtotalintegration}
    dx_i = \sum_j \left[T_{ij}F_j + \frac{1}{\beta}\,\partial_{x_j} T_{ij} \right]\, dt + \sqrt{\frac{2}{\beta}}\,\sum_j T_{ij} \Big(\sqrt{\zeta}\cdot d\mathbf{W}(t) \Big)_j\,,
\end{equation}
which is better suited for numerical integration and theoretical calculations.
Three additional terms appear, which we call "Ito forces" (they are actually velocities) defined as $F_{ito,i}=\frac{1}{\beta}\sum_j \partial_{x_j} T_{ij}$, that read
\begin{align}
\label{eq:fito}
    F_{ito,1}&= \frac{1}{\beta}\frac{ \frac{1}{L_1^2}+\frac{1}{(L_1+L_2)^2} }{4\pi\eta}\,;\\
    F_{ito,2}&= \frac{1}{\beta}\frac{ (L_1 - L_2)(L_1 + L_2) }{4\pi\eta\,L_1^2\,L_2^2}\,;\\
    F_{ito,3}&= -\frac{1}{\beta}\frac{ \frac{1}{L_2^2}+\frac{1}{(L_1+L_2)^2} }{4\pi\eta}\,,
\end{align}
We remark that $\sqrt{\zeta}$ is a matrix such that $\sqrt{\zeta} [\sqrt{\zeta}]^T={\bf T}^{-1}$. For the purpose of numerical integration we adopted the Cholesky decomposition of ${\bf T}^{-1}$ as a representation of it.

\begin{figure}[htb]
    \centering
    \sidesubfloat[]
        {\includegraphics[width=0.95\columnwidth]{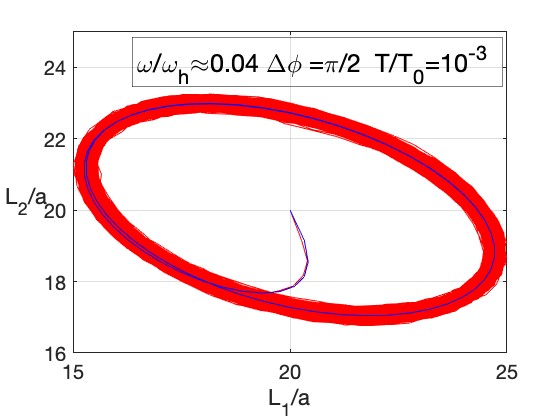}
        \label{}}
    \\    
    \sidesubfloat[]    
        {\includegraphics[width=0.95\columnwidth]{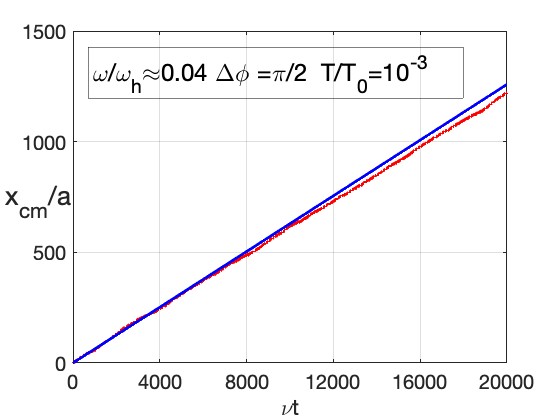}
        \label{}}
    
    \caption{(a) Arms' length in comparison, with (red) and without noise (blue). (b) Position of the center of mass with (red) and without noise (blue). Parameters: $L=20$, $a=1$, $\eta=1$, $\Delta\phi=\pi/2$, $\omega=2\pi/50$, $K=2$, $F_0=10$, (blue) $T=0$ and (red) $T=0.01$.}
    \label{fig:noise}
\end{figure}

We conclude this subsection by showing, see Fig.~\ref{fig:noise}, the effect of the noise, at a relatively small temperature (see next subsection for a discussion of the physical meaning of the units we use in the numerical simulations). The Figure shows both the trajectory of the center of mass of the swimmer, as well as its trajectory in phase space i.e. the plot of $L_1(t)=x_2(t)-x_1(t)$ vs $L_2(t)=x_3(t)-x_2(t)$, both in the case of $T=0$ and $T=10^{-2}$. As we will discuss later,a clear effect of the growth of temperature (i.e. of noise) is an increase of the diffusivity and therefore of transient deviations from the average trajectory, which become hardly distinguishable when the total time is increased and the average swimming motion dominates.

\subsection{Units of the physical parameters}
\label{sec:units}
We give all the results in arbitrary units. Here we discuss a possible conversion of those units in physical units. The candidate conversion to International System units is
\begin{align}
1\,\textrm{space unit} &= 10^{-6} \textrm{m} = 1 \mu\textrm{m},\\
1\,\textrm{time unit} &= 10^{-3} \textrm{s} = 1 \textrm{ms},\\ 
1\,\textrm{force unit} &= 5 \times 10^{-14} \textrm{N}
\\ \textrm{or equivalently} \\
1\,\textrm{mass unit} &= 5 \times 10^{-14} \textrm{Kg}.
\end{align}
With such a conversion table, we get that $T=0.1$ in arbitrary units corresponds to $T \approx 300 ^\circ K$, a viscosity $\eta=1$ in arbitrary units corresponds to $\eta = 5 \times 10^{-5} Pa\,s$ which is $20$ times smaller than water viscosity. A period of $50$, i.e. $\omega=2\pi/50 \approx 0.12$ in abritrary units corresponds to $\omega \approx 20 Hz$. 
A velocity of the center of mass $v=10^{-3}$ corresponds to $v = 1\mu m/s$.

\subsection{Characteristic values of the parameters}

For the numerical study we need to define a few characteristic numbers which can be useful as a reference in the presentation of the results, in order to draw adimensional axis in the graphs. We define $\nu=\omega/2 \pi$ as the forcing frequency, $v_0 =  \omega a F_0^2/(L^2 K^2)$ which is the swimming velocity expected for small forcing amplitudes and in the adiabatic limit $\omega \to 0$ (see the linear theory in Section~\ref{sec:analyt}), $K_0 = F_0/a $ a typical stiffness related to the forcing amplitude $F_0$ and the diameter of the spheres $a$, $\eta_0 = F_0/(6 \pi v_0 a)=(KL)^2/(6 \pi a^2 F_0 \omega) $ a typical viscosity related to forcing, swimming velocity and diameter $a$,
interestingly when $\omega =2\pi/10$ (ie. forcing period $10 \textrm{ms}$ with the units assumed above, one has $\eta_0 \sim 13.5$ which is close to water viscosity in the same units), $T_0 = F_0*a$ a typical thermal energy (temperature having assumed $k_B=1$). When we present results as a function of $\omega$, the above choices for characteristic parameters cannot be used, therefore we introduce other characteristic parameters related to hydrodynamics, i.e. $\nu_h = F_0/(6 \pi \eta a^2)$ a hydrodynamic frequency (close to $0.5$ in most of the other plots) and the corresponding pulsation $\omega_h = 2 \pi \nu_h$, and finally an associated swimming velocity $v_{h} = 2 \pi \nu_h a F_0^2/(L^2 K^2)$.

\subsection{Observables of interest in the numerical integration.}

In the rest of the paper we report several observations obtained by numerical integration of Eqs.~\eqref{eq:langtotalintegration} by a simple Euler scheme, i.e. by replacing $dt$ by the time-step $\Delta t=10^{-2}$, replacing each Wiener increment $dW_i(t)$ by a Gaussian-distributed random number (independent from each other) with zero average and variance $\Delta t$, and replacing $dx_i(t)$ by $x_i(t+\Delta t)-x_i(t)$. We have verified that reducing further the time-step has negligible effect on the observation. Numerical integration is always initialised with $x_1=20$, $x_2=0$ and $x_3=-20$, and then some time ($10^{3}$ time steps) is waited before measuring observables of interest.

In all interesting cases we have analysed the limit cycle in the phase space $L_1(t)=x_2(t)-x_1(t)$ vs. $L_2(t)=x_3(t)-x_2(t)$. Another observable containing interesting information about the trajectory of the swimmer is the center of mass evolution $x_{cm}(t)=(x_1(t)+x_2(t)+x_3(t))/3$. The swimming velocity is measured using the best linear fit $x_{cm}(t)=v t$ over a trajectory of length $10^5$ s  Information about the fluctuations with respect to the average trajectory is obtained through the mean squared displacement $msd(\tau)=\langle [x_{cm}(t_0+\tau)-x_{cm}(t_0)-v \tau]^2 \rangle$. Whenever the we observe $msd(\tau) \sim 2D\tau$ for large $\tau$ we extract $D$ as the diffusivity of the trajectory. If the trajectory is not long enough, exceptions to the linear asymptotic behavior are observed, which are explained in the next section. We also measure the energy consumption 
of the swimmer in terms of the motor forces $F^{act}_i(t)$, that is the Euler-discretized integral
\begin{equation} \label{eq:wdef}
    W(t)=\int_0^t \sum_i dx_i(s) F^{act}_i(s) 
\end{equation}
which is fitted at large times as $W(t) \sim \dot W t$. The number $\dot W$ is taken as a measurement of the average energy consumption rate.

To conclude, we have measured the precision rate of each long trajectory, by the formula
\begin{equation}
p=\frac{v^2}{D},
\end{equation}
which is expected to satisfy the Thermodynamic Uncertainty Relation~\cite{Barato2015,Gingrich2016,Hasegawa2019II,Lee2021,plati2023thermodynamic}
\begin{equation} \label{eq:tur}
    p \le p_{max}=\frac{\dot W}{k_B T}.
\end{equation}

\subsection{Mean squared displacement}

\begin{figure}[htb]
    \centering
    \sidesubfloat[]
    {\includegraphics[width=0.95\columnwidth]{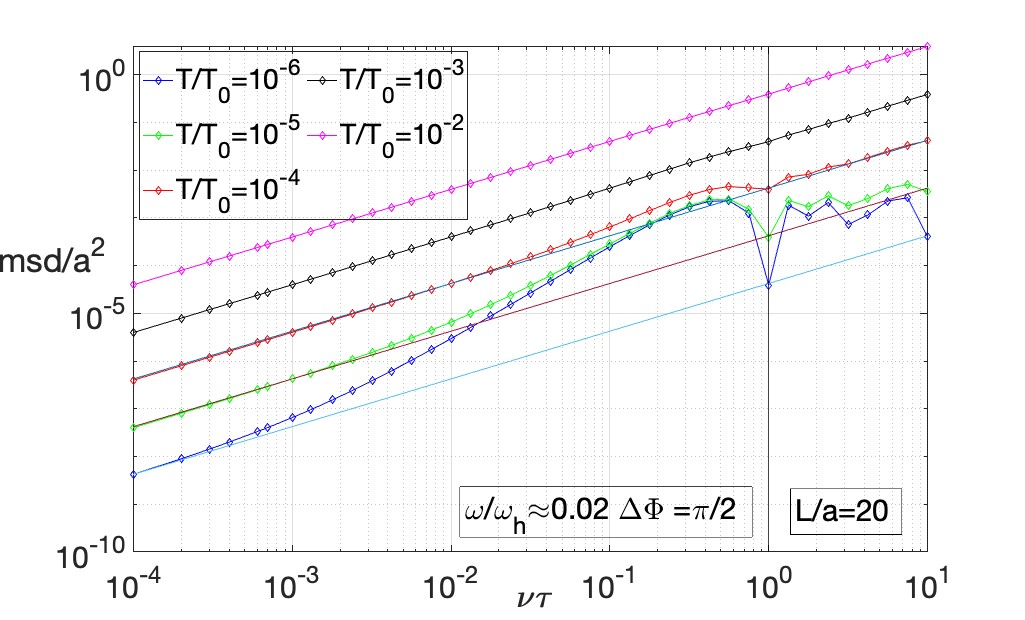}
    \label{} }
    \\
    \sidesubfloat[]
    {\includegraphics[width=0.95\columnwidth]{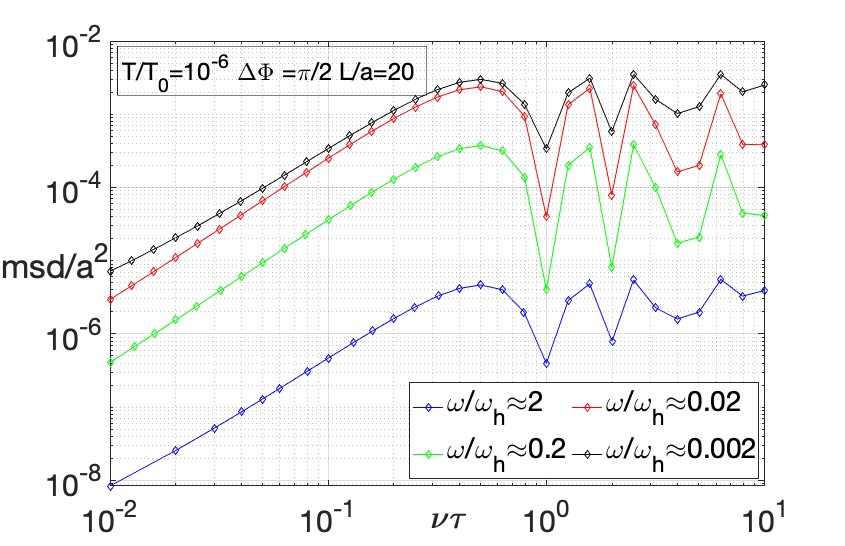}
    \label{}}
    
    \caption{Mean squared displacement as a function of time for several choice of temperatures and frequencies of the active force. Parameters which are not given in the Figure: $L=20$, $F_0=10$, $a=1$, $\eta=1$, $K=2$. All the straight lines in panel (a)  have slope $1$, representing the diffusive regime.}
    \label{fig:msdt}
\end{figure}

We conclude this introductory Section by briefly discussing the behavior of $msd(\tau)$ as a function of the delay $\tau$. As shown in Fig.~\ref{fig:msdt} we observe that when the temperature $T$ is large enough one always has a clear asymptotic diffusive behavior. On the contrary, when the temperature $T$ is too small, the mean-squared-displacement oscillates with a frequency which is equal to the swimming force frequency $\omega/(2\pi)$ and an amplitude which tends to reduce with $\tau$. The oscillations occur around a time-dependent shape which is asymptotically linear in time, therefore it is possible to extract a diffusivity $D$ even in this case.

The reason for such an oscillating behavior is the presence of the oscillating forces which, when the noise is not large enough, dominates on the dynamics and give an observable recurrency in the trajectory.

\section{Numerical study}
\label{sec:numerical}

In this Section we present the results of the numerical simulations. Each of the following subsections is devoted to the effect of one particular physical parameter upon the most relevant observables of our study which are the average velocity of the center of mass, the average consumption rate of energy, the diffusivity and the precision.

\subsection{Changing the active force}

\begin{figure}[htb]
\centering
\sidesubfloat[]
      {\includegraphics[width=0.43\columnwidth]{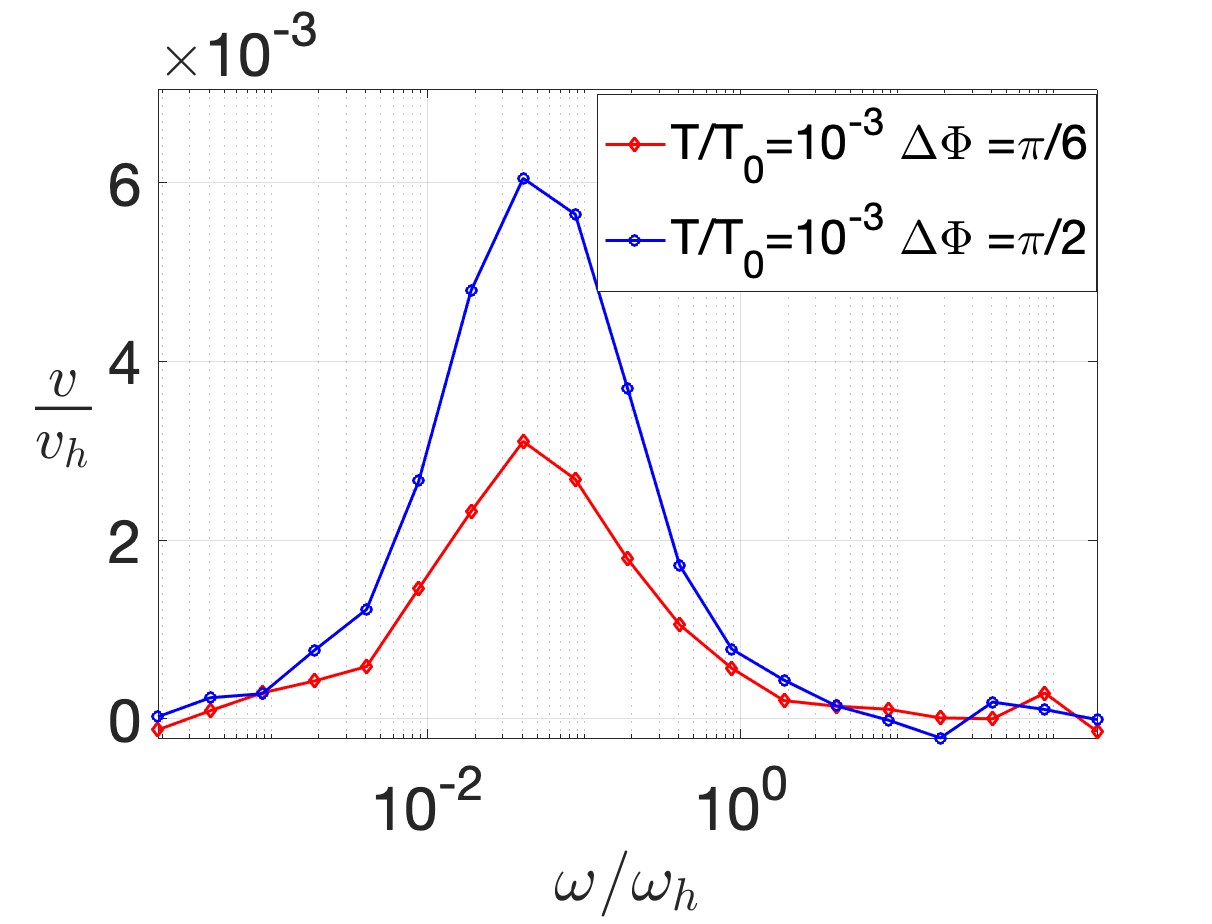} 
      \label{}}
      ~
\hfil 
\sidesubfloat[]
        {\includegraphics[width=0.43\columnwidth]{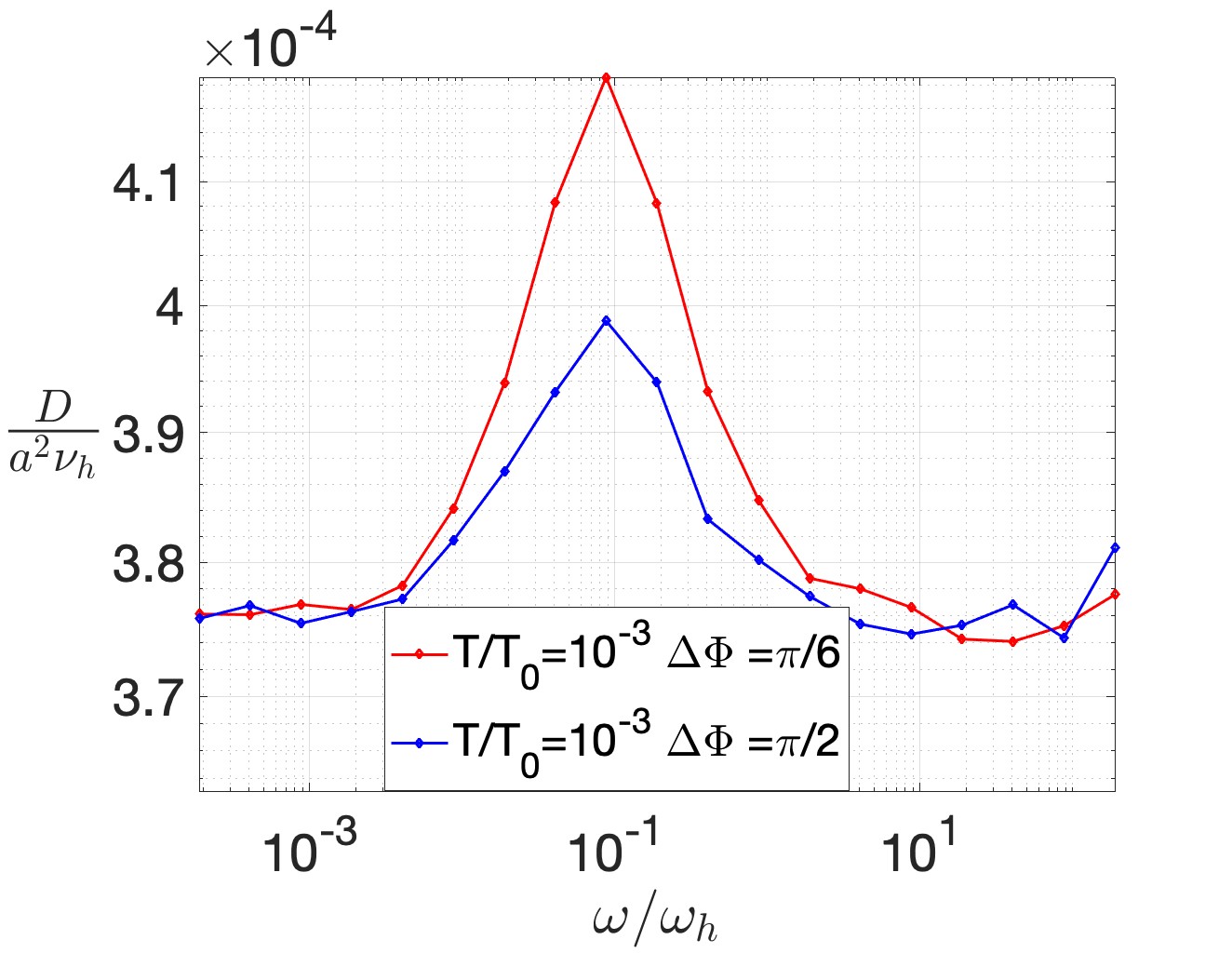}
        \label{}}
\
\sidesubfloat[]
        {\includegraphics[width=0.43\columnwidth]{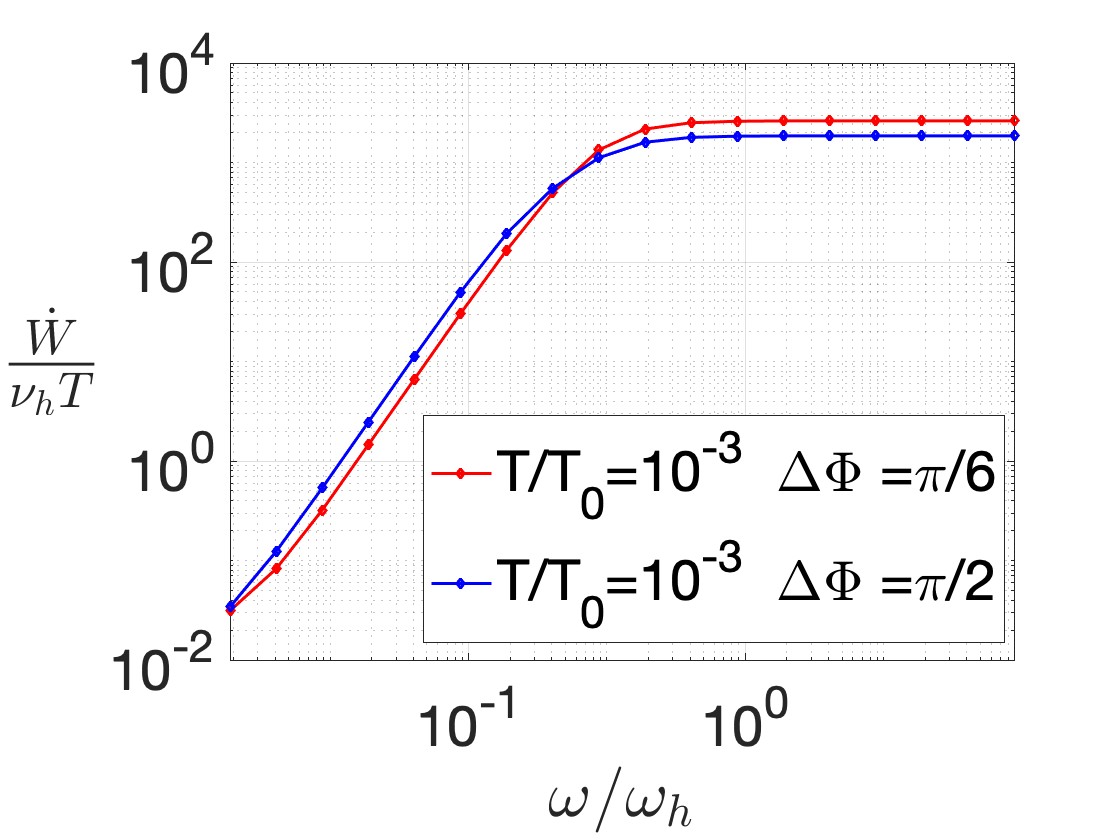}
        \label{}}
        ~
\hfil 
\sidesubfloat[]
        {\includegraphics[width=0.43\columnwidth]{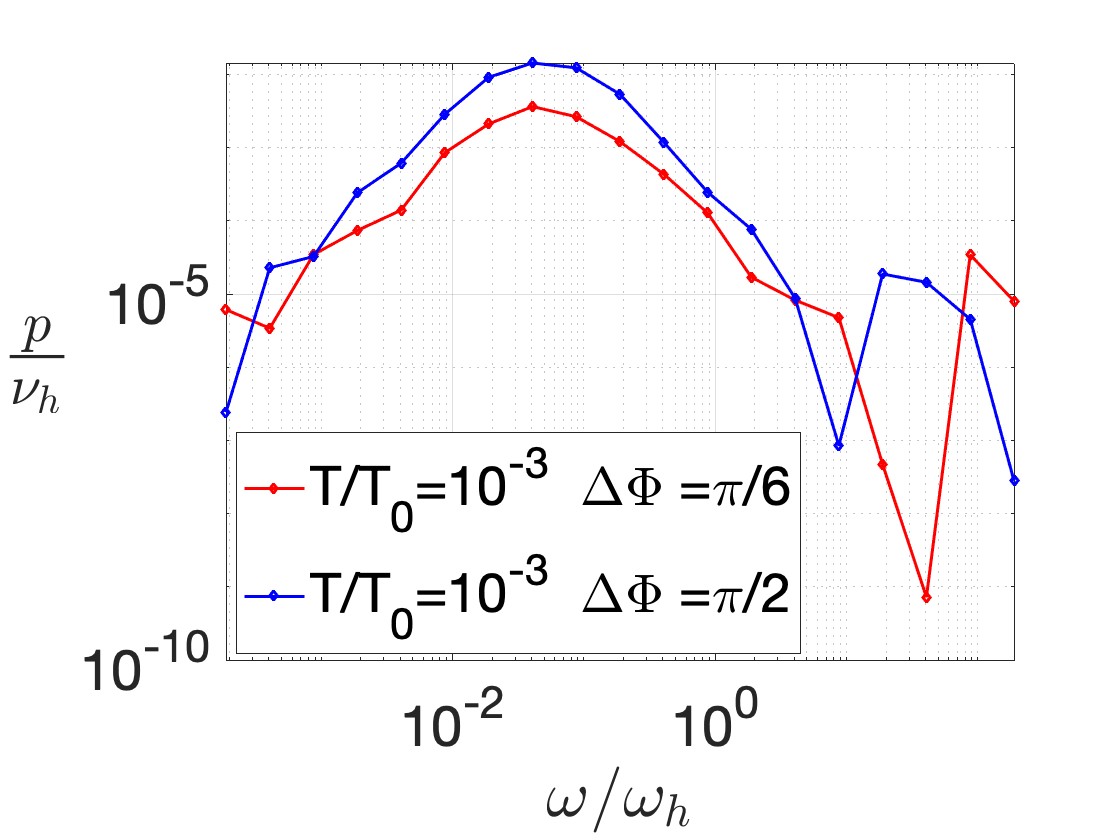}
        \label{}}
\caption{Effect of the frequency of the oscillating active force $\omega$ upon (a) the average swim velocity, (b) the swimming diffusivity, (c) the energy consumption rate and (d) the precision. Parameters: $L=20$, $F_0=10$, $a=1$, $\eta=1$, $K=2$.}
\label{fig:vsomega}
\end{figure}

\begin{figure}[htb]
    \centering
    \sidesubfloat[]
      {\includegraphics[width=0.43\columnwidth]{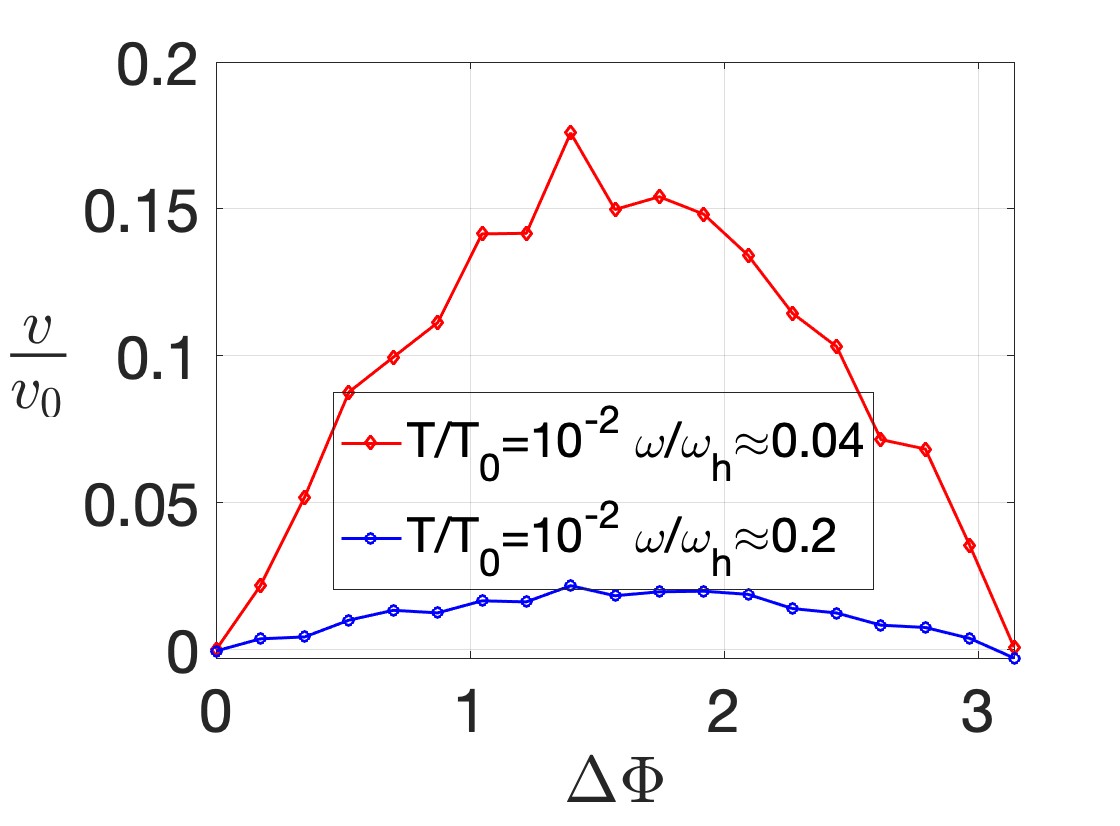} 
      \label{}}
      ~
    \hfil
    \sidesubfloat[]
      {\includegraphics[width=0.43\columnwidth]{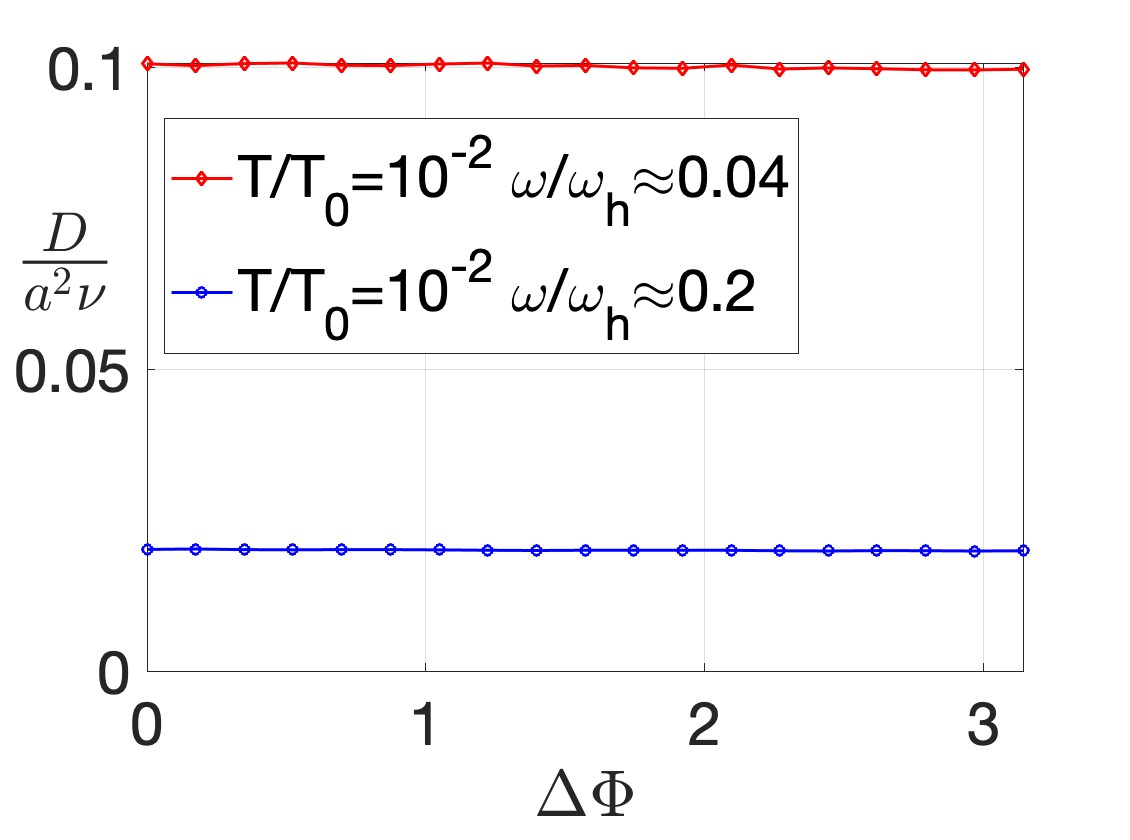}
      \label{}}
    \
    \sidesubfloat[]
      {\includegraphics[width=0.43\columnwidth]{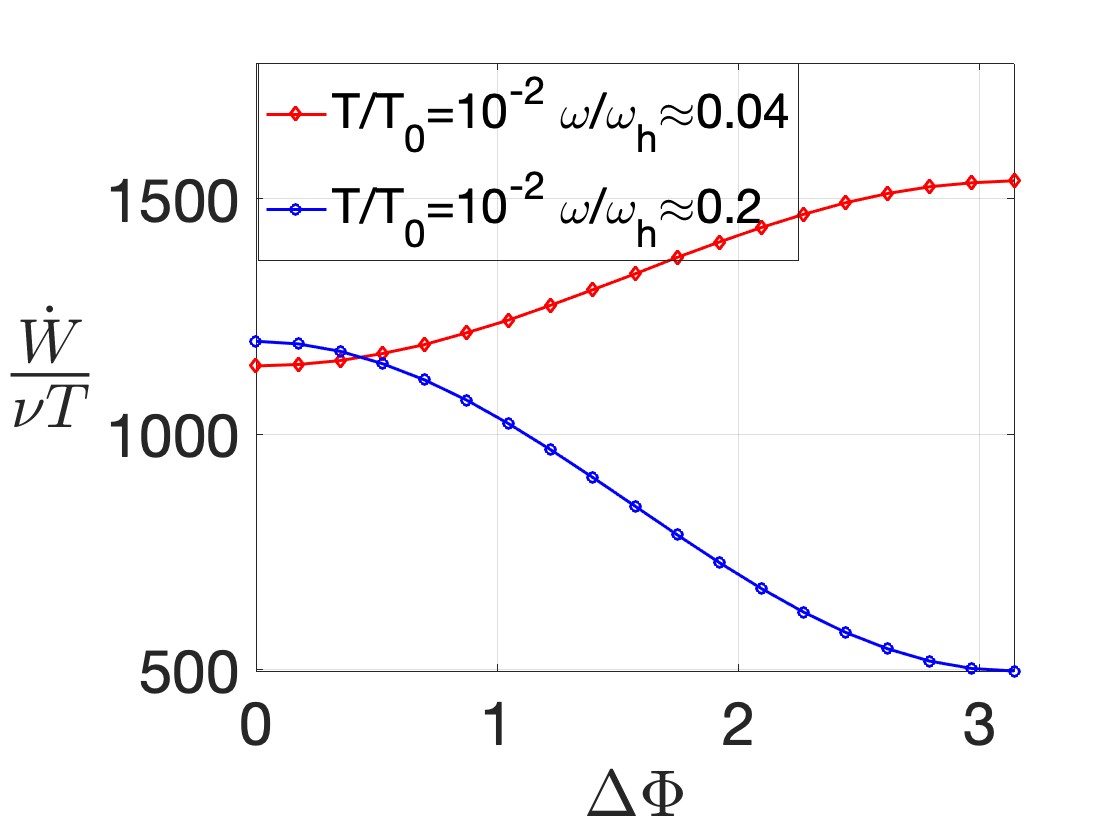}
      \label{}}
      ~
    \hfil
    \sidesubfloat[]
      {\includegraphics[width=0.43\columnwidth]{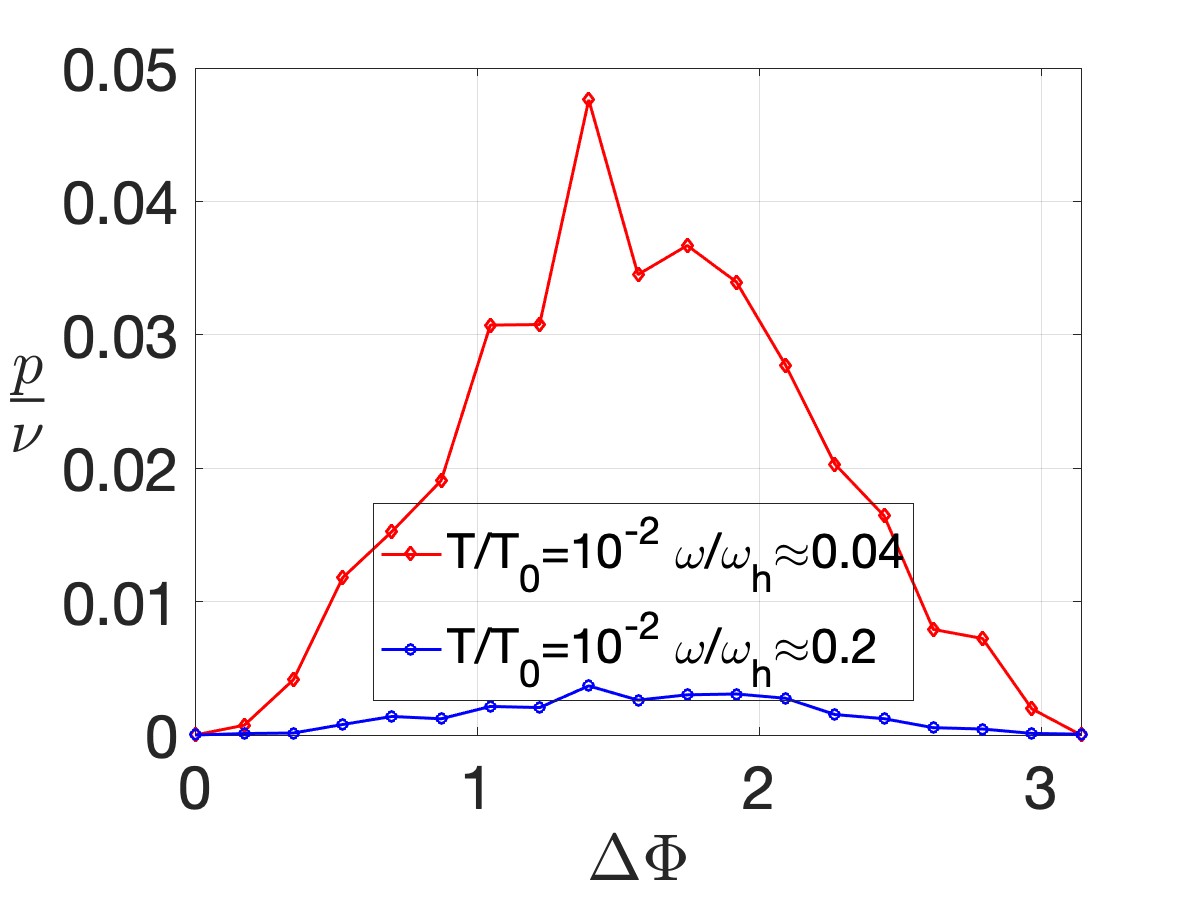}
      \label{}}
    \caption{Effect of the phase difference among the first and last components of the oscillating active force $\Delta \phi$ upon (a) the average swim velocity, (b) the swimming diffusivity, (c) the energy consumption rate and (d) the precision. Parameters: $L=20$, $F_0=10$, $a=1$, $\eta=1$, $K=2$.}
    \label{fig:vsdeltaphi}
\end{figure}

The active force defined in Eq.~\eqref{eq:forcesform} has three parameters: $F_0$, $\omega$ and $\Delta \phi$. In Fig.~\ref{fig:vsomega} we study the effect of $\omega$, while in Fig.~\ref{fig:vsdeltaphi} we consider $\Delta \phi$. Since the theoretical part (Section~\ref{sec:analyt} focuses on linear response, the effect of $F_0$ in this regime is trivial and we do not investigate it. 

Both average velocity and diffusivity have an optimum at a similar frequency $\omega_{opt}$, apparently independent on $\Delta \phi$. Note that the excursion between the peak value and the base value (i.e. at small and large frequencies) is large for $v$ but only of order $20 \%$ for the diffusivity (in fact diffusivity has a finite value even when there is no drift or work). The work rate grows with frequency and saturates at large frequency. The behavior in frequency of work rate and average velocity is qualitatively similar to the response of a driven resonant oscillator: when the perturbation frequency is much smaller than the resonant frequency, the oscillator follows the perturbation (and then the velocity and the work rate decrease when the frequency decreases), when it is much faster the oscillator cannot follow it, therefore the velocity reduces and the work rate becomes independent from the frequency~\cite{yasuda2017elastic}. However the real system is overdamped and the resonant behavior is not obviously related to the parameters of the model. Moreover, we note that this resonant behavior is substantially different from what observed in the original model (kinematically driven and without confining springs) where one simply has $v \sim \omega$, see Eq.~\eqref{eq:avg_vel_gol_2008}. We rationalise this resonant behavior in the linear (small force) limit, discussed in the last Section.

The precision has an optimum at a similar frequency $\sim \omega_{opt}$, since it is dominated by the numerator $v^2$.

The phase difference has quite a small (order few percent points), apparently monotonous, effect on the diffusivity, while it is relevant for the average velocity, having a peak at $\Delta \phi=\pi/2$: clearly there is no swimming when $\Delta \phi=0$ or $\pi$, since in both cases the perturbation force vector becomes symmetric under the time-reveral operation. The effects on both $v$ and $D$ are weakly dependent upon $\omega$. The work rate on the contrary has a dependence on $\Delta \phi$, e.g. decreasing or increasing, that changes with the value of $\omega$. The precision, which is dominated by $v^2$ follows a similar graph with an optimum at $\pi/2$.

\subsection{Changing the properties of the confining potential}

\begin{figure}[htb]
\centering
\sidesubfloat[]
      {\includegraphics[width=0.43\columnwidth]
      {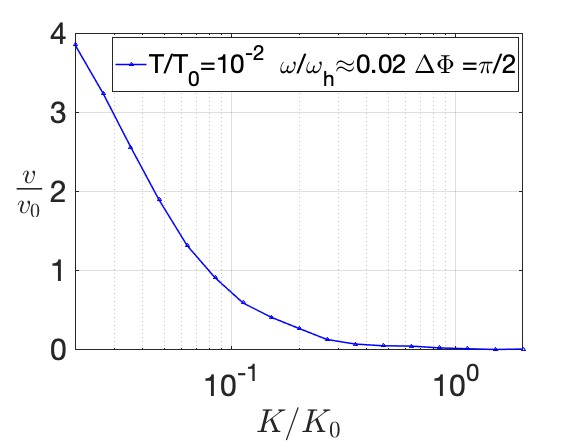}
      \label{}}
      ~
\hfil
\sidesubfloat[]
      {\includegraphics[width=0.43\columnwidth]{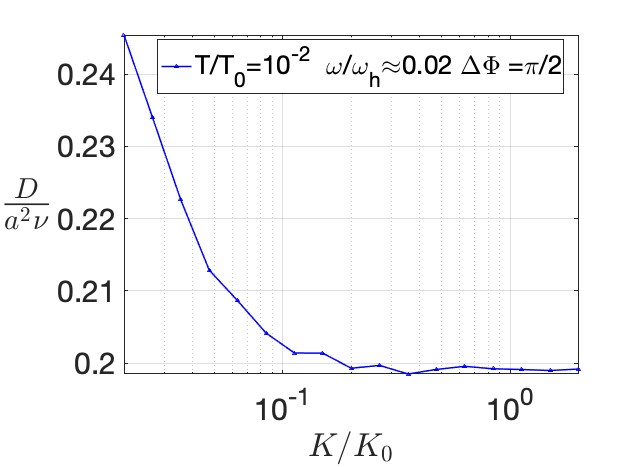}
      \label{}}
\    
\sidesubfloat[]
      {\includegraphics[width=0.43\columnwidth]{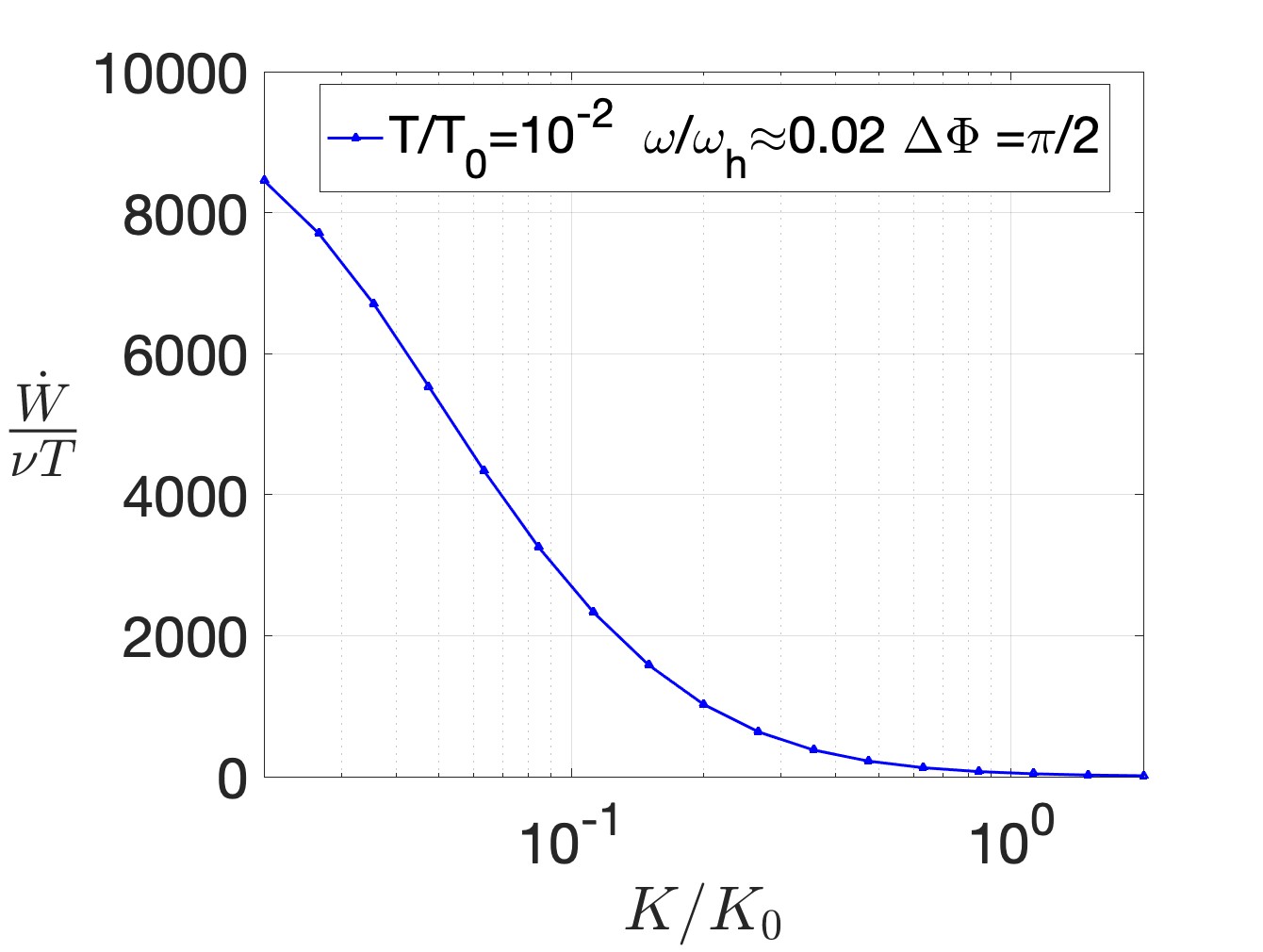}
      \label{}}
      ~
\hfil
\sidesubfloat[]
      {\includegraphics[width=0.43\columnwidth]{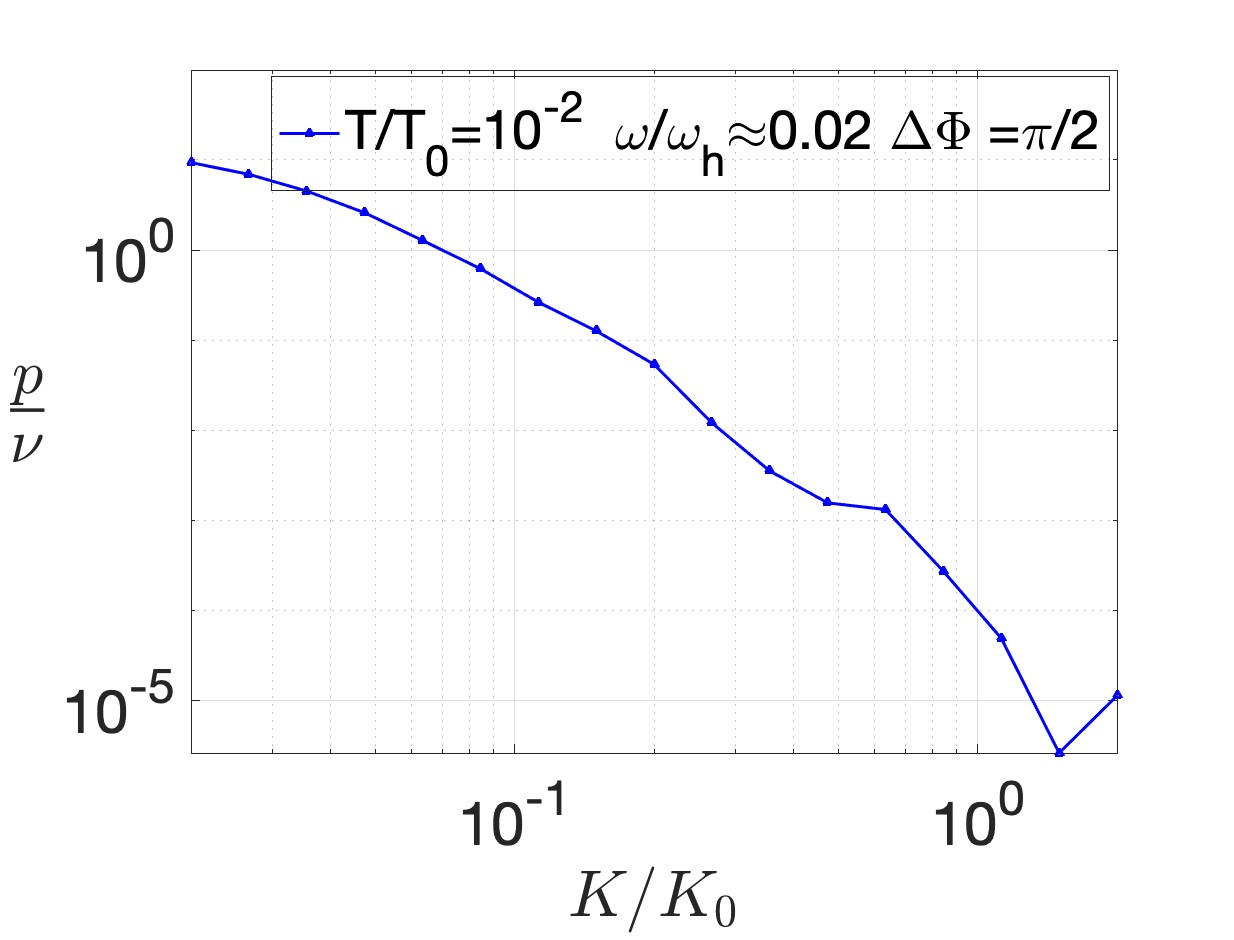}
      \label{}}
    \caption{Effect of the elastic constant of the confining potential $K$  upon (a) the average swim velocity, (b) the swimming diffusivity, (c) the energy consumption rate and (d) the precision. Parameters: $L=20$, $F_0=10$, $a=1$, $\eta=1$.  }
    \label{fig:vsk}
\end{figure}

The effect of the elastic constant $K$ for the confining potential is shown in Fig.~\ref{fig:vsk}. All the quantities of interest decay with $K$. Numerically it is not possible to decrease too much the value of $K$, since it leads to too large excursion of the distances between the particles and therefore to the possibility of two of them to touch each other, breaking the condition of non small distance and to a numerical instability of the mobility matrix which contains inverse powers of the distances. 

\begin{figure}[htb]
\centering
\sidesubfloat[]
    {\includegraphics[width=0.44\columnwidth]{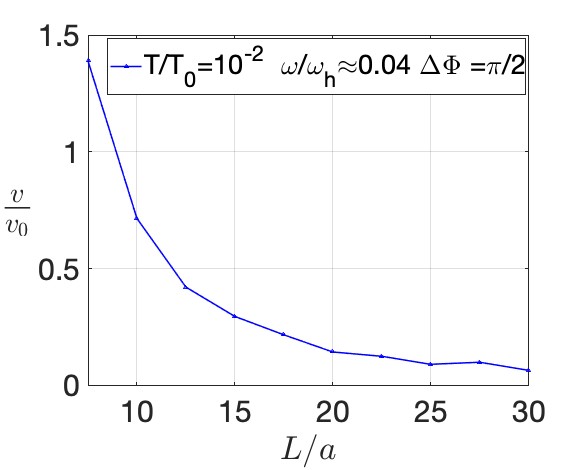}
    \label{}}
    ~
    \hfil
    \sidesubfloat[]
    {\includegraphics[width=0.44\columnwidth]{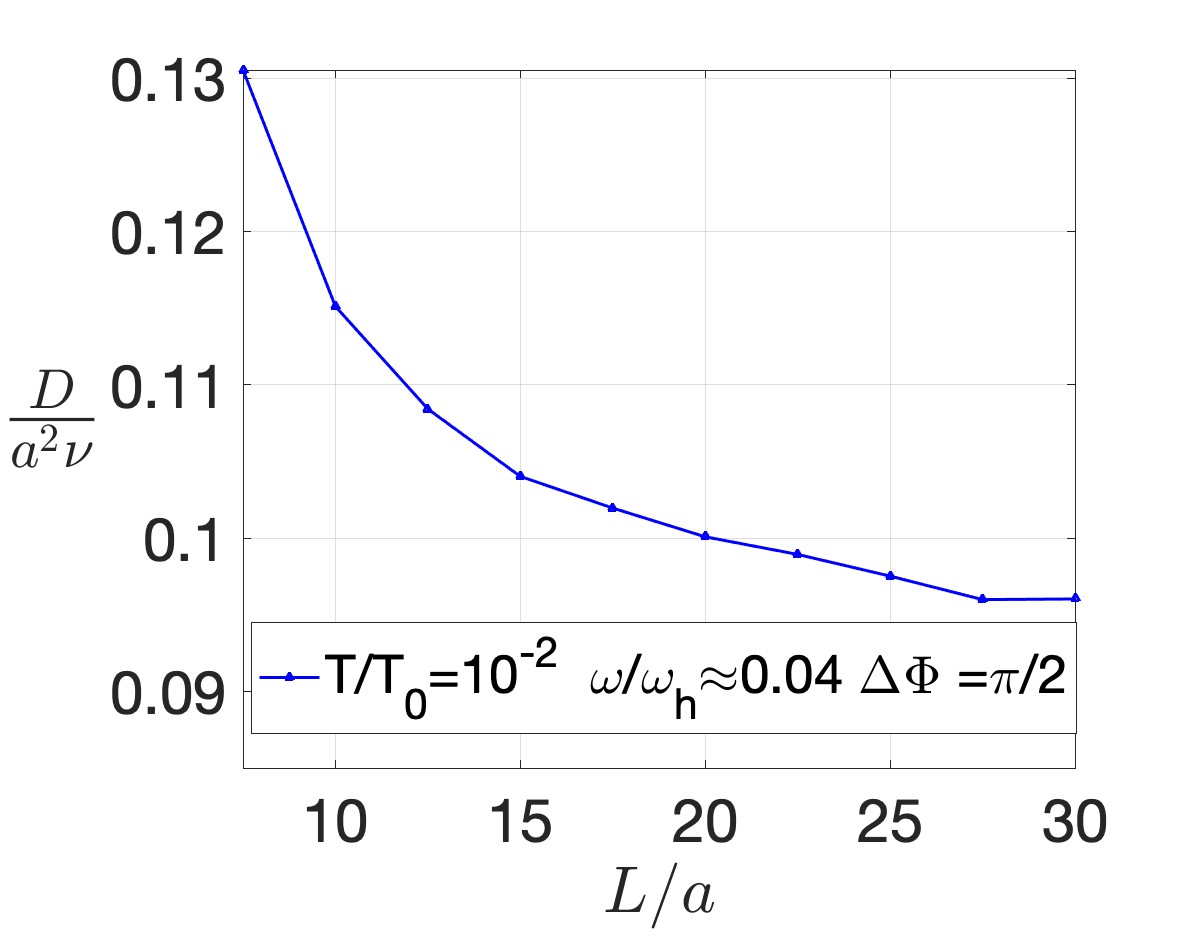}
    \label{}}
    \
    \sidesubfloat[]
    {\includegraphics[width=0.44\columnwidth]{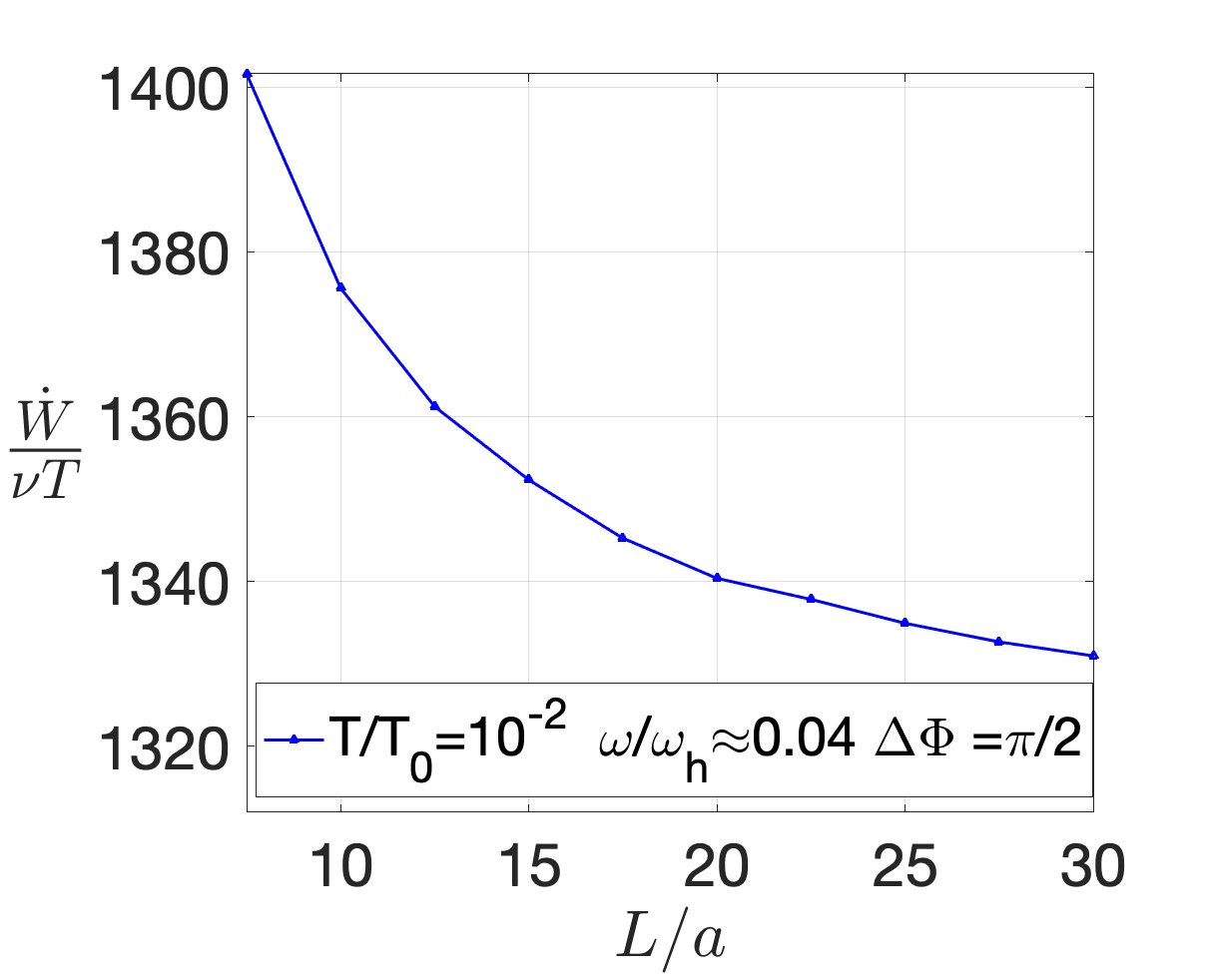}
    \label{}}
    ~
    \hfil
    \sidesubfloat[]
    {\includegraphics[width=0.44\columnwidth]{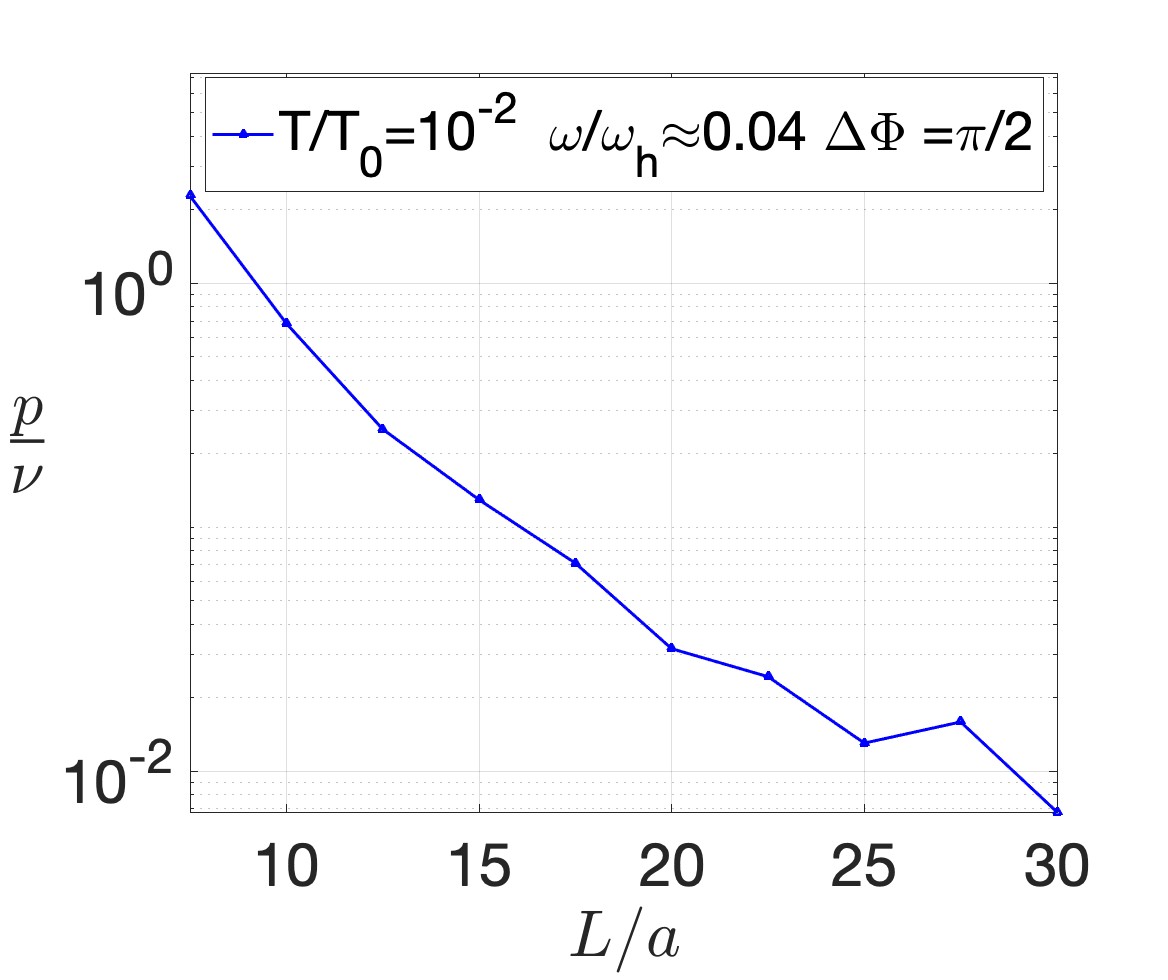}
    \label{}}
    \caption{Effect of the average length of the two arms of the swimmer $L$ upon (a) the average swim velocity, (b) the swimming diffusivity, (c) the energy consumption rate and (d) the precision. Parameters: $a=1$, $F_0=10$, $\eta=1$, $K=2$.  }
    \label{fig:vsL}
\end{figure}

We also consider the effect of changing $L=l_1=l_2$, the rest distance among the swimmer particles.
The consequence of changing $L$ is similar to that of changing $K$, see Fig.~\ref{fig:vsL}. The analogy between these two parameters can be understood in the following way. At fixed $L$ the effect of reducing $K$ is to permit larger excursions of $x_1-x_2$ and $x_2-x_3$ with  with respect to their rest value $L$, but such excursions include both large values (which are irrelevant) as well as small values, where the hydrodynamic interaction is stronger and the swimming efficiency is higher. The same occurs when $L$ is reduced at fixed $K$. Of course this analogy is qualitative, while the quantitative behavior is more complex.

\begin{figure}[htb]
    \centering
    \sidesubfloat[]
        {\includegraphics[width=0.96\columnwidth]{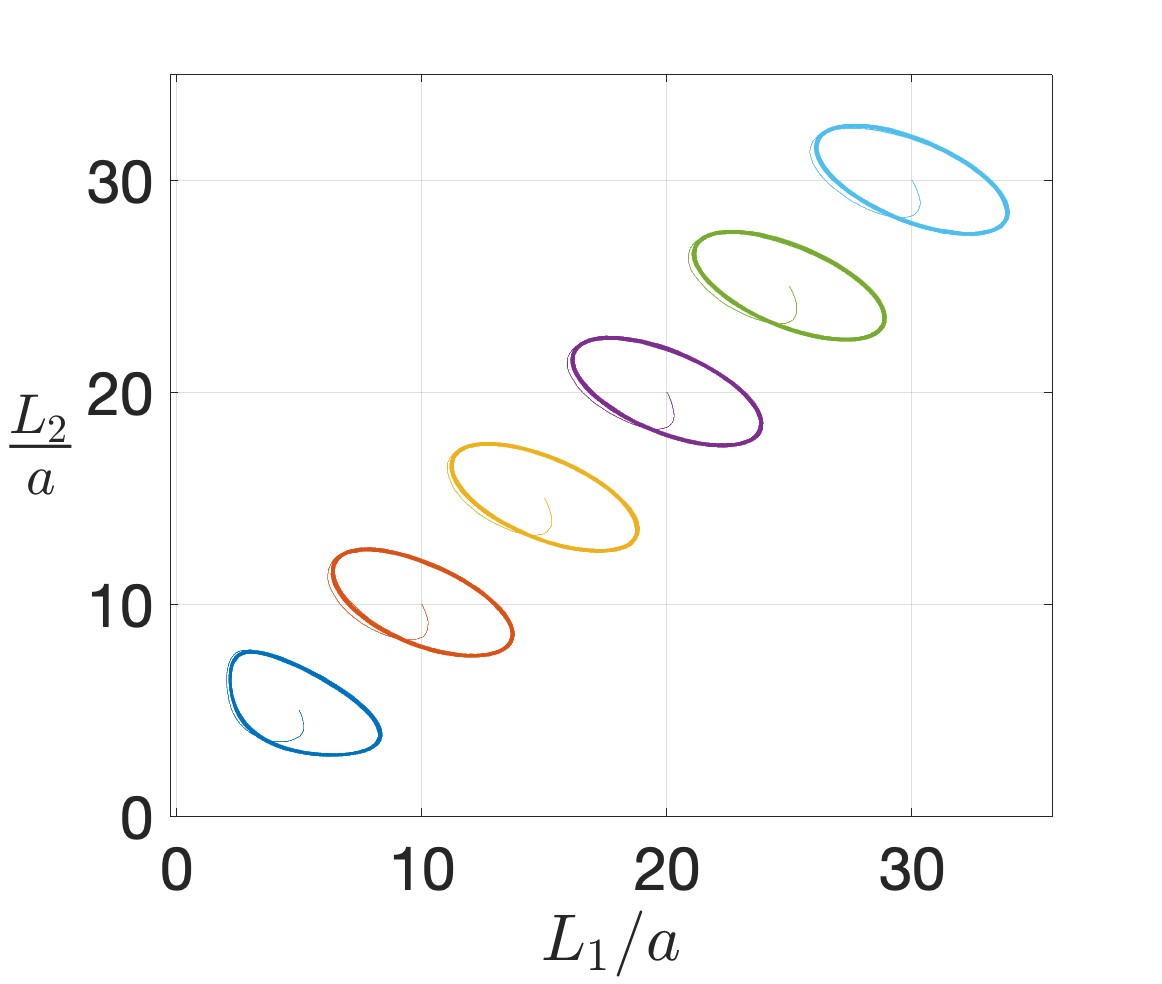}
        \label{}}
    \
    \sidesubfloat[]
        {\includegraphics[width=0.96\columnwidth]{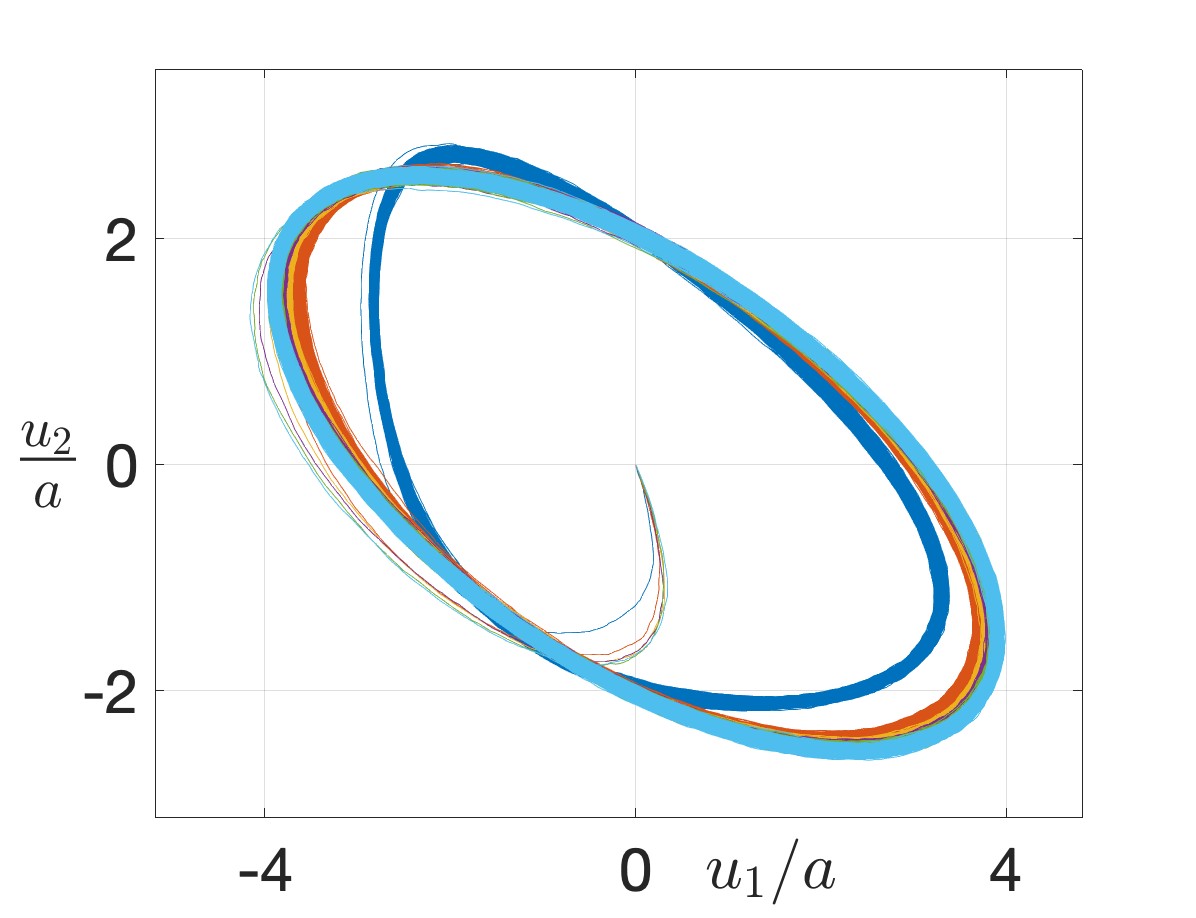}
        \label{}}
    \caption{Effect of the average length of the two arms of the swimmer $L$ upon the orbits in the plane $x_1-x_2,x_2-x_3$ and $u_1,u_2$ where $u_1=x_1-x_2-L$ and $u_2=x_2-x_3-L$. Parameters: $a=1$, $F_0=10$, $\eta=1$, $T=0.01$, $\Delta\phi=\pi/2$, $\omega=2\pi/50$, $K=2$. }
    \label{fig:vsL_orb}
\end{figure}

In Fig.~\ref{fig:vsL_orb} we also show the orbits in the plane of relative distances (with or without shifting by the rest length $L$) when $L$ is varied. This Figure shows a remarkable robustness of the shape of the limit cycle which appears independent of $L$ for $L$ large enough. As already discussed, see Eq.~\eqref{eq:avg_vel_gol_2008}, the average velocity of the swimmer is proportional to the area of the limit cycle with a proportionality factor $\alpha \sim L^{-1}$ therefore the behavior $v \sim L^{-1}$ is compatible with the observed orbital invariance. It is less clear how the observed orbital invariance may be related to the weak dependence of the work rate with $L$. As it will be clear in the Appendix, the explicit dependence of $v$ and $\dot{W}$ on $L=l_1=l_2$ is hard to read explicitly.

\subsection{Changing the properties of the fluid}

\begin{figure}[htb]
\centering
\sidesubfloat[]
      {\includegraphics[width=0.43\columnwidth]{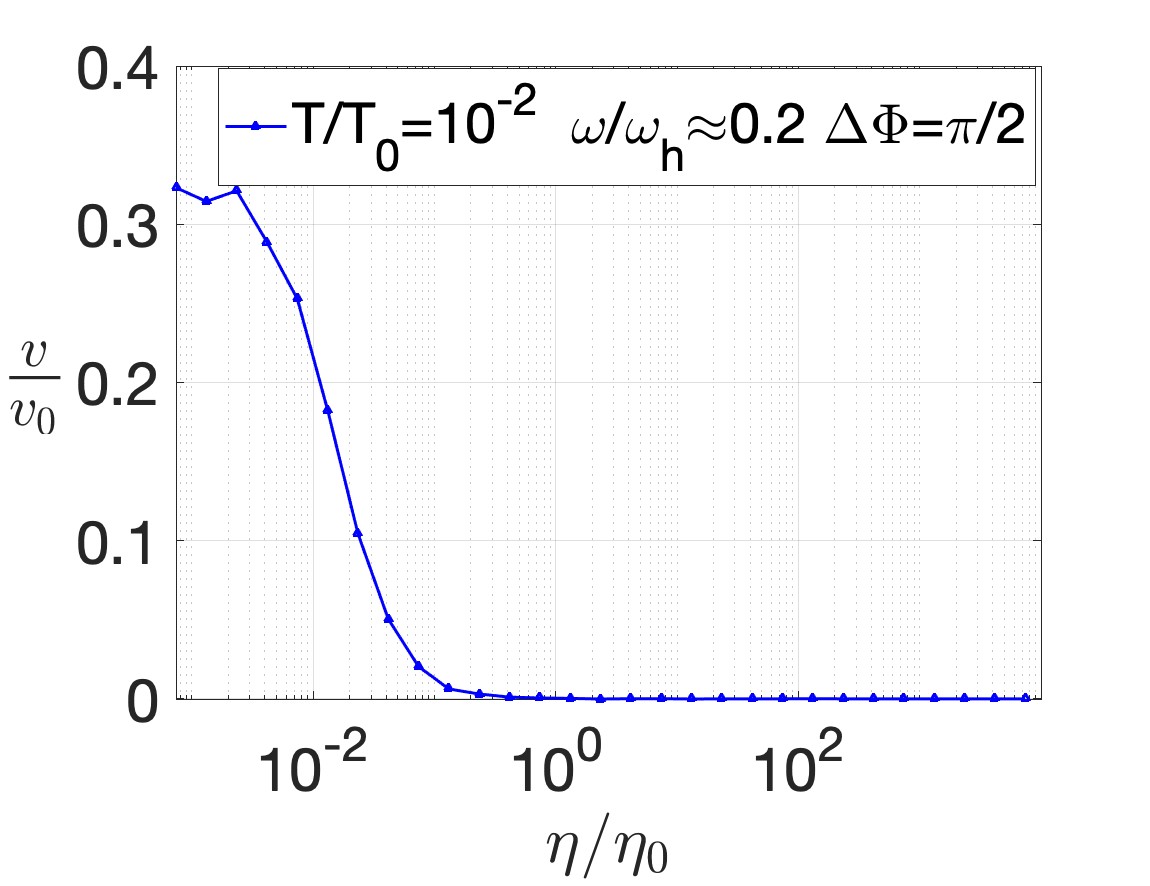} 
      \label{}}
~
\hfil
\sidesubfloat[]
        {\includegraphics[width=0.44\columnwidth]{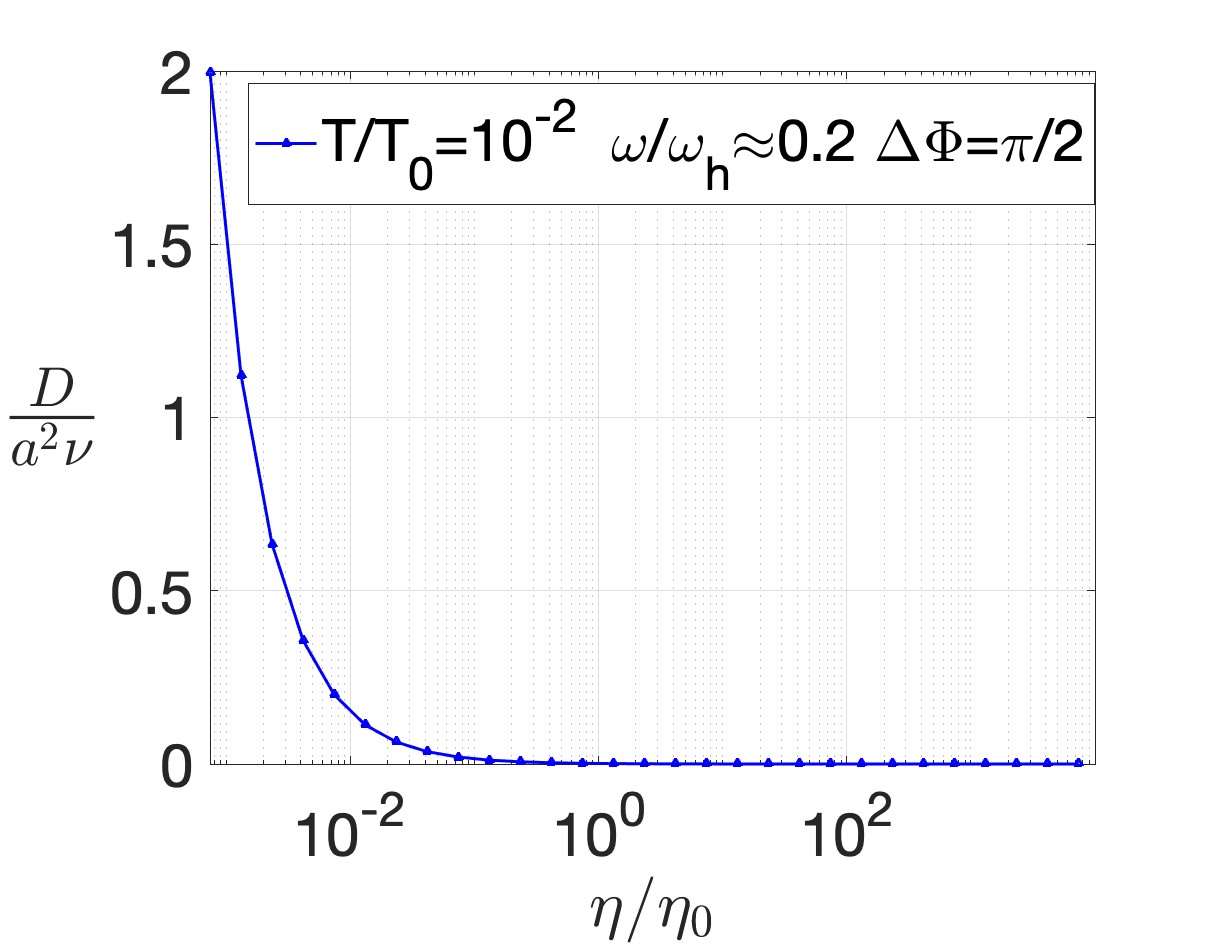}
        \label{}}
\
\sidesubfloat[]
        {\includegraphics[width=0.43\columnwidth]{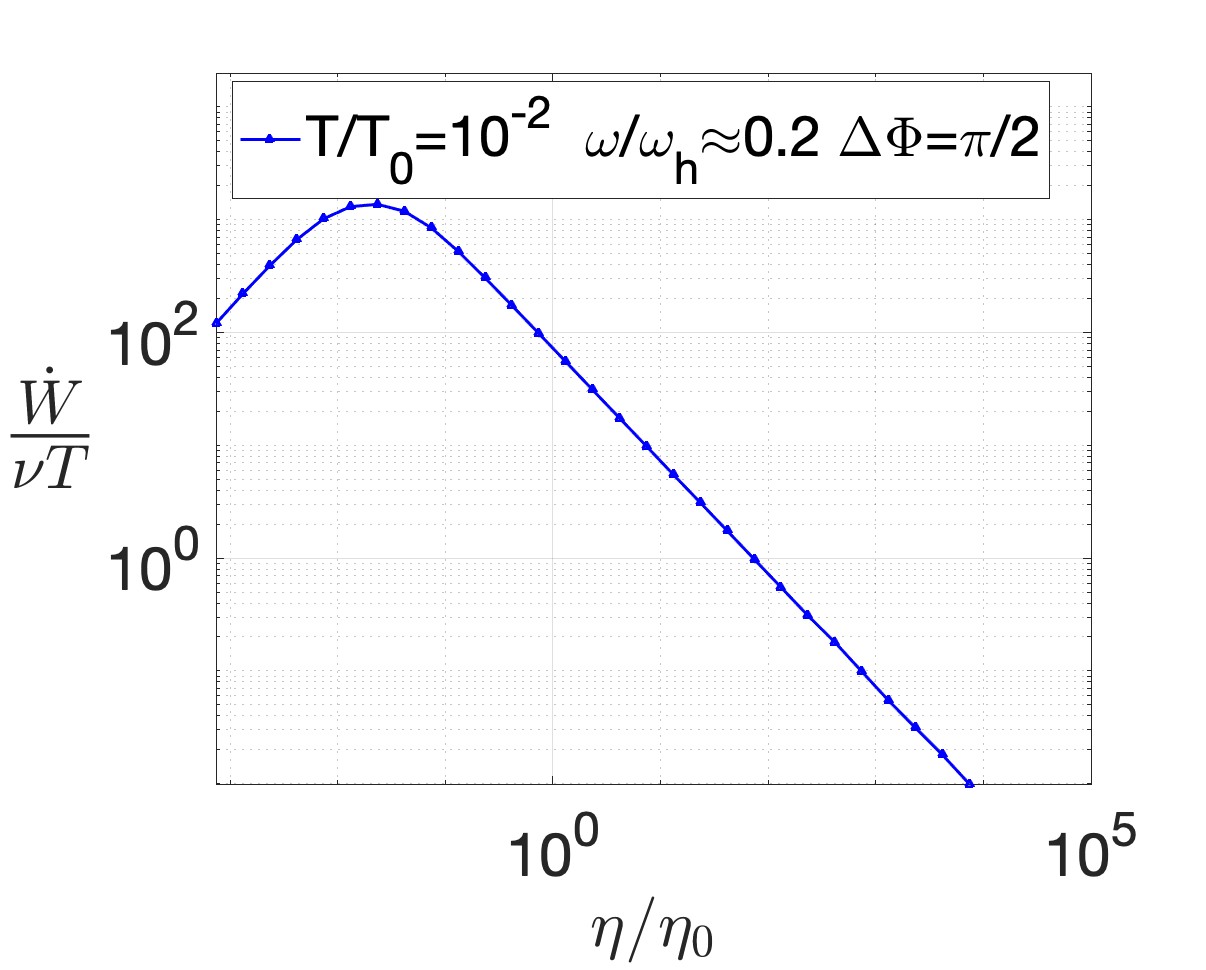}
        \label{}}
~
\hfil
\sidesubfloat[]
        {\includegraphics[width=0.44\columnwidth]{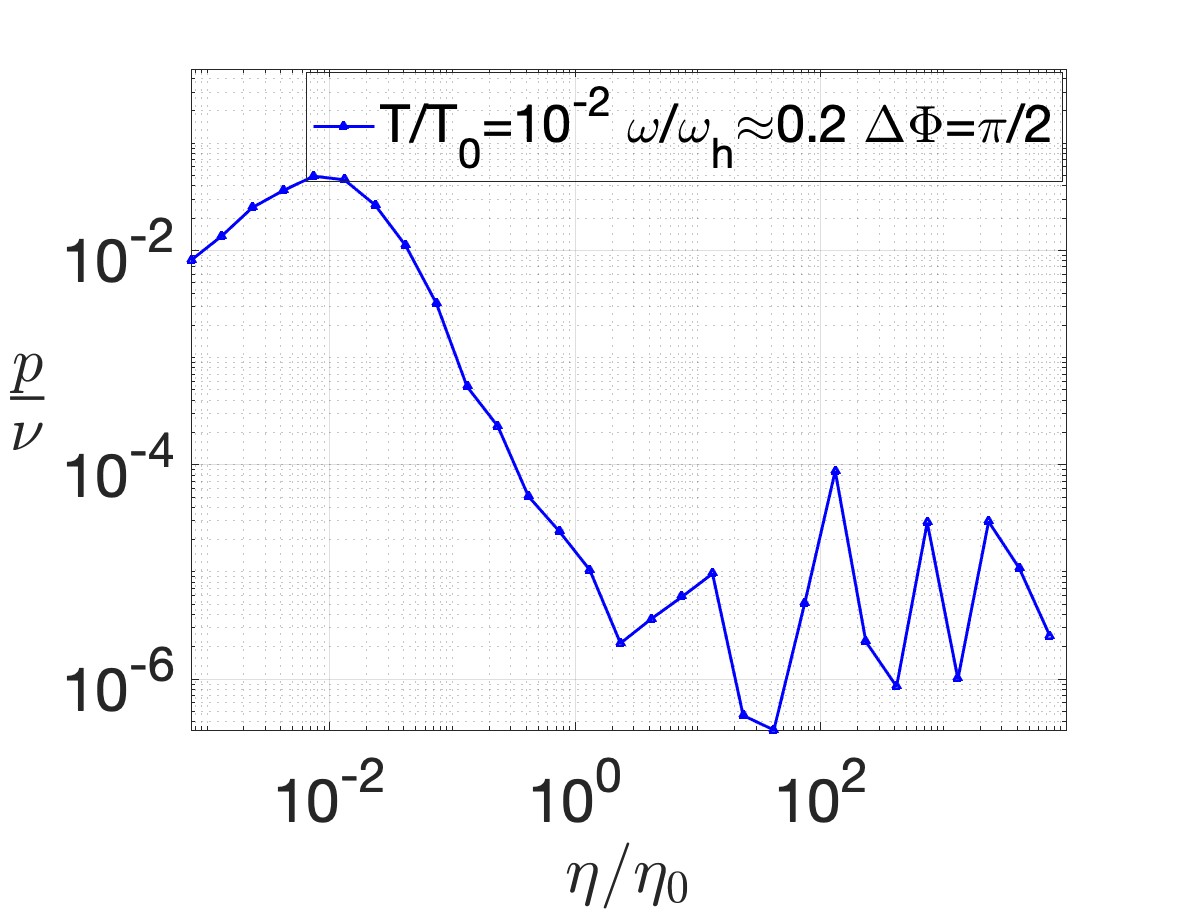}
        \label{}}
    \caption{Effect of the viscosity of the fluid $\eta$ upon (a) the average swim velocity, (b) the swimming diffusivity, (c) the energy consumption rate and (d) the precision. Parameters: $L=20$, $F_0=10$, $a=1$, $K=2$.  }
    \label{fig:vseta}
\end{figure}

\begin{figure}
\centering
\sidesubfloat[]
      {\includegraphics[width=0.43\columnwidth]{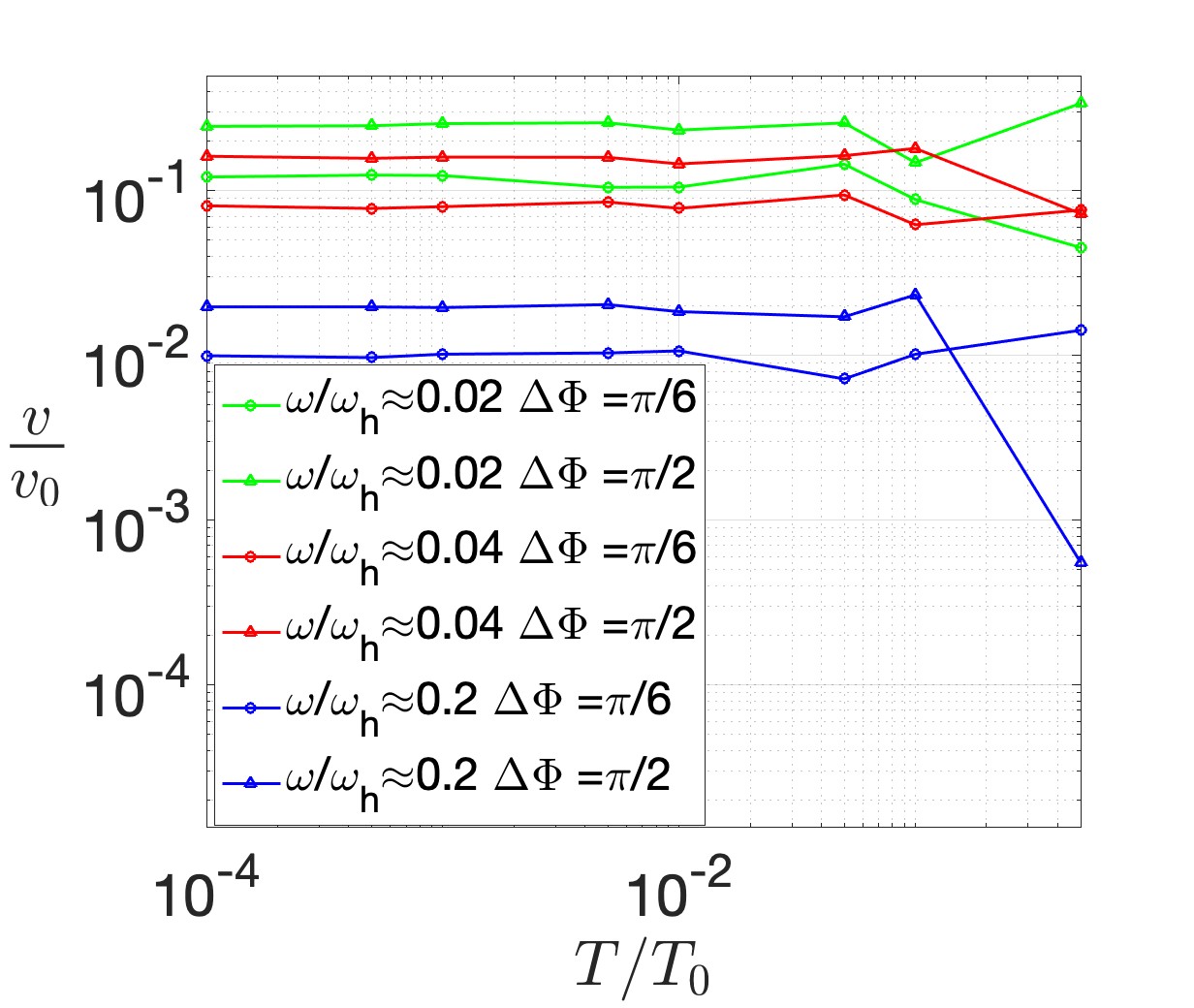}
      \label{}}
      ~
      \hfil
      \sidesubfloat[]
      {\includegraphics[width=0.43\columnwidth]{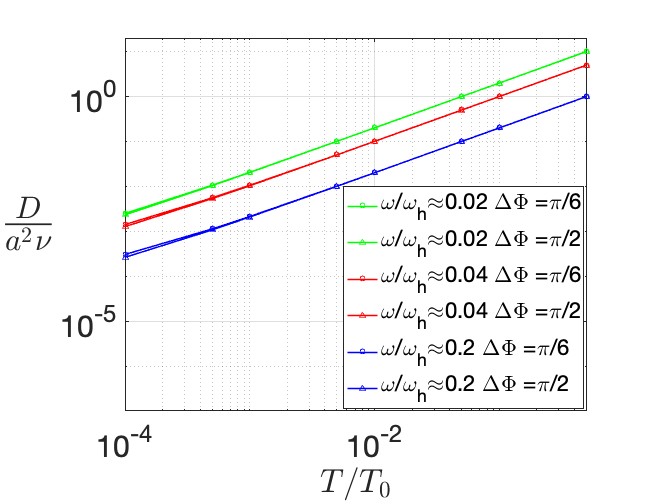}
      \label{}}
      \
      \sidesubfloat[]
      {\includegraphics[width=0.43\columnwidth]{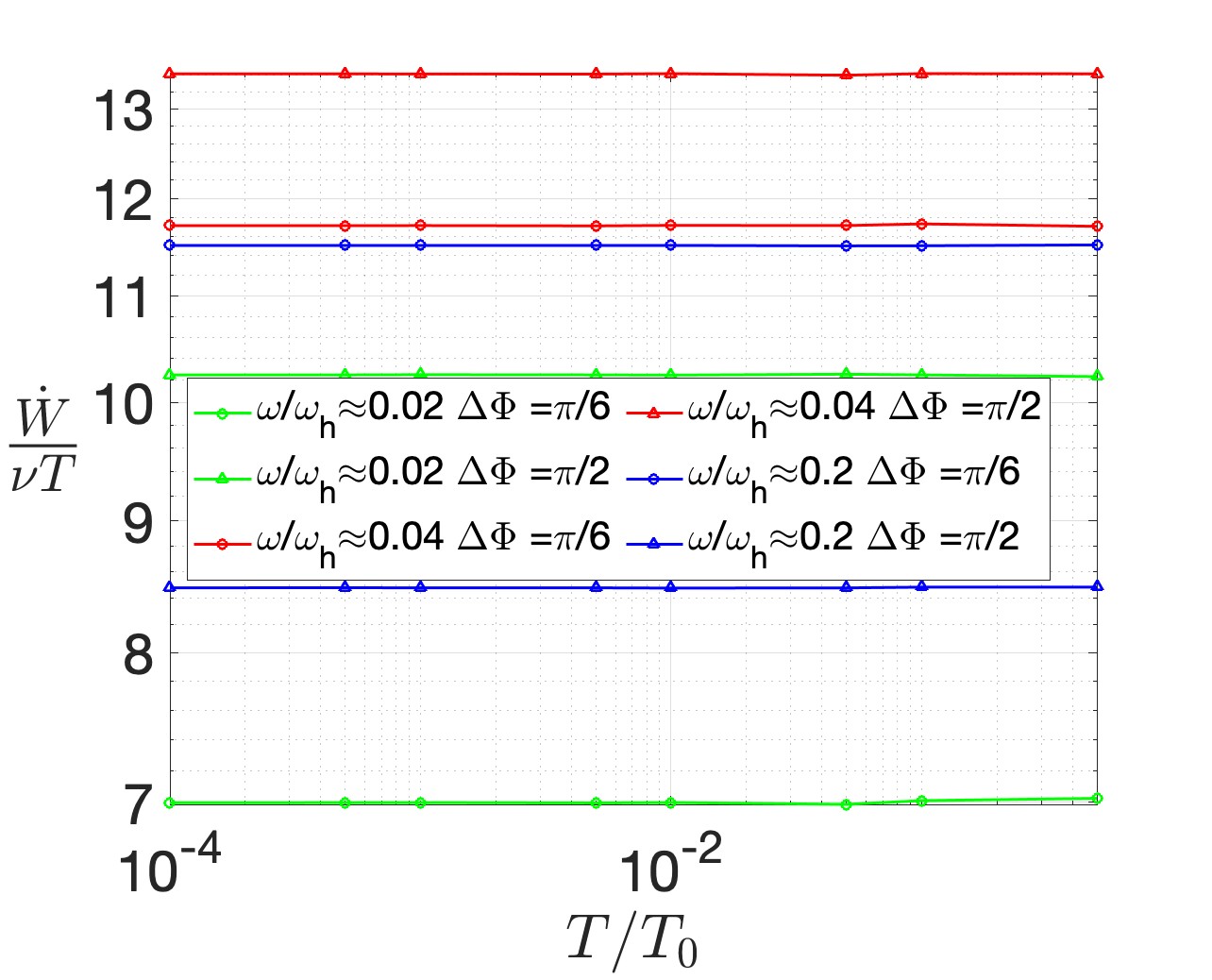}
      \label{}}
      ~
      \hfil
      \sidesubfloat[]
      {\includegraphics[width=0.43\columnwidth]{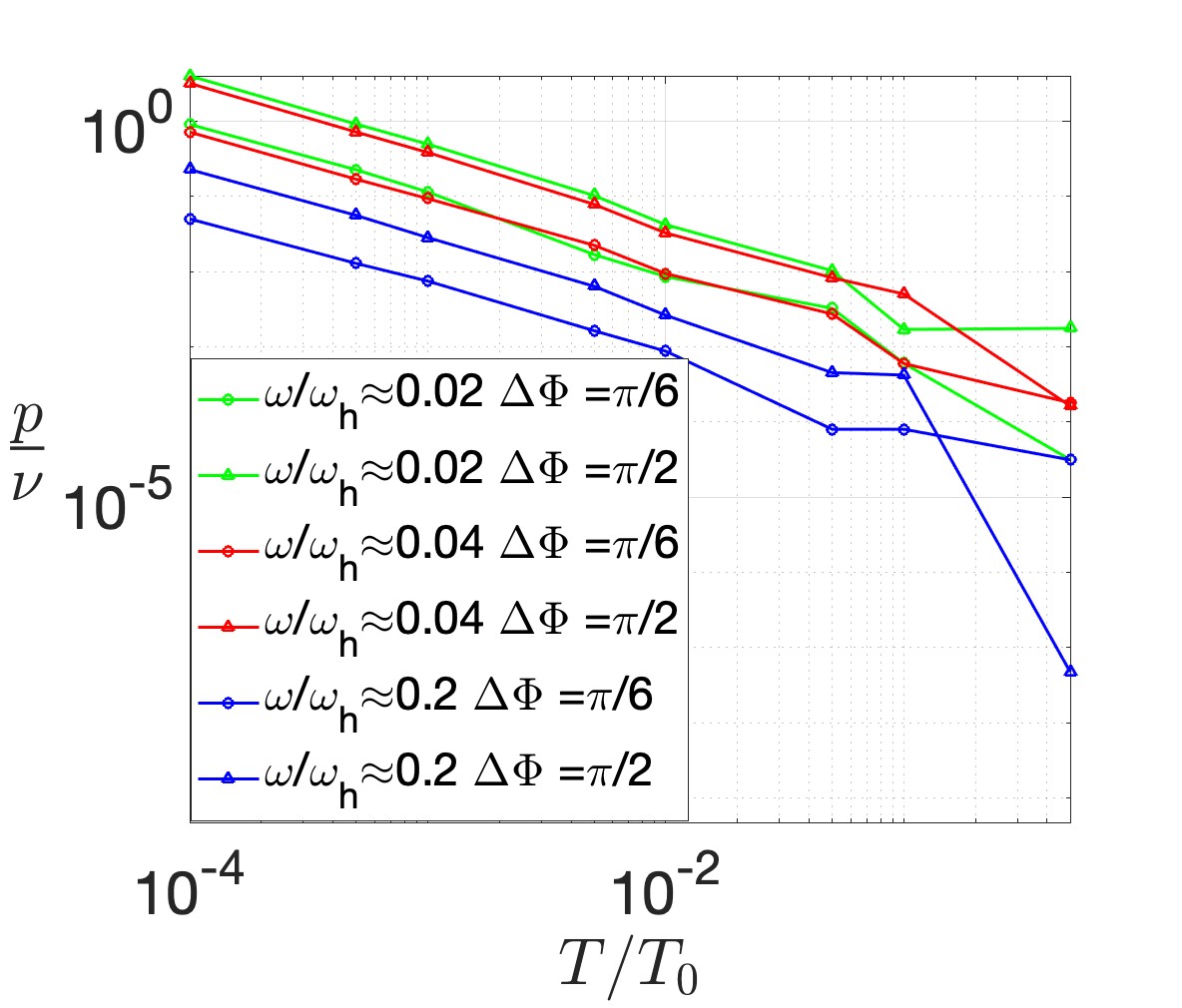}
      \label{}}
    \caption{Effect of the temperature of the fluid $T$ upon (a) the average swim velocity, (b) the swimming diffusivity, (c) the energy consumption rate and (d) the precision. Parameters: $L=20$, $F_0=10$, $a=1$, $\eta=1$, $K=2$.  }
    \label{fig:vsT}
\end{figure}

The fluid is characterized by viscosity $\eta$ and temperature $T$. As shown in the equations of motion~\eqref{eq:langtotalintegration}, there is not a trivial rescaling of time or positions with $\eta$ or $T$, unless in the noiseless limit $T \to 0$ where time can be safely rescaled with $\eta$. Therefore Fig.~\ref{fig:vseta} and Fig.~\ref{fig:vsT}  show the genuine effect of noise on the system. Both velocity and diffusivity decrease with $\eta$, however the first seems to reach a constant value for small viscosities. Work rate and precision have a maximum for a similar viscosity value. The non-monotonous behavior of the work rate is well reproduced by the analysis - obtained in the linear approximation - exposed in the Section~\ref{sec:analyt} and in the Appendix. 
 
Less complicate, at least on the empirical side, is the behavior of the interesting observables with $T$, see Fig.~\ref{fig:vsT}. The average velocity is apparently independent of $T$, and so is the work rate. Diffusivity grows linearly with $T$, as in the simplest scenario of an effective noise which is proportional to the amplitude of the single particle noises, and as a consequence the precision decreases as $1/T$ for the same reason.

\section{Efficiency through the Thermodynamic Uncertainty Relation}
\label{sec:tur}

In view of the Thermodynamic Uncertainty Relation, Eq.~\eqref{eq:tur}, we are interested in the TUR-based efficiency
\begin{equation}
    e_{TUR}=\frac{p}{p_{max}} =\frac{v^2 k_B T}{D \dot W} \le 1,
\end{equation}
which is a figure of merit with respect to the maximum achievable swimming precision.

Let us briefly discuss also energetic efficiency. For an engine its most direct definition is the ratio between energy produced and energy spent. The problem, with a simple swimmer such as ours, is that it obeys, in the steady state, a balance between external forces and hydrodynamic resistance, therefore between spent work and produced energy (the confining potential constitutes an exact difference which vanishes along stationary averages), leading to efficiency $1$. It makes sense, therefore to consider the so-called "low-Re swimming efficiency", $e_L$, given by 
\begin{equation}
e_L=\frac{v \mathcal{F}}{\dot W}
\end{equation}
where $\mathcal{F}$ is the force required to drag a rigid body (with some property shared with our swimmer, e.g. shape, size etc.) at the time-averaged swimming velocity $v$, while $\dot W$ is the average rate of work done by the active forces $\dot W=\langle \sum_i \dot{x}_i F^{act}_i\rangle$ in the stationary swimming regime. The drag force can be put in a general form valid for low-Re regimes, i.e. $\mathcal{F}=6\pi\eta a_{eff} v$ where $\eta$ is the fluid viscosity and $a_{eff}$ is an effective radius which we set to $L$.
\textcolor{blue}{}
In principle $E$ does not have to be less then one, but for many cases - particularly in biology - it is smaller than $1$, typically of order $1\%$ and therefore it is generally regarded as an efficiency or as a figure of merit. It is called sometimes Lighthill efficiency or Froude efficiency~\cite{lauga2009hydrodynamics,michelin2010efficiency}.

We immediately note that there is a simple connection between hydrodynamic efficiency and TUR-based efficiency:
\begin{equation}
e_L=e_{TUR} \frac{6\pi\eta a_{eff} D}{k_B T}
\end{equation}
so that when $D=k_B T/(6 \pi \eta a_{eff})$, one has $e_{TUR}=e_L$. This also implies that the precision of a simple sphere dragged by a constant force is maximum.

\begin{figure}
\centering
\sidesubfloat[]
        {\includegraphics[width=0.43\columnwidth]{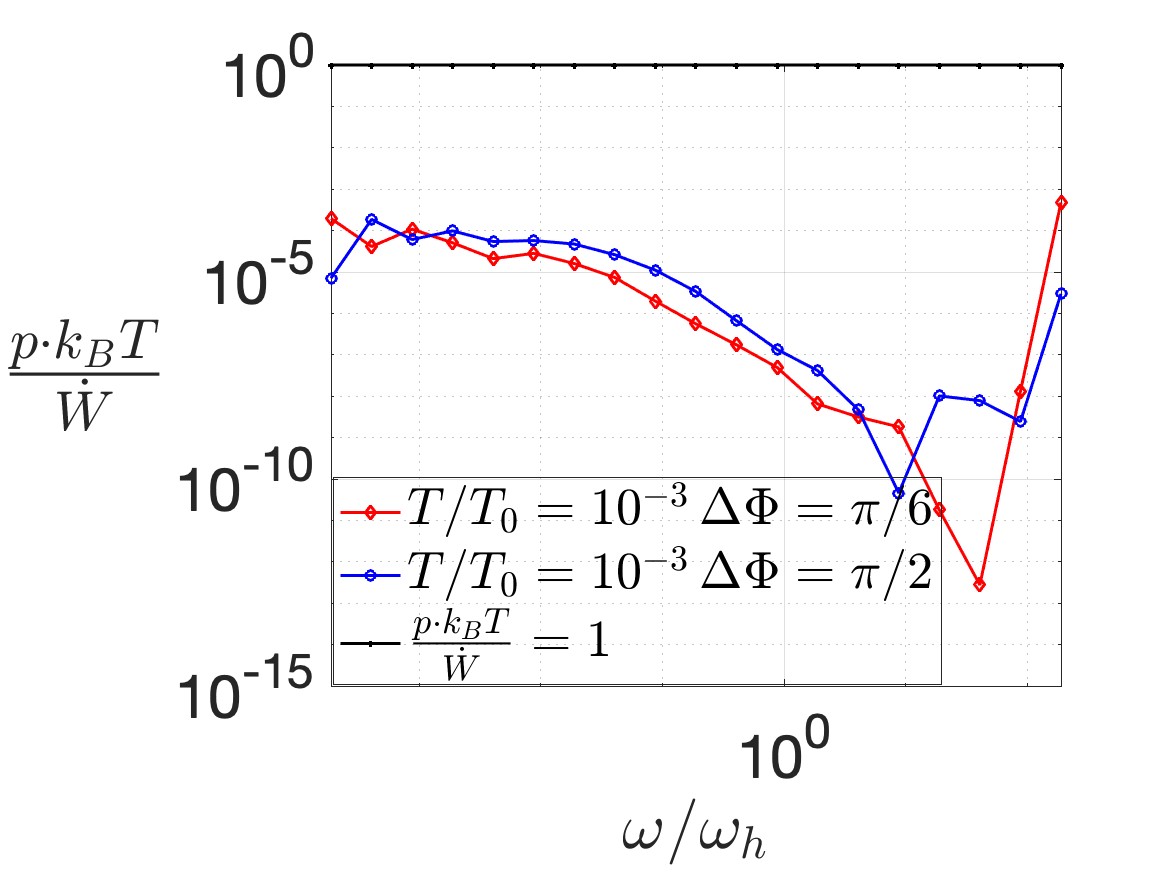}
        \label{}}
~
\hfil
\sidesubfloat[]
        {\includegraphics[width=0.43\columnwidth]{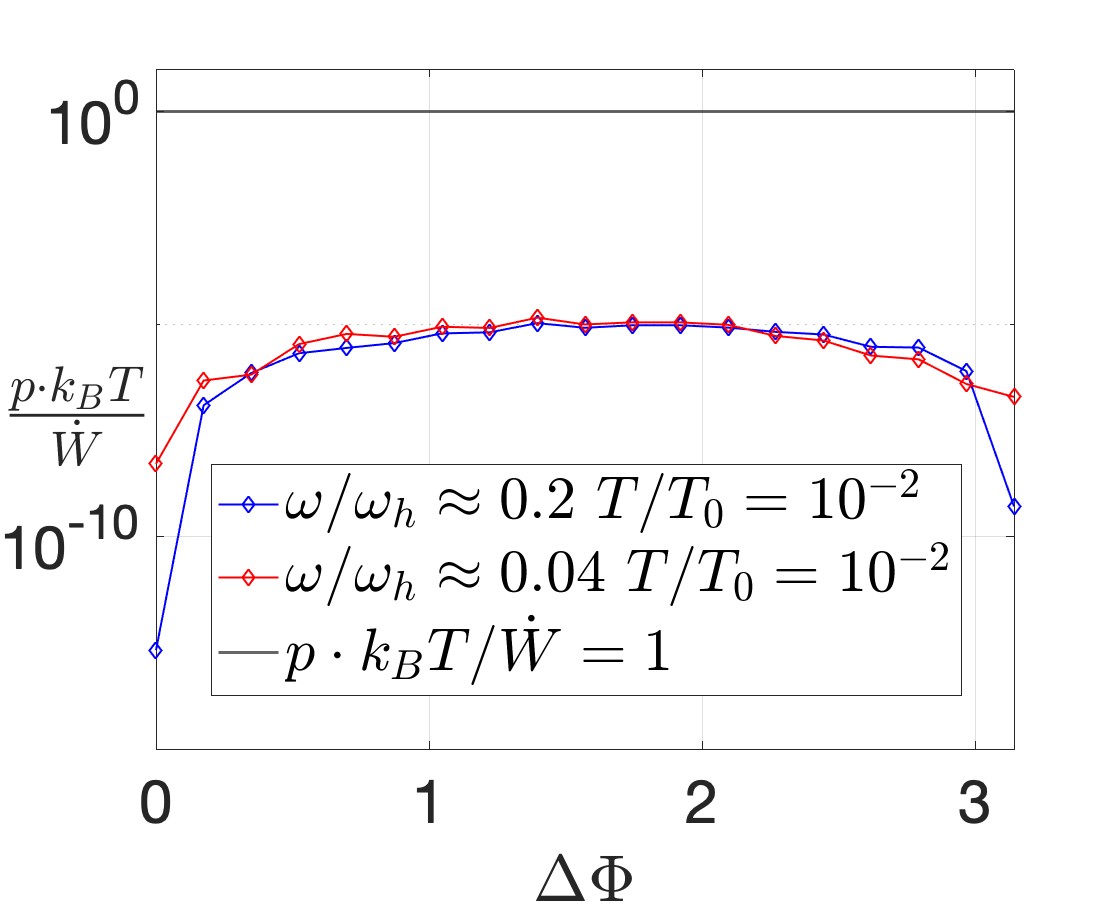}
        \label{}}
\
\sidesubfloat[]
        {\includegraphics[width=0.43\columnwidth]{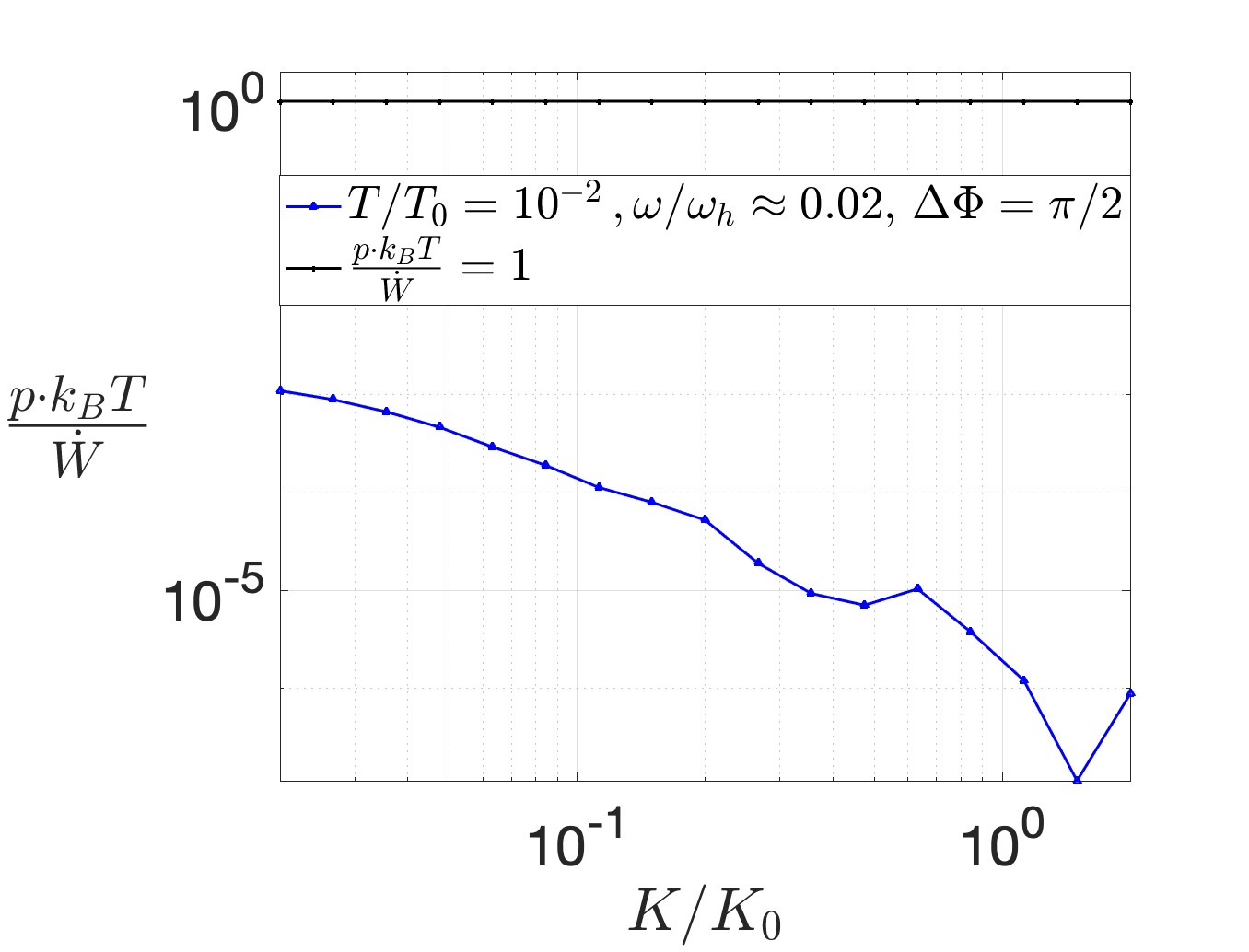}
        \label{}}
~
\hfil
\sidesubfloat[]
        {\includegraphics[width=0.43\columnwidth]{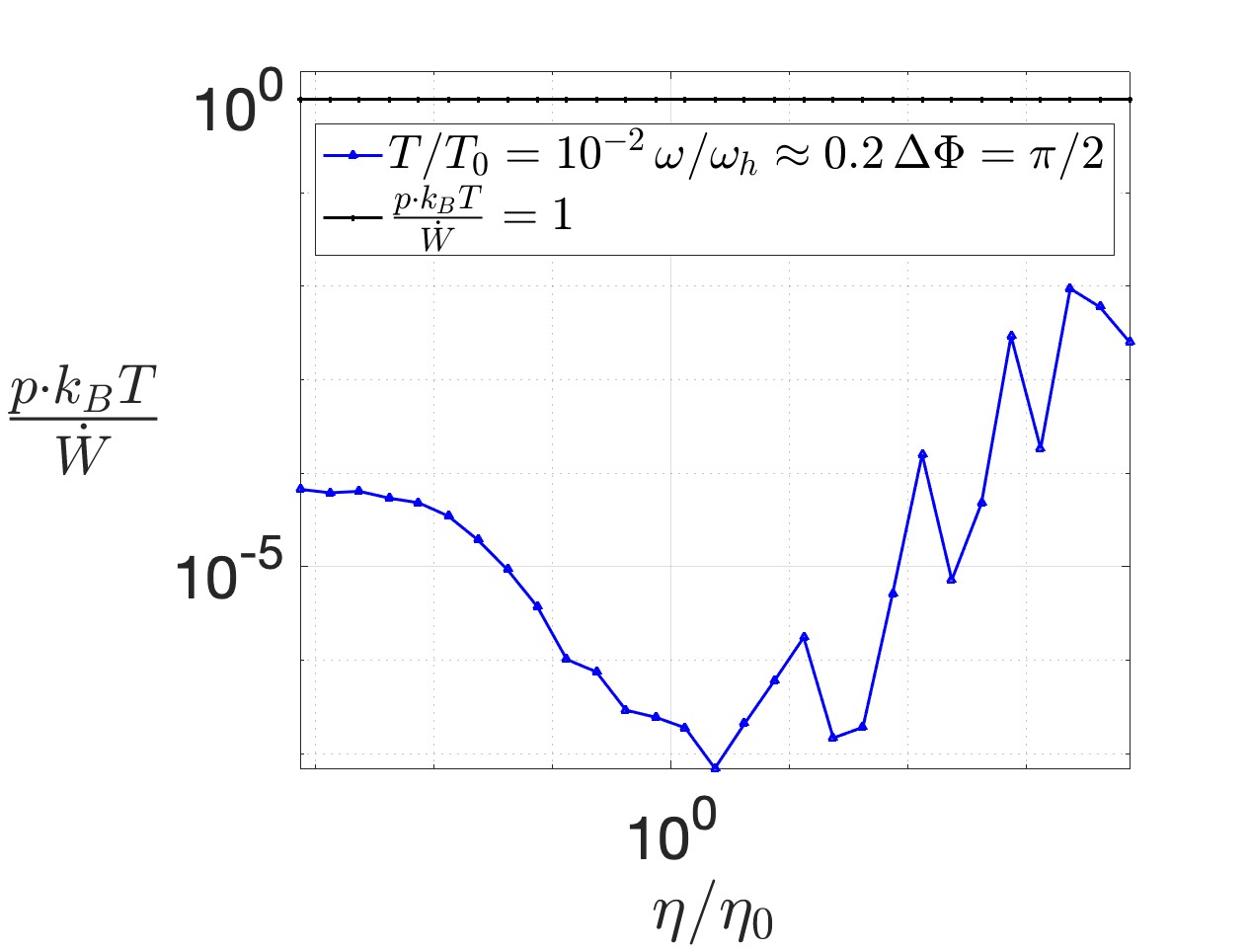}
        \label{}}
\
\sidesubfloat[]
        {\includegraphics[width=0.43\columnwidth]{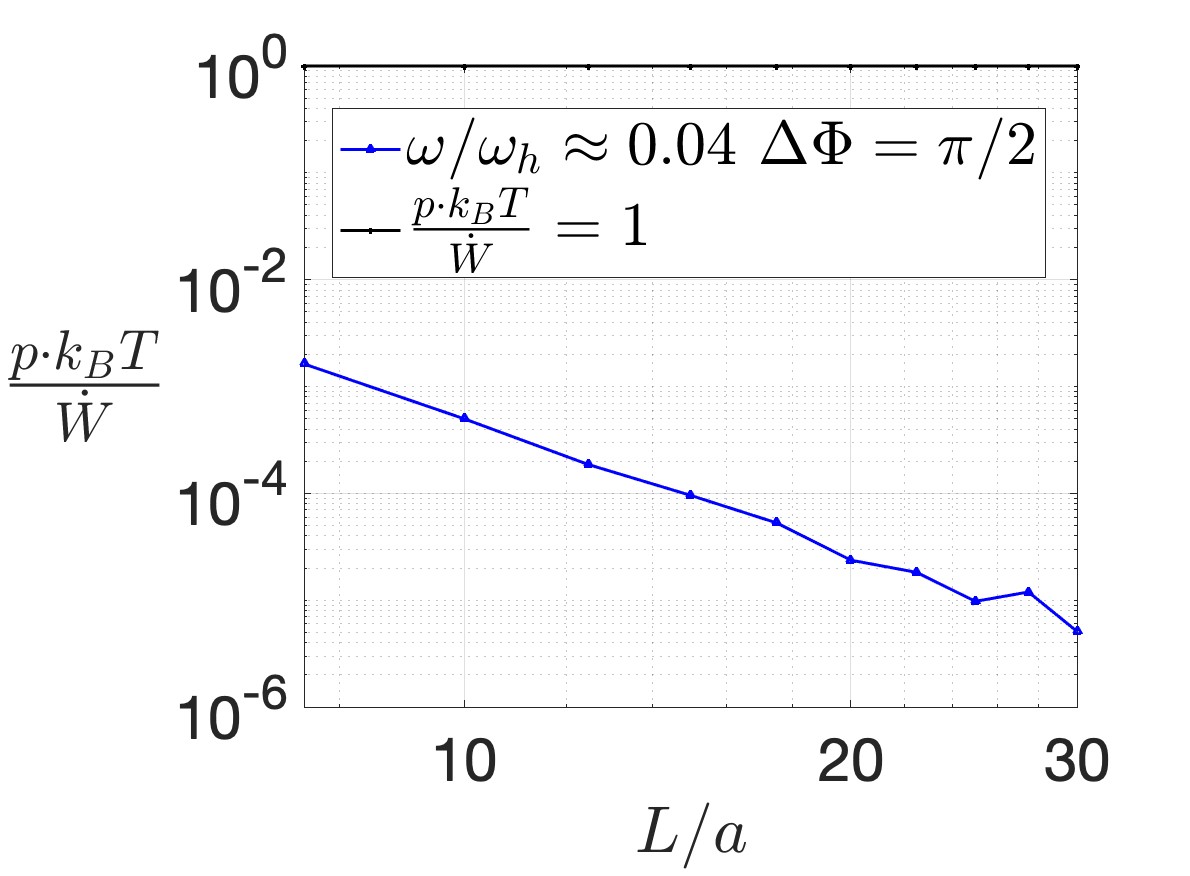}
        \label{}}
~
\hfil
\sidesubfloat[]
        {\includegraphics[width=0.43\columnwidth]{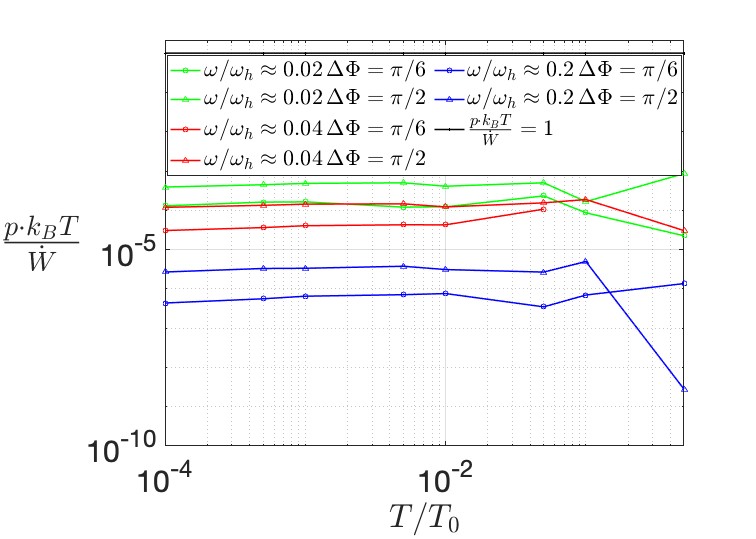}
        \label{}}
        
    \caption{Effect of the physical parameters upon the TUR-based $e_{TUR}$ efficiency. Parameters where not specified: $L=20$, $F_0=10$, $a=1$, $\eta=1$, $T=0.01$, $\Delta\phi=\pi/2$, $\omega=2\pi/50$, $K=2$.}
    \label{fig:eff}
\end{figure}

In figure~\ref{fig:eff} we display the TUR-based efficiency $e_{TUR}$ with its dependence upon the six parameters of the model we have considered so far. Our first remark is that the precision of the model is, in general, smaller than the allowed maximum by several orders of magnitude. The highest observed values of $e_{TUR}$ are obtained for small $K$ and small $L$ and are of the order of $10^{-3}$. 

The efficiency is weakly dependent upon $\Delta \phi$ with an optimum around $\pi/2$. It does not depend evidently on $\omega$ when $\omega<\omega_{opt}$  (we recall that $\omega_{opt}$ is where the velocity and the precision are highest); however the efficiency rapidly decreases with $\omega$ when it is larger than $\omega_{opt}$. It is difficult to validate the apparent growth of $e_{TUR}$ for very high frequencies, but it could be just wide fluctuations induced by the strong noise affecting diffusivity and - as a consequence - precision, see Fig.~\ref{fig:vsomega}.

The efficiency decreases when both $K$ and $L$ are increased, but the effect of $K$ is soft (i.e. $e_{TUR} \sim 1/K$) while the effect of $L$ is relevant, e.g. it decreases by two orders of magnitude increasing $L$ by less than a factor $4$. Such an observation seems to contradict a recent study on the Lighthill efficiency of the original three-beads model~\cite{nasouri2019efficiency}, however a direct comparison is not correct: in the original model, in fact, the $L$ parameter represents the maximum extension of the swimmer's arms, which is externally imposed, while in our case $L$ is the length at rest of the arms, while the real excursion of their length is dictated by the dynamics under the effect of the active forces and the harmonic confinement. 

The efficiency has an almost negligible dependence upon the fluid temperature, however it becomes very noisy for large values of $T$. The effect of the fluid viscosity $\eta$ is surprisingly non-monotonic, with a minimum at values of viscosity between $10$ and $10^2$, which (under the conversion discussed in subsection~\ref{sec:units}) corresponds roughly to water viscosity.

\section{Analytical study for small deformation}
\label{sec:analyt}

In this last Section we discuss an analytic approach mainly focused to the computation of the average velocity of the swimmer model considered here, which - even without noise - is different from the original model for the presence of the confining potential. In the final part of this Section we also discuss zero-order approximations for its diffusivity and, consequently, its thermodynamic precision. A full treatment of the stochastic problem is left to a future study. 

\subsection{Linearised equations for the average motion}

We define $L_1(t)=l_1+u_1(t)$ and $L_2(t)=l_2+u_2(t)$. The equation of motion, Eq.~\eqref{eq:langtotalintegration} after averaging over noise,  reads for small $u_1,u_2$:
\begin{equation}
\label{eq:analyticalexpequation}
    \underline{v} = \mathbf{T}\cdot\underline{F} \approx (\mathbf{T}_0+\mathbf{T}_1)\cdot(\underline{F}^{act}+\underline{F}^{pot}) + (\underline{F}_{Ito,0}+\underline{F}_{Ito,1})\,.
\end{equation}

In fact in the small $u_1, u_2$ limit we can expand the mobility matrix:
\begin{align}
\label{eq:Oseenapprox}
    \mathbf{T} &\approx \mathbf{T}_0+\mathbf{T}_1\\
  \mathbf{T}_0&=   \begin{pNiceMatrix}
        \frac{1}{6\pi\eta a} & \frac{1}{4\pi\eta l_1} & \frac{1}{4\pi\eta (l_1+l_2)} \\
        \frac{1}{4\pi\eta l_1} & \frac{1}{6\pi\eta a} & \frac{1}{4\pi\eta l_2} \\
        \frac{1}{4\pi\eta (l_1+l_2)} & \frac{1}{4\pi\eta l_2} & \frac{1}{6\pi\eta a}
    \end{pNiceMatrix}\\
  \mathbf{T}_1 &=   -    \begin{pNiceMatrix}
        0 & \frac{u_1}{4\pi\eta l_1^2} & \frac{u_1+u_2}{4\pi\eta (l_1+l_2)^2} \\
        \frac{u_1}{4\pi\eta l_1^2} & 0 & \frac{u_2}{4\pi\eta l_2^2} \\
        \frac{u_2}{4\pi\eta (l_1+l_2)^2} & \frac{u_2}{4\pi\eta l_2^2} & 0
    \end{pNiceMatrix}\,.
\end{align}

Also the Ito "forces" can be expanded at first order in $u_1,u_2$:
\begin{align}
\label{eq:itoforcesapprox}
\underline{F}_{ Ito} &\approx \underline{F}_{Ito,0}+\underline{F}_{Ito,1}\\
\underline{F}_{Ito,0}&=\frac{1}{\beta}\frac{1}{4\pi\eta}
    \begin{pNiceMatrix}
        \frac{1}{(l_1+l_2)^2} + \frac{1}{l_1^2} \\
        \frac{1}{l_2^2}-\frac{1}{l_1^2}\\
        \frac{1}{(l_1+l_2)^2} + \frac{1}{l_2^2}
    \end{pNiceMatrix}\\
    \underline{F}_{Ito,1}&=
\frac{1}{\beta}\frac{1}{4\pi\eta}
    \begin{pNiceMatrix}
        -2\,\big( \frac{u_1}{l_1^3} + \frac{u_1+u_2}{(l_1+l_2)^3}\big) \\
        2\,\big( \frac{u_1}{l_1\,l_2^2} - \frac{u_2}{l_2^3} +\frac{u_1}{l_1^3} -\frac{u_2}{l_2\,l_1^2}\big) \\
        2\,\big( \frac{u_2}{l_2^3} + \frac{u_1+u_2}{(l_1+l_2)^3}\big)
    \end{pNiceMatrix}\,.
\end{align}

Interestingly, we note that the Ito "forces" $\underline{F}_{Ito,i}$  contains terms of order $1/l^2$ and $u/l^3$ where $l$ is $l_1$ or $l_2$. All these terms are smaller than the terms $1/l$ and $u/l^2$ contained in the expansion of ${\mathbf T}$, therefore at our level of approximation we can drop the Ito "forces". Also the term $\mathbf{T}_1 \cdot \underline{F}^{pot}$ can be dropped as it is of order $\sim u^2$.

Finally, the above equation can be put in the form of an equation for the time-derivative of the two only relevant degrees of freedom  $u_1,u_2$, i.e.:
\begin{equation}
    \dot u_1 = v_1 - v_2\,, \qquad \dot u_2 = v_2 - v_3\,,
\end{equation}
with time-dependent forces reduced only two components $\underline{f}_{act}^{(2)}(t)=(F^{act}_1(t),F^{act}_3)(t)$,
obtaining
\begin{equation}
\label{eq:equation_dot_u}
    \underline{\dot u}(t)=\big( \mathbf{M}_1(t) + \mathbf{M}_2 \big)\cdot \underline{u}(t) + \mathbf{M}_3\cdot\underline{f}_{act}^{(2)}\,,
\end{equation}
with 
\begin{equation}
    \mathbf{M}_1(t)=-\frac{1}{4\eta\pi}\left(
\begin{array}{cc}
 \frac{-f_3^{act}(t)}{l_{12}^2}-\frac{2 f_1^{act}(t)}{l_1^2} & -\frac{f_3^{act}(t)}{l_{21}^2} \\
 \frac{f_1^{act}(t)}{ l_{12}^2 } & \frac{2f_3^{act}(t)}{l_2^2}+\frac{f_1^{act}(t)}{l_{21}^2} 
\end{array}
\right)
\end{equation}
where $1/l_{12}^2=1/l_1^2-1/(l_1+l_2)^2$ and $1/l_{21}^2=1/l_2^2+1/(l_1+l_2)^2$.
\begin{equation}
    \mathbf{M}_2=-\frac{K}{\eta\pi}\left(
\begin{array}{cc}
 \frac{1}{3a_1} & -\frac{1}{6a_{12}} \\
 -\frac{1}{6a_{12}} & \frac{1}{3a_2}
\end{array}
\right)
\end{equation}
where $1/(3a_1)=1/(3a)-1/(2l_1)$, $1/(3a_2)=1/(3a)-1/(2 l_2)$, $1/6 a_{12}=1/(6a)-1/(4l_1)-1/(4l_2)+1/(4(l_1+l_2))$, and finally
\begin{equation}
    \mathbf{M}_3=\frac{1}{\eta\pi}\left(
\begin{array}{cc}
 \frac{1}{3a_1} & \frac{1}{6a_{12}} \\
 -\frac{1}{6a_{12}} & -\frac{1}{3a_2}
\end{array}
\right).
\end{equation}

\begin{figure}
    \centering
    \includegraphics[width=0.98\columnwidth]{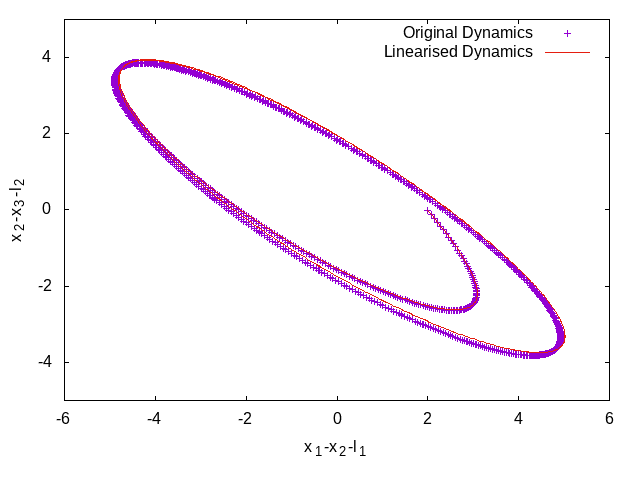}
    \caption{Comparison of the numerical solution of Eq.~\eqref{eq:equation_dot_u} with that of
the original model. The figure shows the limit cycle in which the displacements $u_1,u_2$ stay. Parameters are $L=20$, $F_0=10$, $a=1$, $\eta=1$,  $\Delta\phi=\pi/4$, $\omega=2\pi/50$, $K=2$. }
    \label{fig:compare}
\end{figure}

Before proceeding with the analytical calculations, we have verified the fairness of the linear assumption, by comparing the numerical solution of Eq.~\ref{eq:equation_dot_u} with that of the original model, see for instance Fig.~\ref{fig:compare}. The overlap is almost perfect.

\subsection{Instability without the confining potential}

Let us consider the case where $K=0$, i.e. there is no confining potential.

\begin{figure}
    \centering
    \sidesubfloat[]
        {\includegraphics[width=0.98\columnwidth]{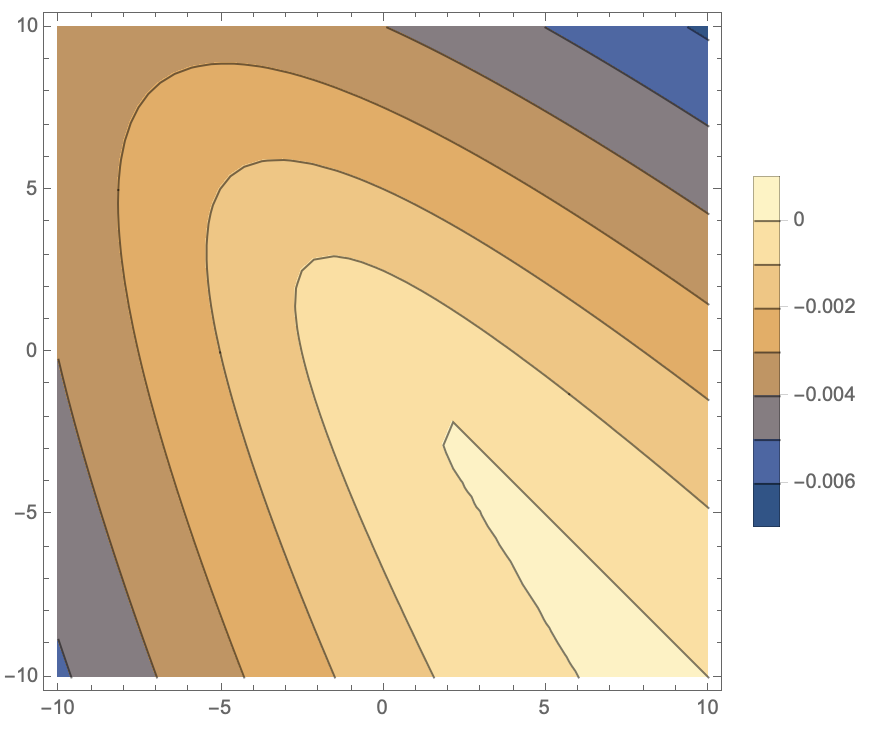}
        \label{}}
    \
    \sidesubfloat[]
        {\includegraphics[width=0.98\columnwidth]{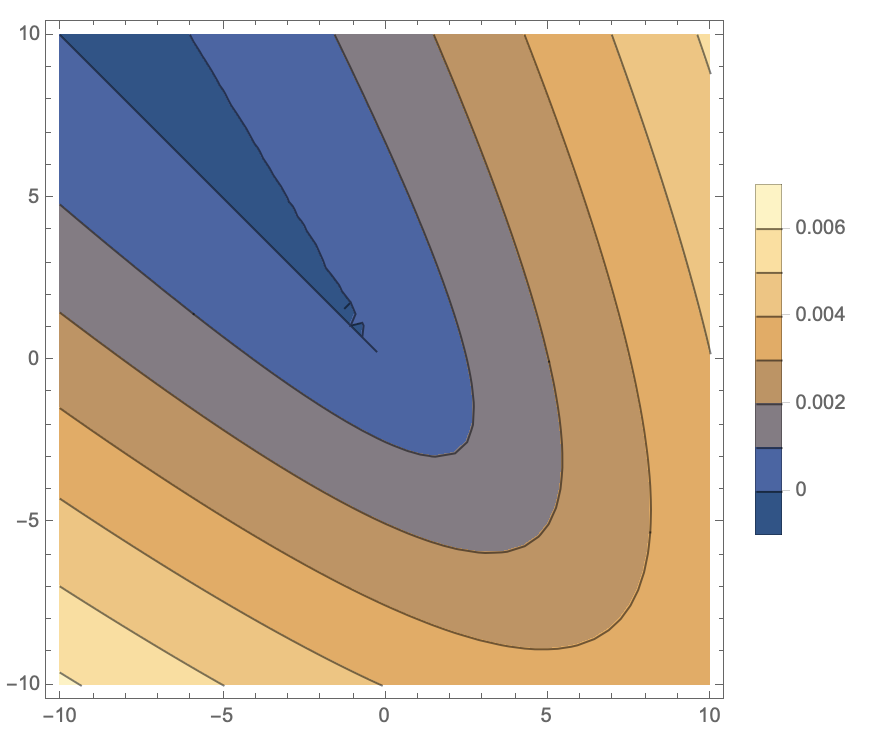}
        \label{}}
    \caption{Eigenvalues of the linearised dynamics, as functions of $F_1^{act},F_3^{act}$ when there is no elastic potential, i.e. when $K=0$. Parameters are $L=20$, $a=1$, $\eta=1$.}
    \label{fig:ei}
\end{figure}

In this case one has that, assuming a limit cycle $\underline{u}_0(t)$ which satisfies $\dot{\underline u}_0={\mathbf M}_1 {\underline u}_0 + {\mathbf M}_3 \underline {f}_{act}^{(2)}(t)$, then small deviations from it $\underline{\delta u}(t)=\underline{u}(t)-\underline{u}_0(t)$ obey the homogeneous equation $\dot{\delta u}=\mathbf{M}_1 \underline {\delta u}$ and therefore the stability of the cycle is dictated by the eigenvalues of $\mathbf{M}_1(t)$ which are, however, time-dependent, i.e. they depend upon the values of $f_1^{act}(t)$ and $f_3^{act}(t)$. In principle the limit cycle stability should be determined by studying the eigenvalues of the associated Poincar\'e map of the cycle which depend upon the eigenvalues along the full period of the force oscillation. Anyway, this problem is simplified here since, along the whole force cycle, one of the two eigenvalues is always positive, while the other is always negative. 
The full analytic formula for the eigenvalues is pretty long and can be found in the Appendix A. It is simplified in the case $l_1=l_2=L$ and takes the form
\begin{equation}
    \lambda_{+-}=\frac{3 F^{act}_1-5F^{act}_3\pm \sqrt{169 F^{act}_1+226 F^{act}_3 F^{act}_1+121 F^{act}_3}}{32 \pi\eta L^2}
\end{equation}
The plot of eigenvalues for a particular choice of the parameter, as a function of $F^{act}_1,F^{act}_3$ can be found in Fig.~\ref{fig:ei}.

\begin{figure}[H]
    \centering
    \sidesubfloat[]
        {\includegraphics[width=0.96\columnwidth]{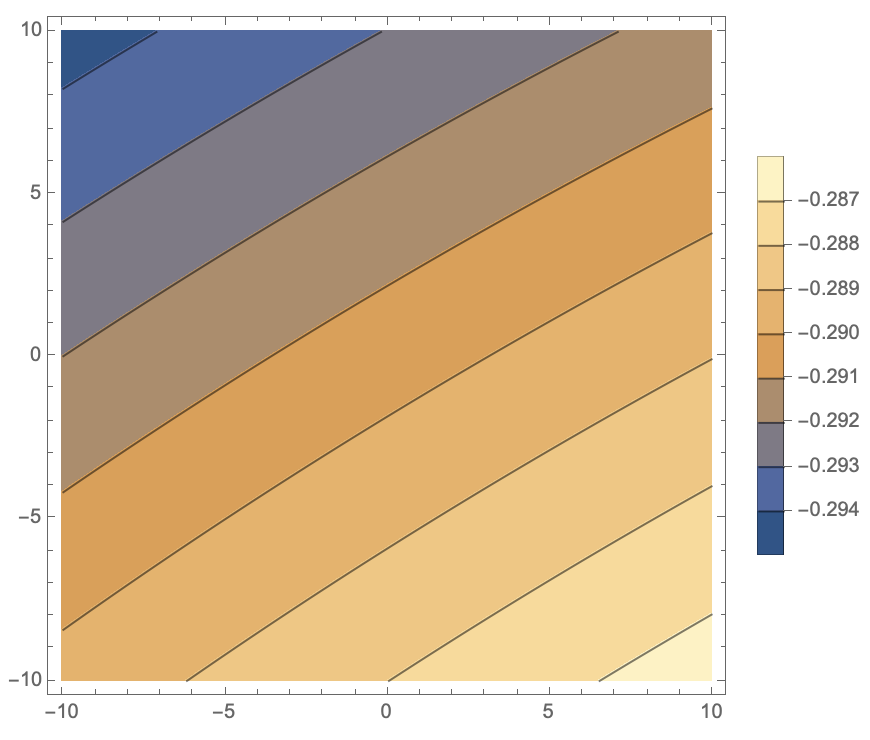}
        \label{}}
    \
    \sidesubfloat[]
        {\includegraphics[width=0.96\columnwidth]{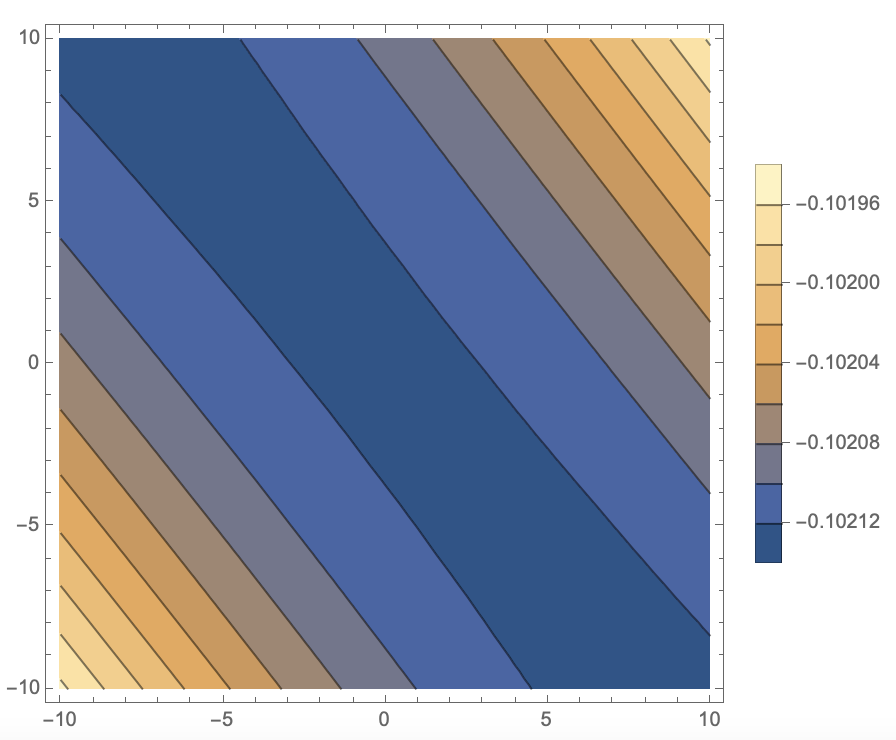}
        \label{}}
    \caption{Eigenvalues of the linearised dynamics, as functions of $F_1^{act},F_3^{act}$ in the presence of elastic potential. Parameters are $L=20$, $a=1$, $\eta=1$, $K=2$.}
    \label{fig:ei2}
\end{figure}

When also $M_2$ is considered, i.e. $K>0$, then the stability is restored as both eigenvalues become negative, as shown in Fig.~\ref{fig:ei2}.

\subsection{Solution for the limit cycle}

A further simplification can be operated on Eq.~\eqref{eq:equation_dot_u}, by considering that when $F^{act}=0$ (and $K>0$) the limit cycle becomes a stable fixed point with $u_1=u_2=0$. This implies that one may expect $u_1,u_2$ to be - for small $F^{act}$ - of the same order of $F^{act}$ and therefore the term ${\bf M}_1(t) \underline{u}$ is of order $(F^{act})^2 \approx \underline{u}^2$ and therefore can be dropped in our small deviations treatment. It is therefore easy to get a solution for the remaning system of equation, the strategy and the detailed results are given in Appendix, here we report the expression for the average velocity of the swimmer in the symmetric case $l_1=l_2=L=\ell a$:
\begin{equation}
\label{eq:avgvelniceform}
    v = \frac{\alpha}{2}\,F_0^2\,\omega\,\sin{(\Delta\phi)}\, Q(K,a\eta\omega,\ell)\,,
\end{equation}
where we recall, for simplicity. the expression for  $\alpha$ in the symmetric case:
\begin{equation}
    \alpha = \frac{7}{12}\frac{a}{L^2}\,.
\end{equation}
and the function $Q(K,a\eta\omega,\ell)$ which takes the form
\begin{equation}
    Q=-\frac{q_1 q_2\left[q_3 a^2 \eta^2  \omega^2+q_1 q_2 K^2 \right]}{[q_4 a^2 \eta^2 \omega^2+ q_1^2 K^2 ][q_5 a^2 \eta^2  \omega^2+q_2^2 K^2 ]}
\end{equation}
with $q_1=4 \ell-7$, $q_2=4 \ell-3$, $q_3=192\pi^2 \ell^2$, $q_4=64 \pi^2 \ell^2$, $q_5=576 \pi ^2 \ell^2$.

\begin{figure}
\centering
\sidesubfloat[]
      {\includegraphics[width=0.44\columnwidth]{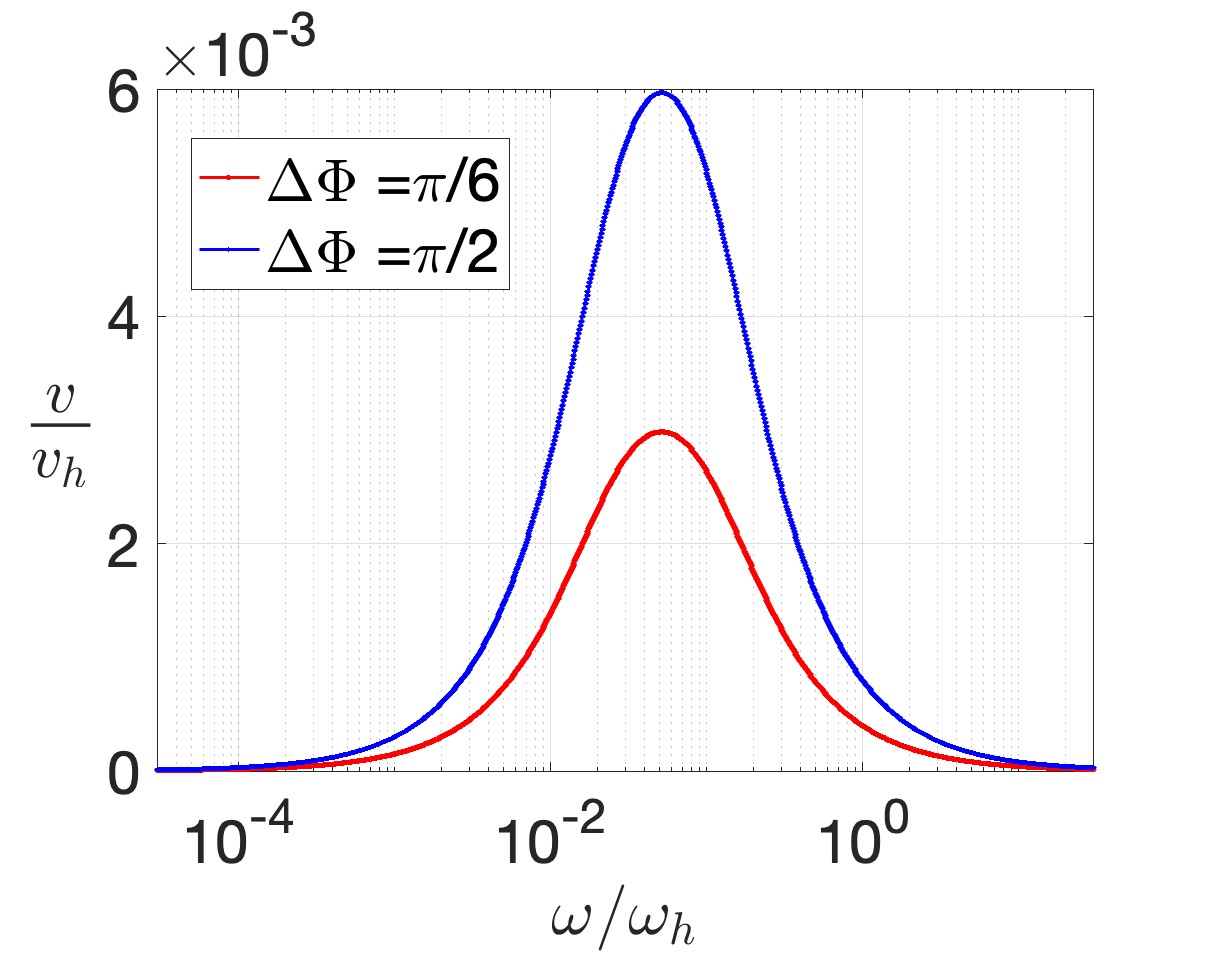}
      \label{}}
~
\hfil
\sidesubfloat[]
      {\includegraphics[width=0.44\columnwidth]{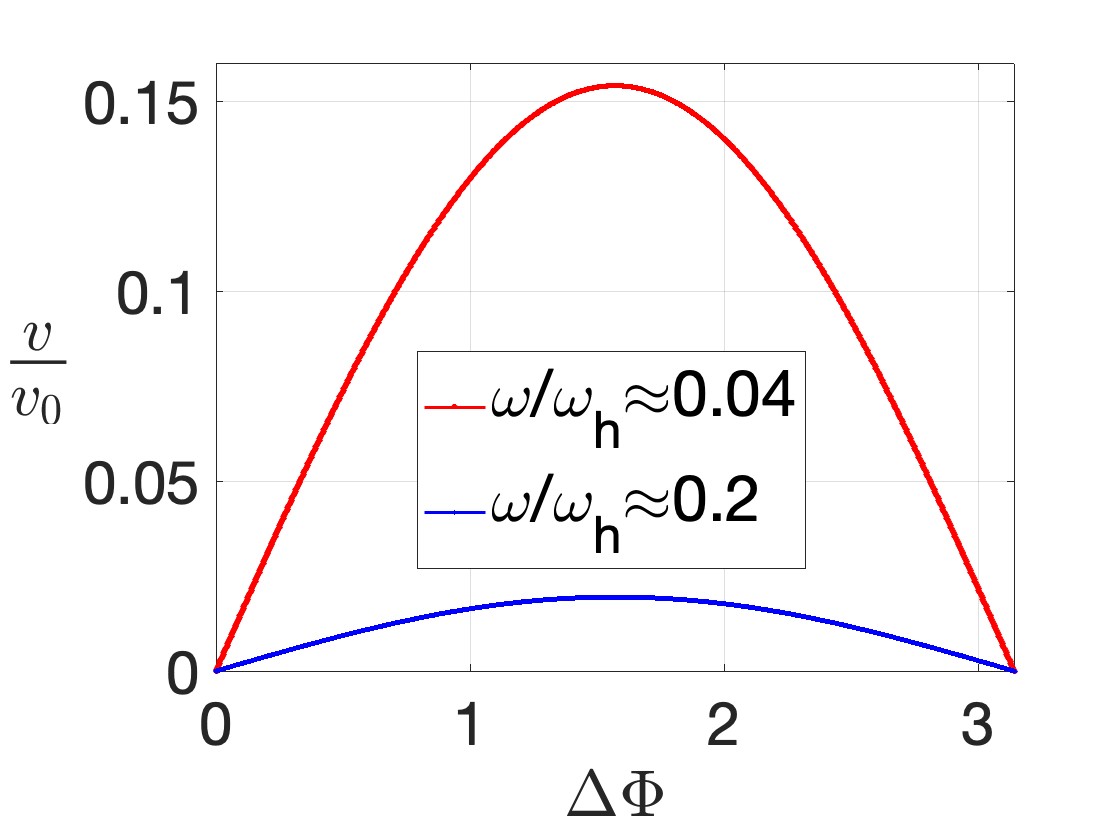}
      \label{}}
\
\sidesubfloat[]
      {\includegraphics[width=0.44\columnwidth]{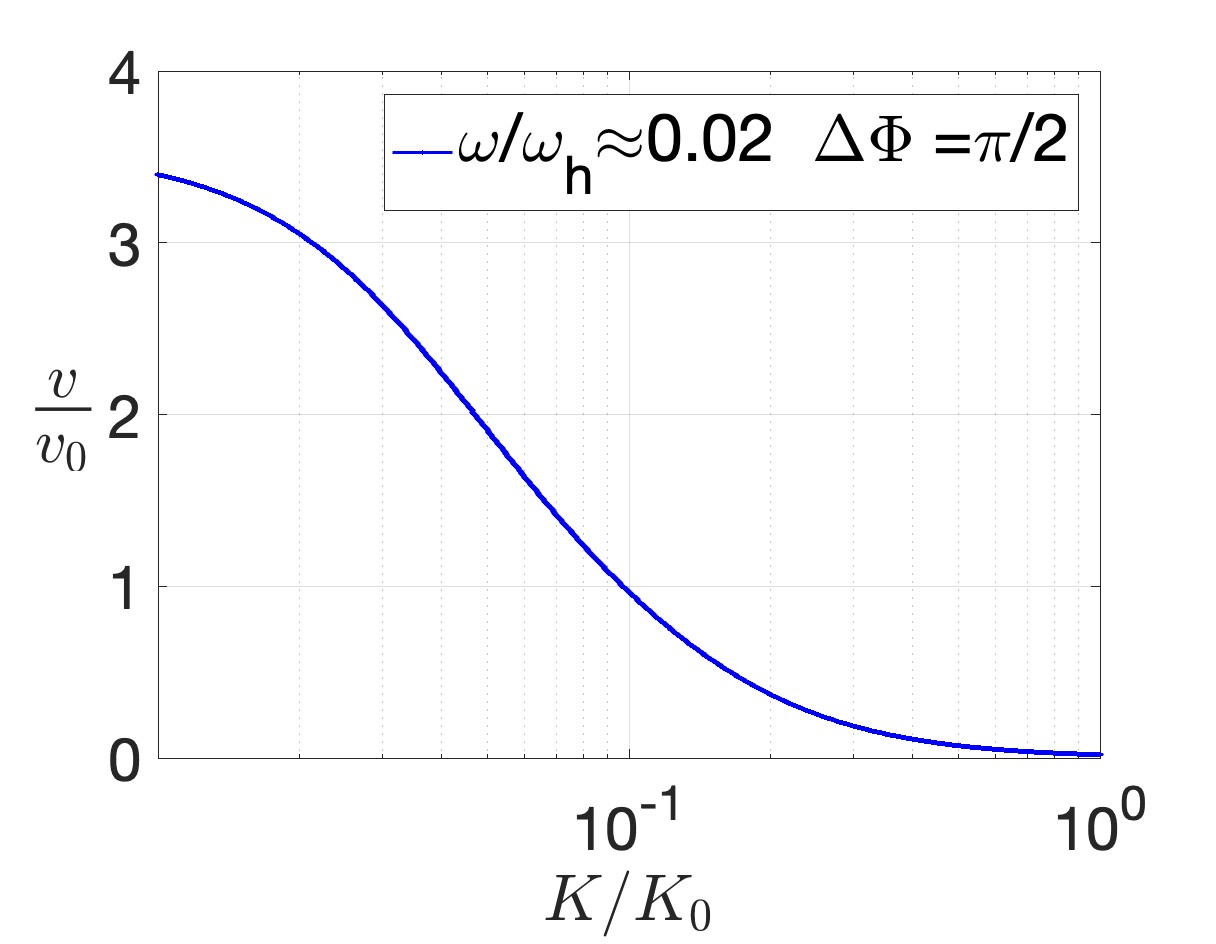}
      \label{}}
~
\hfil
\sidesubfloat[]
      {\includegraphics[width=0.44\columnwidth]{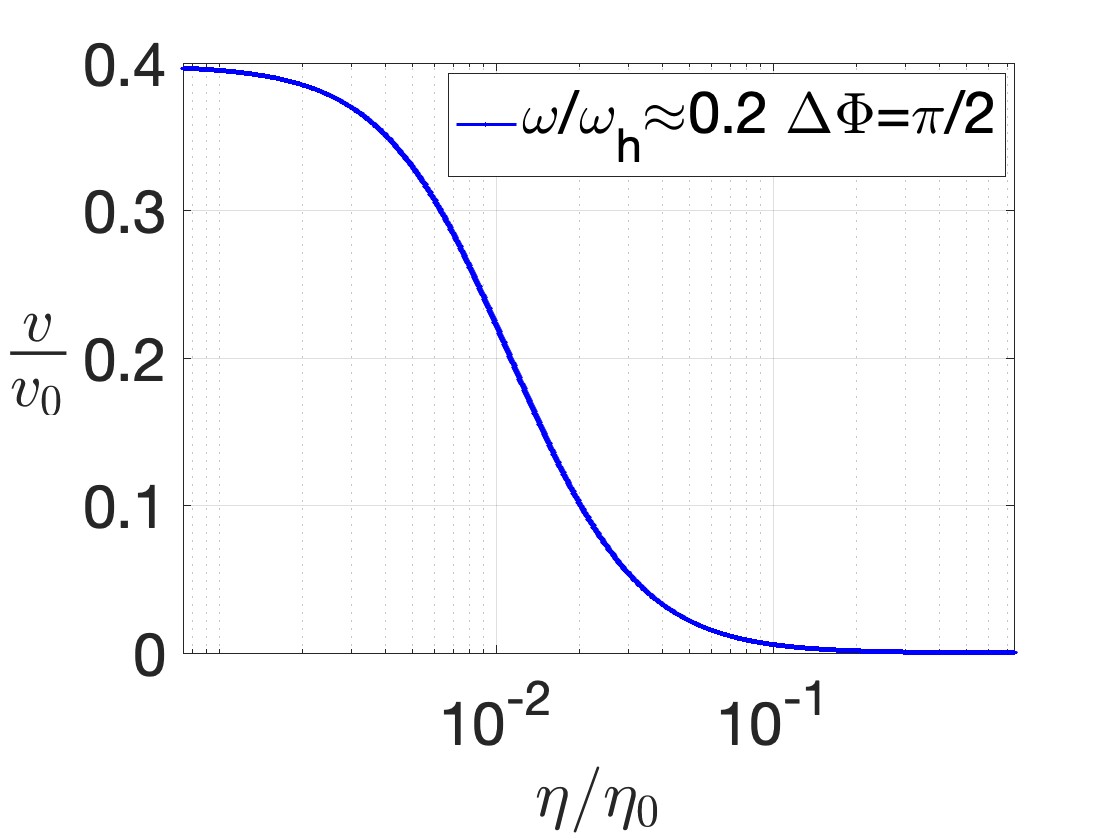}
      \label{}}
\caption{Theoretical behavior of average velocity with respect to  \textit{(a)}  frequency $\omega$,  \textit{(b)} phase difference $\Delta\phi$,  \textit{(c)} stiffness coefficient $K$ and  \textit{(d)}  viscosity $\eta$. Where not specified the parameters are $L=20$, $a=1$, $\eta=1$, $K=2$.}
    \label{fig:analyt}
\end{figure}

The plots of the analytical estimate of $v$ versus the main parameters of the model are shown in Fig.~\ref{fig:analyt}, a comparison with the numerical results in the previous Sections is excellent.

We can also obtain an expression for the average work rate which, putting the internal force condition and the definition o $u_1,u_2$ in the definition of $W$, Eq.~\eqref{eq:wdef}, reads:
\begin{equation} \label{wrate}
    \dot W = \frac{1}{\mathcal{T}}\int_0^\mathcal{T}dt (-F_1(t)\dot{u}_1(t)+F_3(t)\dot{u}_2(t)).
\end{equation}
Explicit formula for all the parameters are shown in Appendix, but in the symmetric case $l_1=l_2=\ell a$ we get
\begin{equation}
\dot W = \frac{8 \pi  a \eta F_0^2 l \omega^2 \left[\cos (\Delta \phi) W_1+ W_2\right]}{ W_3}
\end{equation}
with $W_1=w_1 a^2 \eta^2  \omega^2-w_2 K^2$, $W_2=w_3 a^2 \eta^2 \omega^2+w_4 K^2$, $W_3=w_5 a^4 \eta^4 \omega^4+ w_6 a^2 \eta^2 K^2 \omega^2+ w_7 K^4$ and with $w_1=192 \pi ^2 \ell^2 (4 \ell-9)$, $w_2=(4 \ell-7) (4 \ell-3)(4 \ell-9)$, $w_3=768 (2 l-3)\pi ^2 \ell^2 $, $w_4=4 (2 l-3) (4 \ell-7) (4 \ell-3)$, $w_5=36864 \pi ^4 \ell^4$, $w_6 = 128 \pi ^2 \ell^2 (8 l (10 l-33)+225) $, $w_7=(3-4 l)^2 (7-4 l)^2$.

\begin{figure}[htb]
\centering
\sidesubfloat[]
      {\includegraphics[width=0.45\columnwidth]{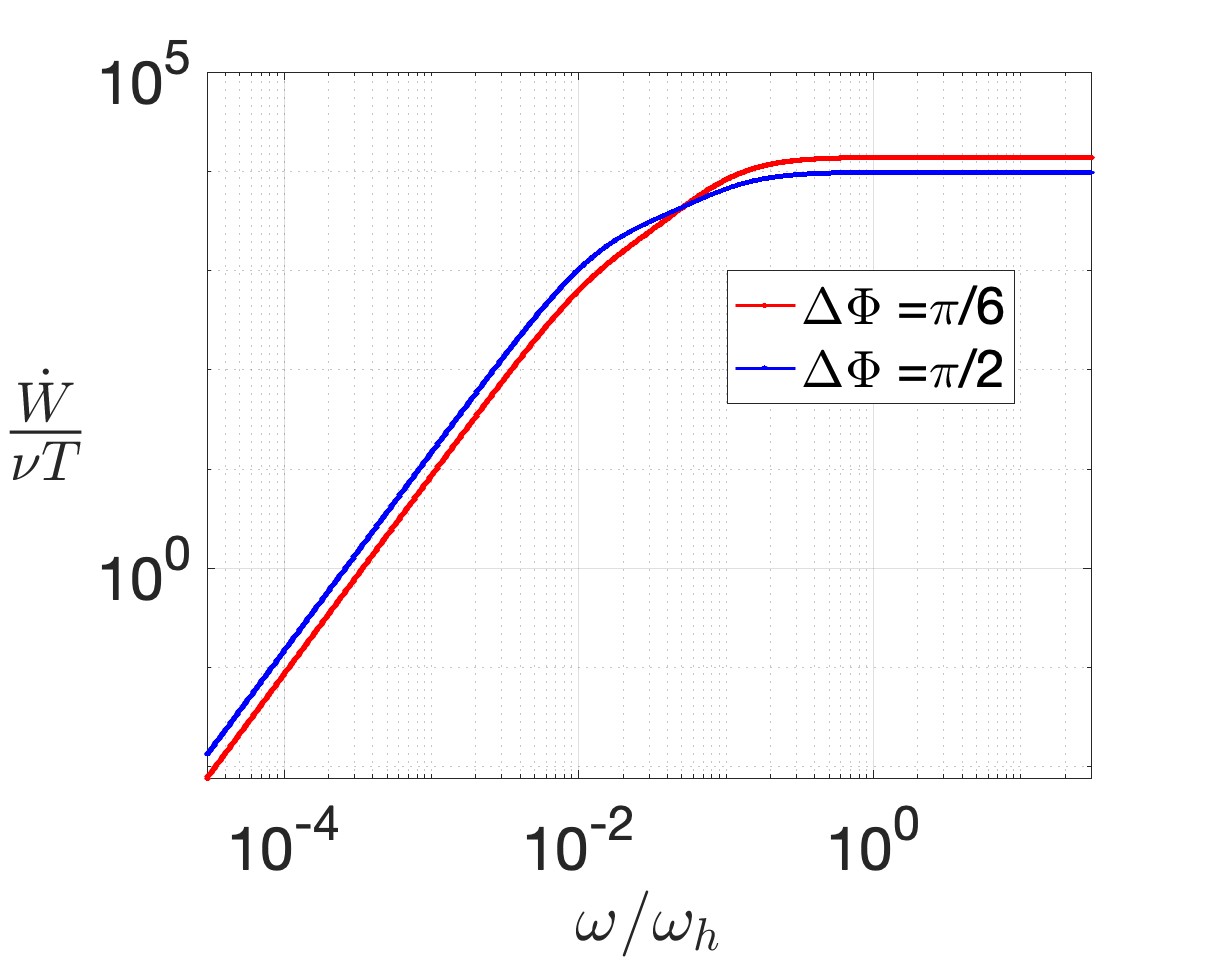}
      \label{}}
~
\hfil
\sidesubfloat[]
        {\includegraphics[width=0.45\columnwidth]{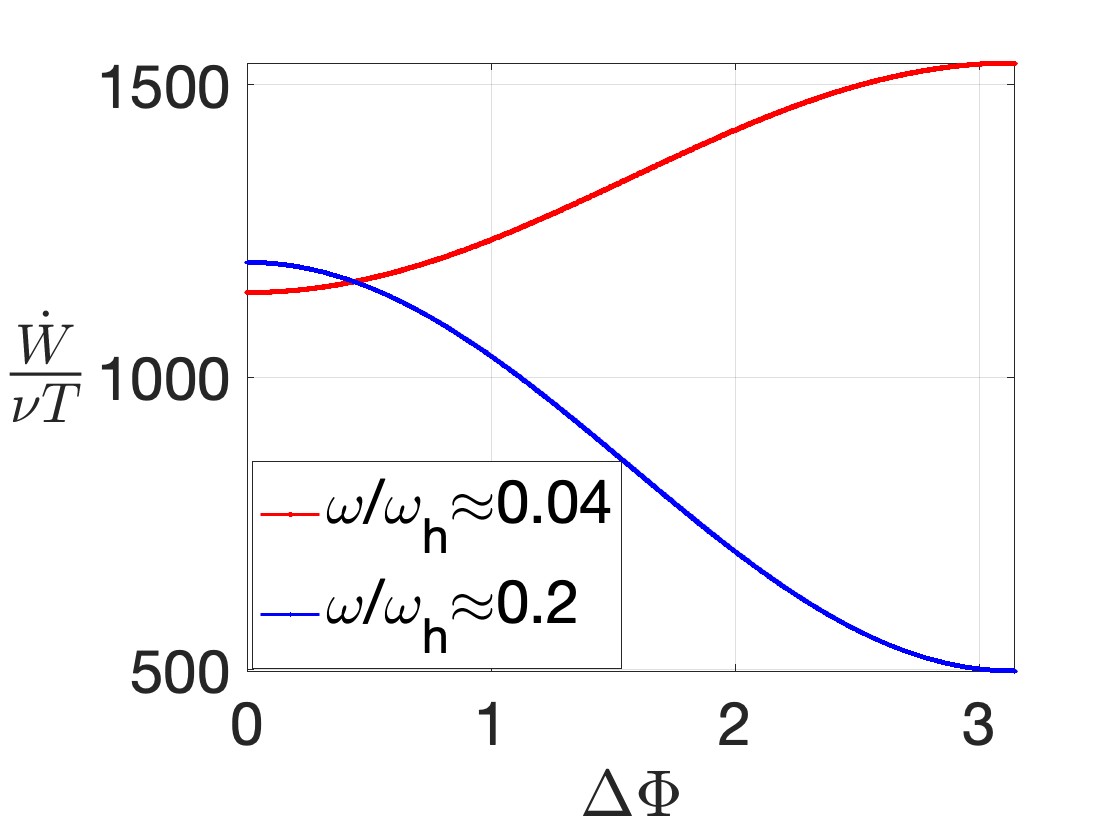}
        \label{}}
\
\sidesubfloat[]
        {\includegraphics[width=0.45\columnwidth]{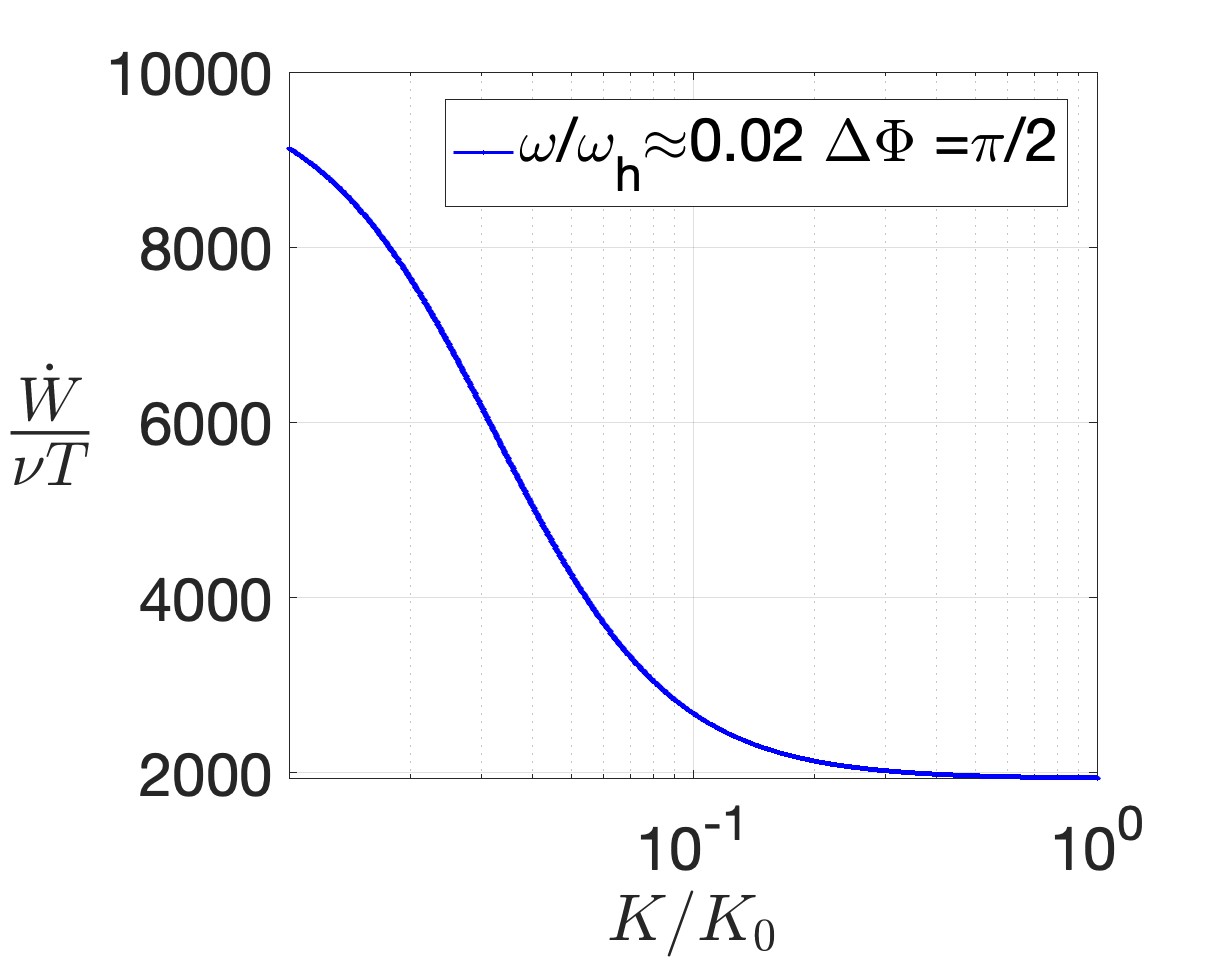}
        \label{}}
~
\hfil
\sidesubfloat[]
        {\includegraphics[width=0.45\columnwidth]{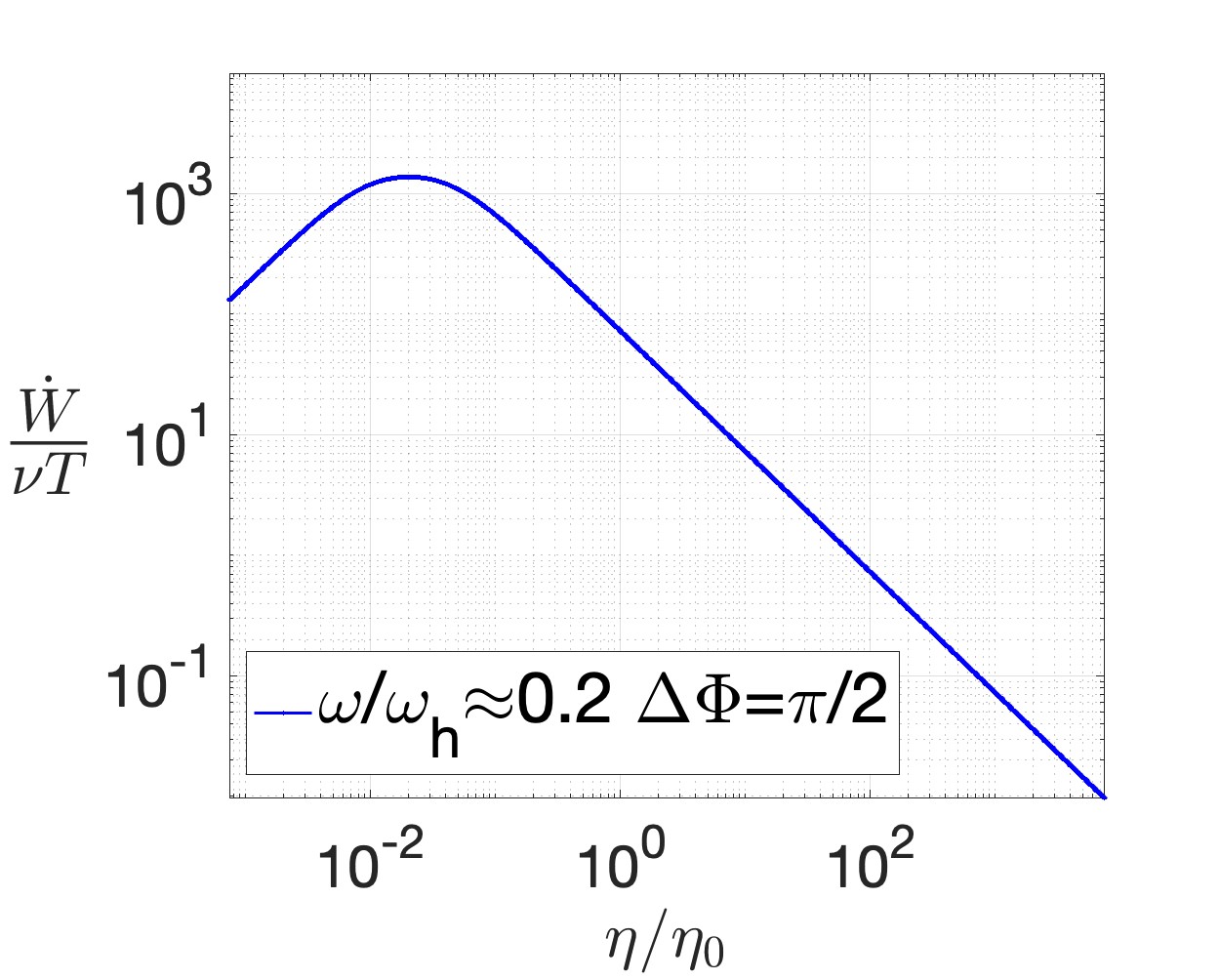}
        \label{}}
    \caption{Theoretical behavior of average work rate with respect to  \textit{(a)}  frequency $\omega$,  \textit{(b)} phase difference $\Delta\phi$,  \textit{(c)} stiffness coefficient $K$ and  \textit{(d)}  viscosity $\eta$. Where not specified the parameters are $L=20$, $a=1$, $\eta=1$, $K=2$.} 
    \label{fig:analytW}
\end{figure}
The plots of the analytical estimate of $\dot W$ versus the main parameters of the model are shown in Fig.~\ref{fig:analytW}. Again, the qualitative comparison with the numerical results in the previous Sections is excellent , the  quantitative one is also quite fair.

\subsection{Estimates for the diffusivity, the precision and the efficiency}

While the swimming velocity in the limit cycle can be estimated by using the average equation, i.e. neglecting noise terms, the diffusivity cannot: in our case where the noise is multiplicative, a consistent estimate becomes a hard job. Numerical evidence suggests that the diffusivity is only weakly dependent upon the parameters of the swimmer: with $\omega$ and $\phi$ it varies less than $10\%$, with $L$ and $K$ it varies less than $50\%$. The fluid properties obviously have a much stronger influence on $D$, but such influence is trivial, in fact it is seen that $D\sim T/\eta$, as it is implied by the coefficient $\approx \sqrt{T_{ij}/\beta} \sim \sqrt{\frac{T}{\eta}}$ in front of the noise, see Eq.~\eqref{eq:langtotalintegration}.  In this Section we discuss a "zero order" approximation for $D$.

The diffusivity of the center of mass, $D$, is deduced from the knowledge of the asymptotic behavior of the mean squared displacement:
\begin{widetext}
\begin{equation}
\label{eq:msdgeneric}
    msd(t)=\big \langle \left[ x_{c_m}(t)-x_{c_m}(0) -\overline{\dot x}_{c_m}\,t \right]^2 \big \rangle = 
    \left \langle \left[ \int_0^t\,ds\,\Delta\dot x_{c_m}(s) \right]^2 \right \rangle \xrightarrow[t\rightarrow\infty]{} 2Dt\,,
\end{equation}
\end{widetext}


We follow the idea of the previous Sections and get a formal expansion (in powers of displacements $u_i$) for the equations of motion including noise 
\begin{equation*}
    \dot x_i = \mathbf{T}_{ij}f_j^{tot}= \underbrace{\mathbf{T}_{ij}^0\,f_j^{el}+\mathbf{T}_{ij}^0\,f_j^{act}+\mathbf{T}_{ij}^1\,f_j^{act}}_{\text{deterministic average velocity}} +
\end{equation*}
\begin{equation}
\label{eq:equationdotx_i}
    \underbrace{
    + \,\big( \mathbf{T}_{ij}^0+\mathbf{T}_{ij}^1 \big)\cdot\underbrace{\sqrt{\frac{2}{\beta}}\,\big( \Sigma_{jk}^0+\Sigma_{jk}^1 \big)\,\eta_k}_{f_j^{n}}
    }_{\Delta\dot x_{c_m}}\,,
\end{equation}
where
\begin{equation}
    \sqrt{\zeta} = \Sigma \sim \Sigma^0 + \Sigma^1 + \hdots\,,
\end{equation}
and leading to
\begin{equation}
\label{eq:msdcondelta}
    msd(t) = \left \langle \left[ \int_0^t\,ds\,\sum_{ijk}\frac{\mathbf{T}_{ij}^0+\mathbf{T}_{ij}^1}{3}\cdot \sqrt{\frac{2}{\beta}}\,\big( \Sigma_{jk}^0+\Sigma_{jk}^1 \big)\,\eta_k \right]^2 \right \rangle.
\end{equation}
The $1$st order terms are time-dependent and make the calculations much more complicate as they imply to integrate the solutions for $u(t)$ including their fluctuations. For this reason, we keep only the $0$-th order term:
\begin{equation*}
    \left \langle \left[ \int_0^t ds \frac{\sum_{ijk} \mathbf{T}_{ij}^0}{3}\cdot \sqrt{\frac{2}{\beta}} \Sigma_{jk}^0\,\eta_k \right]^2 \right \rangle =
\end{equation*}
\begin{equation*}
    = \frac{1}{9}\, \big \langle \int_0^t ds\int_0^t ds' \big( \sum_k \sqrt{2 D_k}\eta_k(s) \; \sum_l \sqrt{2 D_l}\eta_l(s')\big) \big \rangle \,,
\end{equation*}
with
\begin{equation}
    \langle \eta_k(s) \eta_l(s') \rangle = \delta_{kl}\delta(s-s'),
\end{equation}
and
\begin{equation}
    \sqrt{D_k}=\frac{1}{\sqrt{\beta}}\sum_{i}[\mathbf{T}^0\Sigma^0]_{ik}.
\end{equation}
In conclusion, we get
\begin{equation}
\label{eq:msdfinalexpr}
    msd(t) = \frac{1}{9}\sum_k \int_0^t ds\int_0^t ds' 2D_k\,\delta(s-s') = 2 Dt\,,
\end{equation}
with
\begin{equation}
\label{eq:Dexpr}
    D=\frac{1}{9} \sum_k D_k\,.
\end{equation}
In the case $l_1=l_2=L$ and in the limit $a \ll L$, one has $\Sigma^0 = \sqrt{6 \pi \eta a}\, \mathcal{I}$ and therefore $D_k=1/(6 \pi \eta a \beta)\, \mathcal{I}$, leading to 
\begin{equation}
    D=\frac{D_0}{3}
\end{equation}
with
\begin{equation}
    D_0=\frac{1}{6 \pi \eta a \beta},
\end{equation}
the bare diffusivity of a single sphere. With $\eta=1$ and $a=1$, this crude estimate gives $D \approx 0.018 T$ which agrees with the leading value observed in the numerical simulations with those parameters, see Figs.~\ref{fig:vsomega},~\ref{fig:vsdeltaphi},~\ref{fig:vsk},~\ref{fig:vsL}. 

Interestingly, simple calculations show that
\begin{equation}
      D \approx D_0\left[\frac{1}{3}+\frac{5}{6}\frac{a}{L}+O\left(\left(\frac{a}{L}\right)^2\right)\right].
\end{equation}
Therefore, at small but finite $a$ and $L$, the hydrodynamic couplings {\em increases} the diffusivity.

\section{Conclusions and perspectives}
\label{sec:concl}

We have studied numerically and analytically the three beads swimmer model with two important modifications: thermal noise and a confining potential. The latter is physically but also mathematically motivated, since the original model, if solved dynamically (i.e. imposing the driving forces and observing the resulting trajectory), is unstable: the three particles do not remain in a close neighborhood, which is necessary for hydrodynamic interactions to couple their dynamics and prevent reciprocal motion ie. time-reversibility and vanishing of the swimming effect. The introduction of noise makes it possible to measure the swimming precision, entering the Thermodynamic Uncertainty Relation. It has, however, no evident effects on the average dynamics of the swimmer, as it can be deduced by the graphs of the observables in temperature, i.e. Fig.~\ref{fig:vsT} and Fig.~\ref{fig:eff}d. The study also reveal that this model is usually far from the TUR optimal bound, but its precision efficiency can be improved by reducing the confining potential stiffness $K$ or the length at rest of the two swimmer arms, $L$. Future investigations should include a more general class of driving protocols, higher order terms in the perturbative expansion for the diffusivity, and in particular the introduction of noise in the active force, which is perhaps a much stronger - and yet realistic - source of fluctuations for micro-swimmers~\cite{maggi2022thermodynamic}.

\appendix

\section*{Appendix A: linear approximation}

Equation~\eqref{eq:equation_dot_u}, keeping only the linear order in $F_0$, casts into
\begin{equation}
   \underline{\dot u}(t)=\mathbf{M}_2 \cdot \underline{u}(t) + \mathbf{M}_3\cdot\underline{f}_{act}^{(2)}\,,
\end{equation}
which can be solved by setting

\begin{equation}
u_i=A_i \cos(\omega t)+B_i \sin(\omega t) \;\;\;\; (i=1,2)
\end{equation}
Lengthy computations leads to the following expressions for the four coefficients
\begin{align}
A_i &= \frac{\sum_{j=0}^3 A_i^{j} K^{j}(\eta\omega)^{3-j} }{\sum_{j=0}^2 A_{d,i}^{j} K^{2j} (\eta \omega)^{4-2j}}\\
B_i &= \frac{\sum_{j=0}^3 B_i^{j} K^{j}(\eta\omega)^{3-j} }{\sum_{j=0}^2 B_{d,i}^{j} K^{2j} (\eta \omega)^{4-2j}}
\end{align}
with
\begin{widetext}
    \begin{align} \nonumber
A_1^0&=432 \pi ^3 a_1 a_{12}^4 a_{2}^2 \sin(\Delta\phi) \;\;\;\;
A_1^1=-36 \pi ^2 a_{12}^2 a_{2} \left(a_{1}^2 (\cos(\Delta \phi)a_{2}+2 a_{12})+4 \cos(\Delta \phi)a_{12}^2 a_{2}+2 a_{1} a_{12} a_{2}\right)\\ \nonumber
A_1^2&=-12 \pi  a_{1} a_{12}^2 \sin(\Delta \phi)\left(a_{1} a_{2}-4 a_{12}^2\right) \;\;\;\;
A_1^3=-\cos(\Delta \phi)\left(a_{1} a_{2}-4 a_{12}^2\right)^2\\ \nonumber
A_{d,1}^0&=1296 \pi ^4 a_{1}^2 a_{12}^4 a_{2}^2 \;\;\;\;
A_{d,1}^1=72 \pi ^2 a_{12}^2 \left(a_{1}^2 \left(2 a_{12}^2+a_{2}^2\right)+2 a_{12}^2 a_{2}^2\right) \;\;\;\;
A_{d,1}^2=(-4 a12^2 + a1 a2)^2\\ \nonumber
B_1^0&=216 \pi ^3 a_{1} a_{12}^3 a_{2}^2 (2 \cos(\Delta \phi)a_{12}+a_{1}) \;\;\;\;
B_1^1=36 \pi ^2 a_{12}^2 a_{2}^2 \sin(\Delta \phi)\left(a_{1}^2+4 a_{12}^2\right)\\ \nonumber
B_1^2&=-6 \pi  a_{1} a_{12} (2 \cos(\Delta \phi)a_{12}-a_{2}) \left(a_{1} a_{2}-4 a_{12}^2\right) \;\;\;\;
B_1^3=\sin(\Delta \phi)\left(a_{1} a_{2}-4 a_{12}^2\right)^2\\ \nonumber
B_{d,1}^0&=1296 \pi ^4 a_{1}^2 a_{12}^4 a_{2}^2 \;\;\;\;
B_{d,1}^1=72 \pi ^2 a_{12}^2 \left(a_{1}^2 \left(2 a_{12}^2+a_{2}^2\right)+2 a_{12}^2 a_{2}^2\right) \;\;\;\;
B_{d,1}^2=\left(a_{1} a_{2}-4 a_{12}^2\right)^2\\ \nonumber
A_2^0&=-216 \pi ^3 a_{1}^2 a_{12}^3 a_{2}^2 b \;\;\;\;
A_2^1=36 \pi ^2 a_{1} a_{12}^2 \left(a_{2}^2 (2 \cos(\Delta \phi)a_{12}+a_{1})+2 \cos(\Delta \phi)a_{1} a_{12} a_{2}+4 a_{1} a_{12}^2\right)\\ \nonumber
A_2^2&=-6 \pi  a_{1} a_{12} a_{2} \sin(\Delta \phi)\left(a_{1} a_{2}-4 a_{12}^2\right) \;\;\;\;
A_2^3=\left(a_{1} a_{2}-4 a_{12}^2\right)^2\\ \nonumber
A_{d,2}^0&=1296 \pi ^4 a_{1}^2 a_{12}^4 a_{2}^2 \;\;\;\;
A_{d,2}^1=72 \pi ^2 a_{12}^2 \left(a_{1}^2 \left(2 a_{12}^2+a_{2}^2\right)+2 a_{12}^2 a_{2}^2\right) \;\;\;\;
A_{d,2}^2=\left(a_{1} a_{2}-4 a_{12}^2\right)^2\\ \nonumber
B_2^0&=-216 \pi ^3 a_{1}^2 a_{12}^3 a_{2} (\cos(\Delta \phi)a_{2}+2 a_{12}) \;\;\;\;
B_2^1=-72 \pi ^2 a_{1} a_{12}^3 a_{2} \sin(\Delta \phi)(a_{1}+a_{2})\\ \nonumber
B_2^2&=-6 \pi  a_{12} a_{2} (\cos(\Delta \phi)a_{1}-2 a_{12}) \left(a_{1} a_{2}-4 a_{12}^2\right) \;\;\;\;
B_2^3=0\\ \nonumber
B_{d,2}^0&=1296 \pi ^4 a_{1}^2 a_{12}^4 a_{2}^2 \;\;\;\;
B_{d,2}^1=72 \pi ^2 a_{12}^2 \left(a_{1}^2 \left(2 a_{12}^2+a_{2}^2\right)+2 a_{12}^2 a_{2}^2\right) \;\;\;\;
B_{d,2}^2=\left(a_{1} a_{2}-4 a_{12}^2\right)^2.
\end{align}
\end{widetext}

We conclude showing analytical expression for the average velocity
\begin{equation}
    v = \frac{\alpha}{2}\,\overline{\left(u_1\dot{u}_2\,-\,u_2\dot{u}_1 \right)}
\end{equation}
where
\begin{equation*}
    \overline{\left(u_1\dot{u}_2\,-\,u_2\dot{u}_1 \right)} = \frac{1}{\mathcal{T}}\int_{t_0}^{t_0+\mathcal{T}} \left(u_1\dot{u}_2\,-\,u_2\dot{u}_1 \right)\, dt =
\end{equation*}
\begin{equation*}
    =\,\frac{1}{\mathcal{T}}\int_{t_0}^{t_0+\mathcal{T}}\left(u_1\dot{u}_2- \underbrace{(u_2\dot{u}_1 + u_1\dot{u}_2)}_{\text{total derivative}} + u_1\dot{u}_2\,\right) \,dt =
\end{equation*}
\begin{equation}
\label{eq:avgu1dotu2}
    = \,\frac{1}{\mathcal{T}}\int_{t_0}^{t_0+\mathcal{T}}\, 2\,u_1\dot{u}_2\,dt \,.
\end{equation}

Through the above expressions one gets
\begin{widetext}
    \begin{equation}
    v=\frac{\alpha}{2}\,\frac{F_0^2 \omega \left(a_1 a_2-4 a_{12}^2\right) \left(\sin (\Delta \phi) \left(4 a_{12}^2 \left(9 \pi ^2 a_1 a_2 \eta^2 \omega^2+K^2\right)-a_1 a_2 K^2\right)+12 \pi  a_{12}^2 \eta K \omega (a_1-a_2) \cos (\Delta \phi)\right)}{1296 \pi ^4 a_1^2 a_{12}^4 a_2^2 \eta^4 \omega^4+72 \pi ^2 a_{12}^2 \eta^2 K^2 \omega^2 \left(a_1^2 \left(2 a_{12}^2+a_2^2\right)+2 a_{12}^2 a_2^2\right)+K^4 \left(a_1 a_2-4 a_{12}^2\right)^2}
\end{equation}
\end{widetext}
which, in terms of the original lengths at rest $l_1$ and $l_2$ takes the complicate form
\begin{equation}
    v=\frac{\alpha}{2}\,\frac{v_n}{v_d}
\end{equation}
with
\begin{widetext}
\begin{multline} 
v_n=F_0^2 \omega \left[3 a^2 \left(l_1^4-2 l_1^3 l_2-5 l_1^2 l_2^2-2 l_1 l_2^3+l_2^4\right)+4 a l_1 l_2 (l_1+l_2) \left(l_1^2+3 l_1 l_2+l_2^2\right)-4 l_1^2 l_2^2 (l_1+l_2)^2\right]  \times \\ \left\{\sin (\Delta \phi) \left[48 \pi ^2 a^2 \eta^2 l_1^2 l_2^2 \omega^2 (l_1+l_2)^2+K^2 \left(-3 a^2 \left(l_1^4-2 l_1^3 l_2-5 l_1^2 l_2^2-2 l_1 l_2^3+l_2^4\right)-4 a l_1 l_2 (l_1+l_2) \left(l_1^2+3 l_1 l_2+l_2^2\right)+ \right. \right. \right. \\ \left. \left. \left. 4 l_1^2 l_2^2 (l_1+l_2)^2\right)\right]+24 \pi  a^2 \eta K l_1 l_2 \omega (l_2-l_1) (l_1+l_2)^2 \cos (\Delta \phi)\right\} 
\end{multline}
and
\begin{multline}
    v_d=9 a^4 \left[256 \pi ^4 \eta^4 l_1^4 l_2^4 \omega^4 (l_1+l_2)^4+32 \pi ^2 \eta^2 K^2 l_1^2 l_2^2 \omega^2 (l_1+l_2)^2 \left(3 l_1^4+6 l_1^3 l_2+7 l_1^2 l_2^2+6 l_1 l_2^3+3 l_2^4\right)+ \right. \\ \left. K^4 \left(l_1^4-2 l_1^3 l_2-5 l_1^2 l_2^2-2 l_1 l_2^3+l_2^4\right)^2\right]+24 a^3 K^2 l_1 l_2 \left[K^2 (l_1+l_2) \left(l_1^2+3 l_1 l_2+l_2^2\right) \left(l_1^4-2 l_1^3 l_2-5 l_1^2 l_2^2-2 l_1 l_2^3+l_2^4\right)- \right. \\ \left. 16 \pi ^2 \eta^2 l_1^2 l_2^2 \omega^2 (l_1+l_2)^3 \left(3 l_1^2+5 l_1 l_2+3 l_2^2\right)\right]+8 a^2 K^2 l_1^2 l_2^2 (l_1+l_2)^2 \left[80 \pi ^2 \eta^2 l_1^2 l_2^2 \omega^2 (l_1+l_2)^2- \right.\\ \left.K^2 \left(l_1^4-18 l_1^3 l_2-37 l_1^2 l_2^2-18 l_1 l_2^3+l_2^4\right)\right]-32 a K^4 l_1^3 l_2^3 (l_1+l_2)^3 \left(l_1^2+3 l_1 l_2+l_2^2\right)+16 K^4 l_1^4 l_2^4 (l_1+l_2)^4.
\end{multline}
\end{widetext}
For instance if $l_1=l_2=L=\ell a$ , one has
\begin{widetext}
\begin{equation}
v=-\frac{\alpha}{2}\,\frac{F_0^2 (4 \ell-7) (4 \ell-3) \omega \sin (\Delta \phi) \left[192 \pi ^2 a^2 \eta^2 \ell^2 \omega^2+K^2 (4 \ell-7) (4 \ell-3)\right]}{[64 \pi ^2 a^2 \eta^2 \ell^2 \omega^2+K^2 (7-4 \ell)^2][576 \pi ^2 a^2 \eta^2 \ell^2 \omega^2+K^2 (3-4 \ell)^2]}
\end{equation}
\end{widetext}
We also report the expression for the total work rate, Eq.~\eqref{wrate}:
\begin{widetext}
    \begin{multline}
\dot{W}=    \frac{6 \pi  a_{12} \eta F_0^2 \omega^2 \left(a_1 a_2 \cos (\Delta \phi) \left(a_1 a_2 K^2-4 a_{12}^2 \left(K^2-9 \pi ^2 a_1 a_2 \eta^2 \omega^2\right)\right)+a_{12} (a_1+a_2) \left(4 a_{12}^2 \left(9 \pi ^2 a_1 a_2 \eta^2 \omega^2+K^2\right)-a_1 a_2 K^2\right)\right)}{1296 \pi ^4 a_1^2 a_{12}^4 a_2^2 \eta^4 \omega^4+72 \pi ^2 a_{12}^2 \eta^2 K^2 \omega^2 \left(a_1^2 \left(2 a_{12}^2+a_2^2\right)+2 a_{12}^2 a_2^2\right)+K^4 \left(a_1 a_2-4 a_{12}^2\right)^2}.
\end{multline}
\end{widetext}

\section{List of symbols}

\begin{table*}
\begin{tabular}{|l|l|}
\hline
$a$ & radius of the three spheres \\
$\eta$ & viscosity of the host fluid \\
$\bf r_i$ & position vector of $i$th sphere (with $i=1,2,3$) \\
$\bf v_i$ & velocity vector of $i$th sphere (with $i=1,2,3$) \\
$u_i$ & relative velocity between particle $i$ and particle $i+1$ (with $i=1,2$) \\
$\bf f_i$ & internal force vector \\
$\bf f_i^R$ & external (thermal) noise force vector \\
$H_{ij}(\bf r)$ & mobility tensor \\
$x_i$ & position along $x$ axis of the $i$th sphere (with $i=1,2,3$) \\
$x_{cm}$ & position of the center of mass of the swimmer \\
$L_i$ & distance between sphere $i$ and sphere $i+1$ (with $i=1,2$) \\
$\alpha$ & constant (with the dimensions of an inverse length) for the formula of average velocity, Eq.~\eqref{alphadef} \\
$K$ & spring constant \\
$l_i$ & length at rest of spring $i$ joining sphere  $i$ and sphere $i+1$ (with $i=1,2)$ \\
$u_i$ & deformation of spring $i$ joining sphere  $i$ and sphere $i+1$ (with $i=1,2)$ \\
$\bf T$, $T_{ij}$ & mobility coefficient coupling $x$ components of particles $i$ and $j$  \\
$\zeta$ & inverse of matrix $\bf T$ \\
$F_i$ & internal forces acting on particle $i$, components along $x$ \\
$F_i^{act}$ & internal forces acting on particle $i$ of active origin, components along $x$ \\
$F_i^{pot}$ & internal forces  acting on particle $i$ of conservative origin (potential), components along $x$ \\ 
$F_i^R$& noise forces  acting on particle $i$ of conservative origin (potential), components along $x$ \\
$\underline{F},\underline{F}^{act},\underline{F}^{pot},\underline{F}^R$ & lists of total, active, potential and noise forces \\
$v
$ & average speed of the center of mass of the swimmer \\
$\omega$ & angular frequency (pulsation) of the active force (or of the displacement in the old models) \\
$\mathcal{T}$ & period of the active force \\
$\phi_1,\phi_2$ & phase of the active force on sphere $i$ (or of its displacement in the old models) \\
$F_0$ & amplitude of the active force ($F_0=10$ in all numerical results)\\
${\bf D}$ & diffusivity matrix of thermal noise \\
$T$ & temperature \\
$\beta$ & inverse thermal energy \\
$F_{ito,i}$ & Ito "force" on sphere $i$ ($i=1,2,3)$ (dimensionally it is a velocity)\\
$W,\dot W$ & work done by the active forces (total and average rate) \\
$p,p_{max}$ & thermodynamic precision and its maximum according to the TUR \\
$D$, $D_0$ & coefficient of diffusion in the swimming direction and its bare value (thermal Einstein relation) \\
$e_L$, $e_{TUR}$ & swimmer efficiencies with respect an effective hydrodynamic force and to the TUR bound \\
$\nu=\omega/2 \pi$ & forcing frequency \\
$v_0 =  \omega a F_0^2/(L^2 K^2)$ & swimming velocity in the adiabatic limiti for small perturbation (linear theory)\\ 
$K_0 = F_0/a $ & typical stiffness related  \\
$\eta_0 = F_0/(6 \pi v_0 a)=(KL)^2/(6 \pi a^2 F_0 \omega) $ & typical viscosity \\
 $T_0 = F_0*a$ & typical thermal energy \\
 $\nu_h = F_0/(6 \pi \eta a^2)$ & hydrodynamic frequency ($\omega_h = 2 \pi \nu_h$)\\
 $v_{h} = 2 \pi \nu_h a F_0^2/(L^2 K^2)$ & swimming velocity associated to $\omega_h$\\
\hline 
\end{tabular}
\caption{List of symbols used in the text}
\end{table*}

\bibliographystyle{apsrev4-1}
\bibliography{biblioTurnew}

\end{document}